%% file: main.tex
\documentclass[journal,twoside,web]{ieeecolor}

\input{packages.tex}

\input{macros.tex}

\input{acronyms.tex}

\def\BibTeX{{\rm B\kern-.05em{\sc i\kern-.025em b}\kern-.08em
		T\kern-.1667em\lower.7ex\hbox{E}\kern-.125emX}}
\begin{document}
\title{\huge Distributionally Robust Trajectory Optimization Under Uncertain Dynamics via Relative Entropy Trust-Regions}
\author{Hany Abdulsamad{$^{*1}$}, Tim Dorau{$^{*1}$}, Boris Belousov{$^1$}, Jia-Jie Zhu{$^2$}, \IEEEmembership{Member, IEEE}, \\ and Jan Peters{$^{1}$}, \IEEEmembership{Fellow, IEEE}
	\thanks{\{$*$\} Equal Contribution. H. Abdulsamad, T. Dorau, B. Belousov, and J. Peters are with the Intelligent Autonomous Systems Lab, Technische Universität Darmstadt, Darmstadt, Germany. J.-J. Zhu is with the Data-driven Optimization and Control Group, Weierstrass Institute for Applied Analysis and Stochastics, Berlin, Germany. E-mail: hany@robot-learning.de.}}%

\maketitle

\begin{abstract}
	Trajectory optimization and model predictive control are essential techniques underpinning advanced robotic applications, ranging from autonomous driving to full-body humanoid control. State-of-the-art algorithms have focused on data-driven approaches that infer the system dynamics online and incorporate posterior uncertainty during planning and control. Despite their success, such approaches are still susceptible to catastrophic errors that may arise due to statistical learning biases, unmodeled disturbances, or even directed adversarial attacks. In this paper, we tackle the problem of dynamics mismatch and propose a distributionally robust optimal control formulation that alternates between two relative entropy trust-region optimization problems. Our method finds the worst-case maximum entropy Gaussian posterior over the dynamics parameters and the corresponding robust policy. Furthermore, we show that our approach admits a closed-form backward-pass for a certain class of systems. Finally, we demonstrate the resulting robustness on linear and nonlinear numerical examples.
\end{abstract}


\section{Introduction}
\label{sec:intro}
Trajectory optimization~\cite{mayne1966second} is a well-established tool for solving control problems that rely on a model of the system dynamics to optimize a control signal that induces a desired system behavior. However, as systems of interest are getting more complex, involving nonlinear effects and high-dimensional state-action spaces, accurate analytical modeling has become challenging, if not impossible, in some cases. Consequently, data-driven control approaches that employ black- and gray-box statistical models are becoming popular rapidly. However, trajectory optimization may exploit model imperfections that can arise as a result of statistical learning biases, resulting in brittle controllers that may fail at deployment time due to modeling discrepancies. The recent trend of over-reliance on learning from simulated data carries with it similar pitfalls w.r.t. optimization bias.

Advanced trajectory optimization and model predictive control (MPC) techniques successfully use learned probabilistic models to incorporate the uncertainty of data-driven learning ~\cite{kamthe2018data, hewing2018cautious}. However, those approaches are not robust against adversarial disturbances and general model mismatch. An alternative to such probabilistic modeling is the robust control paradigm. Unfortunately, robust methods tend to produce sub-optimal controllers on average because the disturbance model gives too much power to the adversary, forcing the controller to be too conservative. A new class of methods at the intersection of robust and stochastic optimal control is gaining momentum, based on distributionally robust optimization (DRO)~\cite{scarf1958min, delage2010distributionally, van2015distributionally, coulson2019regularized, yang2020wasserstein, zhu20worst, coppens2020data}, as it promises to combine the strengths of both approaches. In the DRO framework, one seeks to find a controller that performs optimally under a worst-case stochastic model chosen from a so-called ambiguity set. Here, the adversary's strength is more limited compared to the classical robust control.

To make distributionally robust optimization practical, one needs to be able to infer the optimal adversary, i.e., find the worst-case stochastic system, in closed form or numerically. So far, closed-form solutions have been obtained only for special choices of the ambiguity set~\cite{rahimian2019distributionally}. For example, when the ambiguity set is given as a relative entropy ball around a nominal distribution~\cite{hu2013kullback, charalambous2007stochastic}. In this paper, we build upon this insight to develop an algorithm for distributionally robust trajectory optimization.

We consider the problem of controlling a discrete-time stochastic dynamical system with transition density $f(\vec{x}\p | \vec{x}, \vec{u}, \vec{\theta})$ for which system uncertainty is encoded in the parameter distribution $p(\vec{\theta})$ and the ambiguity set is given by the KL ball $\mathcal{B}_\delta(\hat{p}) = \{p \,| \kl (p \,||\, \hat{p}) \leq \delta \}$ centered around the nominal parameter distribution $\hat{p}(\vec{\theta})$, that we assume is available after a data-driven model learning phase. We seek a time-varying stochastic policy $\pi_{t}(\vec{u} | \vec{x})$ that minimizes the worst-case expected cost $\max_{p \in \mathcal{B}_\delta(\hat{p})} \, J(\pi, p)$. In this setting, we develop an iterative trust region algorithm that alternates between optimizing the worst-case distribution $p$ and the corresponding distributionally robust policy $\pi$. We derive optimality conditions for $p$ and $\pi$ and an efficient forward-backward procedure in the style of differential dynamic programming is provided for each optimization step. The resulting method applies to nonlinear systems via iterative local linearization. Empirical validation on uncertain linear and nonlinear dynamical systems demonstrates the robustness of the optimized policies against adversarial disturbances. 

Our approach brings together several strands of research. First, we rely on distributionally robust optimization to find the worst-case parameter distribution. Second, our problem formulation is based on an iterative scheme of relative entropy policy search, a trust-region algorithm for policy optimization~\cite{peters2010relative}. Third, we employ iterative linearization and approximate integration to enable applications to nonlinear uncertain dynamical systems. Below, we highlight related work from these areas and point out key differences.

\section{Related Work}
Distributionally robust optimization finds numerous applications in control. However, methods differ in ambiguity set representations, uncertainty, system modeling assumptions, and optimization algorithms. For example, in \cite{van2015distributionally}, the problem of controlling a linear system under distributionally robust chance-constraints was tackled using a moment-based ambiguity set.
The moment-based representation was also employed in~\cite{coppens2020data} to derive high-probability guarantees for the stability of a linear system with multiplicative noise. For Wasserstein-based ambiguity sets, a general formulation was given in~\cite{yang2020wasserstein}, which, however, requires solving a semi-infinite problem numerically while we obtain a closed-form solution instead. Furthermore, a Wasserstein-based ambiguity set on the noise distribution was used to solve a data-enabled control problem in~\cite{coulson2019regularized}. For the linear-quadratic case, a relaxed version was solved in~\cite{kim2020minimax} via a modified Riccati equation. The model, however, only included ambiguity over the additive noise while the rest of the dynamics were assumed known and time-invariant. In this paper, in contrast, we employ a time-varying linearized probabilistic dynamics model and allow ambiguity in the distribution over all model parameters.

Ambiguity sets based on relative entropy were also employed in several prior works. Distributionally robust optimization of stochastic nonlinear partially observable systems with relative entropy constraints was studied in a general abstract setting in~\cite{charalambous2007stochastic}, where the ambiguity set is formulated on the space of path measures. In contrast, we consider ambiguity sets on the space of parameter distributions of the underlying dynamical system. In~\cite{petersen2000minimax}, a formulation for nonlinear systems was presented where the uncertainty in the additive disturbance was ambiguous. Through Lagrangian duality, a connection to risk-sensitive control was furthermore established. This connection was recently used to derive a model predictive control algorithm~\cite{nishimura2021rat} that builds upon the iterative linear exponential quadratic Gaussian method (iLEQG)~\cite{farshidian2015risk}. That approach employs cross-entropy methods (CEM) to optimize the risk-sensitivity parameter and obtain the worst-case distribution, as opposed to our algorithm, which uses a principled approach based on our trust-region optimization to solve the resulting DRO problem. Finally, our DRO formulation captures the uncertainty in the whole trajectory rather than considering only the noise distribution. This feature gives the adversary additional freedom in allocating the disturbance budget along the trajectory to critical time-steps.

Our approach builds upon relative entropy policy search (REPS)~\cite{peters2010relative} by extending it with an adversarial optimization with respect to the ambiguous distribution over the dynamics parameters. A risk-sensitive formulation based on REPS involving the entropic risk measure in a model-free setting was considered in~\cite{nass2019entropic}. In this paper, we instead work in a model-based setting \cite{levine2013guided} and optimize a controller under the worst-case distribution instead of the nominal one. Furthermore, we introduce an additional approximate state propagation step to accommodate for uncertainty in the dynamics parameters.

Stochastic optimal control with linearized dynamics \cite{mayne1966second, todorov2005generalized} is a powerful technique for controlling nonlinear systems. A locally optimal controller can be found via dynamic programming techniques by using first-order approximations of the dynamics along a given trajectory. The updated controller is used to generate a new reference trajectory, and the process is iterated until convergence.

Since the linearized dynamics is only a good approximation around the linearization point, it is important to limit the optimism in the controller updates. One successful approach has been to enforce a relative entropy bound between the trajectory distributions of successive iterations. This technique is used for trajectory optimization in the context of guided policy search (GPS)~\cite{levine2013guided} and is dubbed maximum-entropy iterative linear quadratic Gaussian. Our approach relies on a similar formulation to optimize a time-variant policy while considering uncertain dynamics.

In contrast to the aforementioned works, a distributionally robust approach not only considers the stochasticity captured by the probabilistic model but also accounts for what is typically called \textit{ambiguity}, meaning uncertainty about the probabilistic model itself.

\section{Distributional Robustness}
Robustness analysis studies the sensitivity of an optimization objective
\begin{equation*}
	\min_{\vec{x} \in \mathcal{X}} \max_{\vec{\theta} \in \Theta} ~ J(\vec{x}, \vec{\theta})
\end{equation*}
with a decision variable $\vec{x} \in \mathcal{X}$ w.r.t. to a parameter set $\vec{\theta} \in \mat{\Theta}$. The solution $\vec{x}^{*}$ is a conservative point that minimizes the worst-case objective w.r.t. $\vec{\theta}$ and delivers an upper-bound on the objective $J$ \cite{rahimian2019distributionally}. Furthermore, robust optimization assumes all parameters $\vec{\theta} \in \Theta$ are equally probable.

In contrast, stochastic optimization assumes the parameters $\vec{\theta}$ are random variables drawn from a known distribution $p(\vec{\theta})$. Thus, the objective $J$ can be minimized under a distributional risk measure or an expectation operator for example
\begin{equation*}
	\min_{\vec{x} \in \mathcal{X}} ~ \mathbb{E}_{p(\vec{\theta})} \left[ J(\vec{x}, \vec{\theta})  \right].
\end{equation*}
Distributionally robust optimization is a paradigm that combines the concepts of worst-case solutions and distributional uncertainty in one general framework. As in stochastic optimization the parameters $\vec{\theta}$ are assumed to be random variables, however, the knowledge about the distribution $p(\vec{\theta})$ is uncertain. An example of a distributionally robust optimization can be written as
\begin{equation*}
	\min_{\vec{x} \in \mathcal{X}} \max_{p \in \mathcal{P}} ~ \mathbb{E}_{p(\vec{\theta})} \left[J(\vec{x}, \vec{\theta})  \right],
\end{equation*}
where $\mathcal{P}$ is a set over distributions, commonly referred to as the \emph{ambiguity set}, and contains the worst-case distribution $p^{*}(\vec{\theta})$ that upper-bounds the expected loss. The generality of this formulation becomes evident when we consider two different scenarios. In one scenario, the set $\mathcal{P}$ may contain a single distribution, which recovers the stochastic optimization problem. In another, $\mathcal{P}$ may contain all possible distributions with a support on $\vec{\theta}$, thus delivering the classical robust optimization formulation.

The motivation behind distributional robustness considerations pertains to data-driven stochastic learning applications, where the distribution $p(\vec{\theta})$ is hard to estimate due to limited data. Defining an ambiguity set and optimizing for the worst case is a robust approach to combat statistical learning biases and avoid catastrophic results.

Finally, the choice of the ambiguity set remains an important decision, as it directly influences the overall solution and its usefulness. Sets that are very broadly defined can lead to over-powered biases that cripple the optimization, while a restrictive set definition can undermine the robustness objective. Related literature includes a wide spectrum of possible definitions \cite{rahimian2019distributionally}. We focus on ambiguity sets defined using discrepancy measures w.r.t. a nominal distribution $\hat{p}(\vec{\theta})$
\begin{equation*}
	\mathcal{P} := \mathcal{B}_\delta(\hat{p}) = \{p \, | D(p, \hat{p}) \leq \delta \},
\end{equation*}
where $D$ is a measure of the distance or divergence between an arbitrary distribution $p$ and the reference $\hat{p}$. The parameter $\delta$ is the radius around $\hat{p}(\vec{\theta})$, which effectively bounds the worst-case scenario or the severity of the worst possible case. More specifically, in this paper, we rely on the Kullback-Leibler divergence as a measure due to its tractable computational properties and its compatibility with trust-region stochastic optimization.

\section{Problem Statement}
\label{sec:problem}
In the scope of this paper, we concentrate on finite-horizon Markov decision processes (MDP) with a state space $\mathcal{X} \subseteq \mathds{R}^{d}$, an action space $\mathcal{U} \subseteq \mathds{R}^{m}$, and a time horizon $T$. We assume a probabilistic state transition density $p(\vec{x}\p, \vec{\theta}| \vec{x}, \vec{u}) = f(\vec{x}\p | \vec{x}, \vec{u}, \vec{\theta}) p(\vec{\theta})$, where $p(\vec{\theta})$ is a distribution over the dynamics parameters. The policy $\pi_{t}(\vec{u} | \vec{x})$, a time-variant conditional density, induces the state distribution $\mu_{t}(\vec{x})$ according to transition dynamics.

In this setting, the stochastic optimal control objective can be written as
\begin{equation}
	\begin{aligned}[t]
		J(\pi_{t}, p) = & \sum_{t=1}^{T-1} \iint c_{t}(\vec{x}, \vec{u}) \mu_{t}(\vec{x}) \pi_{t}(\vec{u} | \vec{x}) \dif \vec{x} \dif \vec{u} \\
		& + \int c_{T}(\vec{x}) \mu_{T} (\vec{x})  \dif \vec{x},
	\end{aligned}
\end{equation}
where $c(\vec{x}, \vec{u})$ is the cost function. This objective is constrained by the following integral equation describing the evolution of $\mu_{t}(\vec{x})$ over time
\begin{equation*}
	\mu_{t+1}(\vec{x}\p) = \iiint \mu_{t}(\vec{x}) \pi_{t}(\vec{u} | \vec{x}) f(\vec{x}\p | \vec{x}, \vec{u}, \vec{\theta}) p(\vec{\theta}) \dif \vec{u} \dif \vec{x} \dif \vec{\theta}.
\end{equation*}

The distributionally robust trajectory optimization can then be written as a minimax problem over the distributions $\pi_{t}(\vec{u} | \vec{x})$ and $p(\vec{\theta})$
\begin{subequations}\label{eq:rsoc}
	\begin{alignat}{3}
		& \minimize_{\pi_{t}} & \quad & J(\pi_{t}, p^{*}), \\
		& \st                 & \quad & \int \pi_{t}(\vec{u} | \vec{x}) \dif \vec{u} = 1, \quad \forall \vec{x}, \forall t < T,
	\end{alignat}
\end{subequations}
where $p^{*}(\vec{\theta})$ is the worst-case distribution given by
\begin{subequations}
	\begin{alignat}{3}
		p^{*} := & \argmax_{p} & \quad & J(\pi_{t}, p), \\
		& \st         & \quad & \kl (p(\vec{\theta})  \,||\, \hat{p}(\vec{\theta})) \le \delta, \\
		&             & \quad & \int p(\vec{\theta}) \dif \vec{\theta} = 1,
	\end{alignat}
\end{subequations}
where $\hat{p}(\vec{\theta})$ is the nominal parameter distribution and $\delta$ controls the size of the corresponding KL-based distributional ambiguity set.

The former robust optimization problem is typically hard to solve for general nonlinear dynamical systems $p(\vec{x}\p | \vec{x}, \vec{u}, \vec{\theta})$ and arbitrary forms of $p(\vec{\theta})$. Therefore, our approach is to solve the nested optimization via an iterative sequential programming approach which is regularized using an additional trust-region imposed on the outer policy optimization problem in Equation~\eqref{eq:rsoc}.

\SetEndCharOfAlgoLine{}
\RestyleAlgo{boxruled}
\begin{algorithm}[!t]
	\SetKwInput{Input}{input~}
	\SetKwInput{Output}{output~}
	\SetKwInput{Initialize}{initialize~}
	\SetKwRepeat{Do}{do}{while}
	
	\Input{$\hat{\mu}_{1}, c_{t}, f, \hat{p}, \delta, \varepsilon, K$}
	\Initialize{$\pi^{1}_{t}$}
	\For{$k \gets1$ \KwTo $K$}
	{
		$p^{k+1} \leftarrow$\bf{worstCaseParamDist}$(\hat{p}, \pi^{k}_{t}, \hat{\mu}_{1}, c_{t}, f, \delta)$ \\
		$\pi^{k+1}_{t} \leftarrow$\bf{robustPolicyUpdate}$(\pi^{k}_{t}, p^{k+1}, \hat{\mu}_{1}, c_{t}, f, \varepsilon)$
	}
	$p^* \leftarrow p^{K+1}$, \quad $\pi^{*}_{t} \leftarrow \pi^{K+1}_{t}$ \\
	
	\Output{$\pi^{*}_{t}, p^{*}$}
	\caption{Dist. Robust Trajectory Optimization}
	\label{alg:global}
\end{algorithm}

\section{Trust-Region Distributionally \\ Robust Control}
Introducing a trust region over $\pi_{t}$ has multiple advantages. On the one hand, it regularizes the policy optimization step, which is crucial for the convergence of the overall minimax problem. On the other hand, it offers a tractable maximum-entropy stochastic optimal control framework for dealing with nonlinear dynamics through successive linearization around a local trajectory distribution \cite{levine2013guided, arenz2016optimal}.

The resulting overall approach alternates between updating the parameter and policy distribution. For every iteration $k$, we compute the updated worst-case distribution $p^{k+1}$ given the ambiguity set $\mathcal{B}_\delta$ around the nominal $\hat{p}$, and policy $\pi^{k}_{t}$
\begin{equation}
	p^{k+1} = \argmax_{p \in \mathcal{B}_\delta(\hat{p})} \quad J(\pi_{t}^{k}, p), \label{eq:worst_param}
\end{equation}
then we compute the updated robust policy $\pi^{k+1}_{t}$ under $p^{k+1}$ in a trust-region $\mathcal{B}_\varepsilon$ around the old policy $\pi_{t}^{k}$
\begin{equation}
	\pi_{t}^{k+1} =  \argmin_{\pi_{t} \in \mathcal{B}_\epsilon(\pi_{t}^{k})} \quad J(\pi_{k}, p^{k+1}) \label{eq:worst_policy}.
\end{equation}
These steps can also be seen as trust-region versions of the proximal updates performed by the mirror descent algorithm~\cite{beck2003mirror}.
Algorithm~\ref{alg:global} offers a schematic view of the optimization. The following sections provide further details.


\SetEndCharOfAlgoLine{}
\RestyleAlgo{boxruled}
\begin{algorithm}[!t]
	\SetKwInput{Input}{input~}
	\SetKwInput{Output}{output~}
	\SetKwInput{Initialize}{initialize~}
	\SetKwRepeat{Do}{do}{while}
	
	\Input{$\hat{p}, \pi^{k}_{t}, \hat{\mu}_{1}, c_{t}, f, \delta$}
	\Initialize{$\beta$}
	
	$q_{t} \leftarrow${paramForPass}$(\hat{p}, \pi^{k}_{t}, f, \hat{\mu}_{1})$
	
	\While{$F$ \normalfont is not at minimum}
	{
		\Repeat{$\kl (q_{t} \,||\, \mu_{t}) \approxeq 0$}
		{
			$p^{k+1}_{t}, V^{\theta}_{t} \leftarrow$\bf{paramBackPass}$(q_{t}, \pi^{k}_{t}, c_{t}, f, \beta)$ \\
			$\mu_{t} \leftarrow$\bf{paramForPass}$(p^{k+1}_{t}, \pi^{k}_{t}, f, \hat{\mu}_{1})$ \\
			$q_{t} \leftarrow \lambda \mu_{t} + (1 - \lambda) q_{t}$ \\
		}
		${\displaystyle \frac{\partial F}{\partial \beta}} \leftarrow$\bf{computeBetaGradient}$(p^{k+1}_{t}, \hat{p}, \delta)$ \\
		$\beta \leftarrow \beta - \eta {\displaystyle \frac{\partial F}{\partial \beta}}$ \\
	}
	\Output{$p^{k+1}_{t}, \mu_{t}$}
	\caption{Worst-Case Parameter Optimization}
	\label{alg:parameter_step}
\end{algorithm}

\subsection{Worst-Case Parameter Distribution}\label{sec:param}
The parameter distribution optimization \eqref{eq:worst_param} for a single iteration $k$ is given by
\begin{subequations}
	\begin{align}
		& \maximize_{p_{t}} &  &
		\begin{aligned}[t]
			& \sum_{t=1}^{T-1} \iint c_{t}(\vec{x}, \vec{u}) \, \mu_{t}(\vec{x}) \, \pi_{t}^{k}(\vec{u} | \vec{x}) \dif \vec{x} \dif \vec{u} \\
			& + \int c_{T}(\vec{x}) \, \mu_{T}(\vec{x}) \dif \vec{x},
		\end{aligned} \\
		& \st               &  &
		\begin{aligned}[t] \label{eq:worst_forward}
			\iiint & \mu_{t}(\vec{x}) \, \pi_{t}^{k}(\vec{u} | \vec{x}) \, f(\vec{x}\p | \vec{x}, \vec{u}, \vec{\theta}) \\
			& \times p^{k+1}_{t}(\vec{\theta}) \dif \vec{u}  \dif \vec{x}  \dif \vec{\theta} = \mu_{t+1}(\vec{x}\p),
		\end{aligned} \\
		&                   &  & \sum_{t=1}^{T-1} \kl (p^{k+1}_{t}(\vec{\theta}) \,||\, \hat{p}(\vec{\theta})) \le \delta, \label{eq:worst_kl} \\
		&                   &  & \int p^{k+1}_{t}(\vec{\theta}) \dif \vec{\theta} = 1, \quad \mu_{1}(\vec{x}) = \hat{\mu}_{1}(\vec{x}).
	\end{align}
\end{subequations}
Notice that we have moved to a time-variant worst-case parameter distribution $p_{t}(\vec{\theta})$. Although this formulation is more general, it is crucial to make this assumption in order to disentangle the adversary's influence over time and restrict it to future time steps. This modification makes sense when considering that the robust policy $\pi_{t}(\vec{u} | \vec{x})$ is likewise time-variant and only influences the current and future time steps.

By solving the former primal problem using the method of Lagrangian multipliers \cite{boyd2004convex}, we arrive at the optimal worst-case parameter distribution $p_{t}^{k+1}$
\begin{equation}\label{eq:worst_dist}
	p_{t}^{k+1}(\vec{\theta}) \propto \hat{p}(\vec{\theta}) \exp \left[ - \frac{1}{\beta} W_{t}(\vec{\theta)} \right],
\end{equation}
a softmax distribution with a temperature $\beta \leq 0$ that corresponds to the trust-region constraint in Equation~\eqref{eq:worst_kl} and a parameter value function $W_{t}(\vec{\theta})$
\begin{equation*}
	W_{t} = \iiint V^{\theta}_{t+1}(\vec{x}\p) \, \mu_{t}(\vec{x}) \, \pi^{k}_{t}(\vec{u} | \vec{x}) \, f(\vec{x}\p | \vec{x}, \vec{u}, \vec{\theta}) \dif \vec{u} \dif \vec{x} \dif \vec{x}\p,
\end{equation*}
where $V^{\theta}_{t+1}(\vec{x}\p)$ is the Lagrangian function associated with Equation~\eqref{eq:worst_forward} and acts as an adversarial state-value function under the last policy $\pi_{t}^{k}(\vec{u} | \vec{x})$.

By plugging the solution in Equation~\eqref{eq:worst_dist} back into the primal problem, we retrieve the dual $F$ as a function of $\mu$, $V^{\theta}$ and $\beta$
\begin{equation*}
	F =
	\begin{aligned}[t]
		& \sum_{t=1}^{T-1} \iint c_{t}(\vec{x},\vec{u}) \mu_{t}(\vec{x}) \pi^{k}_{t}(\vec{u} | \vec{x}) \dif \vec{u} \dif \vec{x} \\
		& + \int c_{T}(\vec{x}) \mu_{T}(\vec{x}) \dif \vec{x} + \int V_{1}^{\theta}(\vec{x}) \hat{\mu}_{1}(\vec{x}) \dif \vec{x} \\
		& - \sum_{t=1}^{T-1} \int V_{t}^{\theta}(\vec{x}\p) \mu_{t}(\vec{x}\p) \dif \vec{x}\p - \int V_{T}^{\theta}(\vec{x}\p) \mu_{T}(\vec{x}\p) \dif \vec{x}\p - \beta \delta \\
		& - \beta \sum_{t=1}^{T-1} \log \int \hat{p}(\vec{\theta}) \exp \left[ - \frac{1}{\beta} W_{t+1}(\vec{\theta}) \right] \dif \vec{\theta}.
	\end{aligned}
\end{equation*}

We set the partial derivative of the dual w.r.t.  $\mu_{t}(\vec{x})$ to zero and get a backward recursion for computing $V^{\theta}_{t}(\vec{x})$
\begin{align*}
	V^{\theta}_{t}(\vec{x}) & =
	\begin{aligned}[t]
		\iiint & V^{\theta}_{t+1}(\vec{x}\p) \pi^{k}_{t}(\vec{u} | \vec{x}) f(\vec{x}\p | \vec{x}, \vec{u}, \vec{\theta}) \\
		& \times p_{t}^{k+1}(\vec{\theta}) \dif \vec{\theta} \dif \vec{u} \dif \vec{x}\p \! + \! \int \! c_{t}(\vec{x}, \vec{u}) \pi^{k}_{t}(\vec{u} | \vec{x}) \dif \vec{u} \\
	\end{aligned}
\end{align*}
where $V_{T}(\vec{x}) = c_{T}(\vec{x})$. Similarly, setting the partial derivative of the dual w.r.t. $V^{\theta}$ to zero delivers a forward recursion for $\mu_{t}(\vec{x})$
\begin{equation*}
	\mu_{t+1} \! = \! \iiint \! \mu_{t}(\vec{x}) \, \pi_{t}^{k}(\vec{u} | \vec{x}) \, f(\vec{x}\p | \vec{x}, \vec{u}, \vec{\theta}) \, p^{k+1}_{t}(\vec{\theta})  \dif \vec{u}  \dif \vec{x} \dif \vec{\theta},
\end{equation*}
where the initial state distribution $\mu_{1}(\vec{x}) = \hat{\mu}_{1}(\vec{x})$ which is assumed given.

Finally, the optimal temperature $\beta$ that satisfies the trust-region in Equation~\eqref{eq:worst_kl} is optimized numerically via gradient descent on the dual where
\begin{equation*}
	\beta^{i+1} = \beta^{i} - \eta_{i} \sum_{t=1}^{T-1} \kl (p^{k+1}_{t}(\vec{\theta}) \,||\, \hat{p}(\vec{\theta})) + \eta_{i} \delta,
\end{equation*}
and $\eta_{i}$ is some step size. This process iterates over $\mu$, $V^{\theta}$ and $\beta$ until convergence, see Algorithm~\ref{alg:parameter_step}. Given the circular dependency between $V^{\theta}$, $\mu$ and $p$, we update $\mu$ through a barycentric interpolation scheme, analogous to \cite{abdulsamad2017state}.

\subsection{Worst-Case Robust Policy}\label{sec:policy}
Imposing a trust-region constraint on the robust stochastic optimal control formulation in \eqref{eq:rsoc} results in the following optimization problem
\begin{subequations}
	\begin{align}
		& \minimize_{\pi_{t}} &  &
		\begin{aligned}[t]
			& \sum_{t=1}^{T-1} \iint c_{t}(\vec{x}, \vec{u}) \, \mu_{t}(\vec{x}) \pi^{k+1}_{t}(\vec{u} | \vec{x}) \dif \vec{x} \dif \vec{u} \\
			& + \int c_{T}(\vec{x}) \mu_{T}(\vec{x}) \dif \vec{x},
		\end{aligned} \\
		& \st                 &  &
		\begin{aligned}[t] \label{eq:policy_forward}
			\iiint & \mu_{t}(\vec{x}) \, \pi^{k+1}_{t}(\vec{u} | \vec{x}) \, f(\vec{x}\p | \vec{x}, \vec{u}, \vec{\theta}) \\
			& \times p^{k+1}_{t}(\vec{\theta}) \dif \vec{u} \dif \vec{x} \dif \vec{\theta} = \mu_{t+1}(\vec{x}\p),
		\end{aligned} \\
		&                     &  & \sum_{t=1}^{T-1} \int \mu_{t}(\vec{x}) \kl (\pi^{k+1}_{t} \,||\, \pi^{k}_{t}) \dif \vec{x} \le \varepsilon, \label{eq:policy_kl} \\
		&                     &  & \int \pi^{k+1}_{t}(\vec{u} | \vec{x}) \dif \vec{u} = 1, \quad \mu_{1}(\vec{x}) = \hat{\mu}_{1}(\vec{x}).
	\end{align}
\end{subequations}

\SetEndCharOfAlgoLine{}
\RestyleAlgo{boxruled}
\begin{algorithm}[t!]
	\SetKwInput{Input}{input~}
	\SetKwInput{Output}{output~}
	\SetKwInput{Initialize}{initialize~}
	\SetKwRepeat{Do}{do}{while}
	
	\Input{$\pi^{k}_{t}, p_{t}^{k+1}, \hat{\mu}_{1}, c_{t}, f, \varepsilon$}
	\Initialize{$\alpha$}
	
	\While{$G$ \normalfont is not at maximum}
	{
		$\pi_{t}^{k+1}, V_{t}^{\pi} \leftarrow$\bf{policyBackPass}$(p^{k+1}, c_{t}, f, \alpha)$ \\
		$\mu_{t} \leftarrow$\bf{policyForPass}$(p_{t}^{k+1}, \pi_{t}^{k+1}, f, \hat{\mu}_{1})$ \\
		${\displaystyle \frac{\partial G}{\partial \alpha}} \leftarrow$\bf{computeAlphaGradient}$(\pi_{t}^{k+1}, \pi_{t}^{k}, \varepsilon, \mu_{t})$ \\
		$\alpha \leftarrow \alpha + \rho {\displaystyle \frac{\partial G}{\partial \alpha}}$ \\
	}
	\Output{$\pi^{k+1}_{t}, \mu_{t}$}
	\caption{Dist. Robust Policy Optimization}
	\label{alg:policy_step}
\end{algorithm}

By formulating the Lagrangian and solving for the robust policy $\pi_{t}^{k+1}$, we find
\begin{equation}
	\pi_{t}^{k+1}(\vec{u} | \vec{x}) \propto \pi^{k}_{t}(\vec{u} | \vec{x}) \exp \left[ -\frac{1}{\alpha} Q_{t}(\vec{x}, \vec{u}) \right], \label{eq:policy_dist}
\end{equation}
where $Q_{t}(\vec{x}, \vec{u})$ is the state-action value function
\begin{equation*}
	Q_{t}= c_{t}(\vec{x}, \vec{u}) + \iint V^{\pi}_{t+1}(\vec{x}\p) f(\vec{x}\p | \vec{x}, \vec{u}, \vec{\theta}) p^{k+1}_{t}(\vec{\theta}) \dif \vec{\theta} \dif \vec{x}\p.
\end{equation*}
The temperature parameter $\alpha \geq 0$ and function $V^{\pi}_{t}(\vec{x})$ are the Lagrangian variables associated with Equation~\eqref{eq:policy_kl} and Equation~\eqref{eq:policy_forward}.

Substituting Equation~\eqref{eq:policy_dist} back into the primal delivers the policy dual function $G$
\begin{equation*}
	\begin{aligned}
		G = & \int c_{T}(\vec{x}) \mu_{T}(\vec{x}) \dif \vec{x} + \int V^{\pi}_{1}(\vec{x}) \hat{\mu}_{1}(\vec{x}) \dif \vec{x} \\
		& - \sum_{t=1}^{T-1} \int V^{\pi}_{t}(\vec{x}\p) \mu_{t}(\vec{x}\p) \dif \vec{x}\p - \int V^{\pi}_{T}(\vec{x}\p) \mu_{T}(\vec{x}\p) \dif \vec{x}\p - \alpha \varepsilon \\
		& - \alpha \sum_{t=1}^{T-1} \pi_{t}^{k}(\vec{u} | \vec{x}) \exp \left[ - \frac{1}{\alpha} Q_{t+1}^{\pi}(\vec{x}, \vec{u}) \right] \dif \vec{u} \dif \vec{x}.
	\end{aligned}
\end{equation*}

By setting the derivatives of $G$ w.r.t. $\mu_{t}(\vec{x})$ to zero, we arrive at an optimality condition in form of a backward recursion for calculating $V^{\pi}_{t}(\vec{x})$
\begin{equation}
	V^{\pi}_{t}(\vec{x}) = \alpha \log \int \pi^{k}_{t}(\vec{u} | \vec{x}) \exp \left[ - \frac{1}{\alpha} Q_{t}(\vec{x}, \vec{u}) \right] \dif \vec{u},
\end{equation}
where $V^{\pi}_{T} = c_{T}(\vec{x})$. On the other hand, the derivatives of $G$ w.r.t. $V^{\pi}(\vec{x})$ lead to a forward recursion for $\mu_{t}(\vec{x})$ that fulfills the propagation constraint
\begin{equation*}
	\mu_{t+1} \! = \! \iiint \! \mu_{t}(\vec{x}) \, \pi^{k+1}_{t}(\vec{u} | \vec{x}) \, f(\vec{x}\p | \vec{x}, \vec{u}, \vec{\theta}) \, p^{k+1}_{t}(\vec{\theta}) \dif \vec{u} \dif \vec{x} \dif \vec{\theta}.
\end{equation*}
Similar to the optimization in Section~\ref{sec:param}, the temperature $\alpha$ is optimized via gradient ascent on the policy dual with
\begin{equation*}
	\alpha^{i+1} = \alpha^{i} + \rho_{i} \sum_{t=1}^{T-1} \int \mu_{t}(\vec{x}) \kl (\pi^{k+1}_{t} \,||\, \pi^{k}_{t}) \dif \vec{x} - \rho_{i} \varepsilon,
\end{equation*}
where $\rho_{i}$ is an adaptive step size. We refer to \cite{nocedal2006numerical} for the convergence properties of trust-region optimization and
specific rules for choosing and adapting the size $\varepsilon$. Algorithm~\ref{alg:policy_step} gives an outline of the overall optimization procedure.

\section{Practical Realization Conditions}\label{sec:assumptions}
The recursive optimality conditions in Section~\ref{sec:param} and Section~\ref{sec:policy} offer a general solution to the optimization problems without any guarantees for computational tractability. In this section, we discuss the assumptions necessary so that the proposed forward and backward passes are feasible.

\subsection{Linearized Quadratic Systems}
Firstly, we assume a linear dynamics with a Gaussian additive noise
\begin{equation*}
	f(\vec{x}\p | \vec{x}, \vec{u}, \vec{\theta}) = \N(\vec{x}\p | \mat{\Theta} \vec{\tau}, \mat{\Sigma}_{x\p}),
\end{equation*}
where $\mat{\Theta} = \begin{bmatrix} \mat{A}, & \mat{B}, & \vec{c} \end{bmatrix}$ is the aggregate linear parameter matrix and $\vec{\tau} = \begin{bmatrix} \vec{x}, & \vec{u}, & \vec{1} \end{bmatrix}^{\top}$ is the combined state-action vector. Moreover, the cost functions $c(\vec{x}, \vec{u})$ are presumed quadratic in state and action. Finally, the nominal parameter distribution is a Gaussian density  $\hat{p}(\vec{\theta}) = \N(\vec{\theta} | \vec{\mu}_{\theta}, \mat{\Sigma}_{\theta})$. Under these assumptions, the following holds
\begin{enumerate}
	\item The state- and parameter-value functions $V^{\pi}$ and $V^{\theta}$ start at time $T$ as quadratic functions and remain as such during the backward recursion due to the functional compatibility with the Gaussian probabilistic dynamics.
	\item The resulting policy is a time-variant linear Gaussian $\pi_{t}(\vec{u} | \vec{x}) = \N(\vec{u} | \mat{K}_{t} \vec{x} + \vec{k}_{t}, \mat{\Sigma}_{u, t})$, where $(\mat{K}_{t}, \vec{k}_{t})$ are the linear feedback matrix and affine offset.
	\item The optimal time-variant worst-case distribution $p_{t}$ is a Gaussian density.
	\item Propagation of the state through probabilistic dynamics results in a non-Gaussian distribution, due to the expectation over $p_{t}(\vec{\theta})$. We circumvent this 	issue by approximating the forwards recursion via spherical cubature.
\end{enumerate}

This setting can be extended to support nonlinear dynamical systems and non-convex costs via local approximations. This approach mirrors the iterative linearization and quadratization scheme used in dynamic differential programming (DDP) \cite{mayne1966second} and iterative linear quadratic regulator (iLQR) \cite{todorov2005generalized}. However, in this approach, a more principled regularization is achieved through the trust-region constraint on the policy \cite{levine2013guided}. Note that this extension requires a new reference nominal distribution $\hat{p}^{k}(\vec{\theta})$ for every linearization iteration $k$, which we assume is given by an external statistical learning process.

\subsection{Cubature-Based State Propagation}
We discuss the details of the approximate cubature forward recursion. The state propagation adheres to the probabilistic dynamics constraint
\begin{equation} \label{eq:forward_pass}
	\mu(\vec{x}\p) = \iiint \mu(\vec{x}) \pi(\vec{u} | \vec{x}) f(\vec{x}\p | \vec{x}, \vec{u}, \vec{\theta}) p(\vec{\theta})  \dif \vec{u} \dif \vec{x} \dif \vec{\theta},
\end{equation}
in which we omit the superscripts and subscripts for brevity. Under the assumptions introduced in the previous section, the expected dynamics can be written as
\begin{align*}
	p(\vec{x}\p | \vec{x}, \vec{u}) & = \int f(\vec{x}\p  | \vec{x}, \vec{u}, \vec{\theta}) p(\vec{\theta}) \dif \vec{\theta} \\
	& = \int \N (\vec{x}\p | \mat{\Theta} \vec{\tau}, \mat{\Sigma}_{x}) \N (\vec{\theta} | \vec{\mu}_{\theta}, \mat{\Sigma}_{\theta}) \dif \vec{\theta} \\
	& = \N (\vec{x}\p | \mat{M}_{\theta} \vec{\tau}, \mat{\Sigma}_{x} + (\vec{\tau}^{\top} \otimes \mat{I}_{\theta})^{\top} \mat{\Sigma}_{\theta} (\vec{\tau} \otimes \mat{I}_{\theta})),
\end{align*}
where $\mat{M}_{\theta}$ is defined as $\vec{\mu}_{\theta} = \vecop (\mat{M}_{\theta})$ with $\vecop$ denoting the vectorization operator, the operator $\otimes$ stands for the Kronecker product and $\mat{I}_{\theta}$ is the identity matrix with size equal to the dimension of $\vec{\theta}$. We write the covariance as $\mat{\Sigma}(\vec{\tau}) = \mat{\Sigma}_{x} + (\vec{\tau}^{\top} \otimes \mat{I}_{\theta})^{\top} \mat{\Sigma}_{\theta} (\vec{\tau} \otimes \mat{I}_{\theta})$, the second term of which depends on both state and action through $\vec{\tau}$. This leads to the integral in Equation \eqref{eq:forward_pass} being non-Gaussian. We use the cubature transform as described in \cite{solin2010cubature}, which constitutes a variant of the unscented transform \cite{wan2001unscented}, to approximate the propagated state distribution. Therefore, we rewrite the dynamics equivalently as
\begin{equation*}
	\vec{x}\p = \mat{M}_{\theta} \vec{\tau} + \sqrt{\mat{\Sigma}(\vec{\tau})} \, \vec{\xi}, \quad \vec{\xi} \sim \N (\vec{0}, \mat{I}),
\end{equation*}
where the matrix square root indicates the lower triangular Cholesky factor. If we include the noise $\vec{\xi}$ in the augmented state $\hvec{\tau} := \begin{bmatrix} \vec{x}, & \vec{u}, & \vec{\xi} \end{bmatrix}^{\top}$, the cubature computation resembles propagating an augmented distribution $p(\hvec{\tau})$ through a nonlinear deterministic dynamics function where
\setlength\arraycolsep{2pt}
\begin{align*}
	p(\hvec{\tau}) & = \mu(\vec{x} | \vec{m}, \mat{\Sigma}_{x}) \, \pi(\vec{u} | \mat{K} \vec{x} + \vec{k}, \mat{\Sigma}_{u}) \, p(\vec{\xi} | \vec{0}, \mat{I}_{x}) \\
	& = \N \left(
	\begin{bmatrix} \vec{x} \\ \vec{u} \\ \vec{\xi} \end{bmatrix} \! | \!
	\begin{bmatrix} \vec{m} \\ \mat{K} \vec{m} + \vec{k} \\ \mat{0} \end{bmatrix} \! , \! \begin{bmatrix} \mat{\Sigma}_{x} & \mat{\Sigma}_{x} \mat{K}^{\top} & \mat{0}     \\ \mat{K} \mat{\Sigma}_{x} & \mat{\Sigma}_{u} + \mat{K} \mat{\Sigma}_{x} \mat{K}^{\top} & \mat{0} \\
		\vec{0}          & \mat{0}                         & \mat{I}_{x}\end{bmatrix}
	\right).
\end{align*}
This reformulation allows us to apply standard cubature rules to obtain the new approximate state distribution $\mu(\vec{x}\p)$.

\subsection{Existence of The Worst-Case Distribution}\label{sec:existence}
Given the assumptions in Section~\ref{sec:assumptions}, it is possible to compute the worst-case parameter distribution $p_{t}^{*}(\vec{\theta}) = \N (\vec{\theta} | \mat{\Omega}_{t}^{-1} \vec{\omega}_{t}, \mat{\Omega}_{t}^{-1})$ in closed-form
\begin{equation*}
	\begin{gathered}
		\vec{\omega}_{t} = \hmat{\Lambda}_{\theta} \hvec{\mu}_{\theta} - \frac{1}{\beta} (\vec{s}_{xu, t}^{\top} \otimes \mat{I}_{x})^{\top} v^{\theta}_{t+1}, \\
		\mat{\Omega}_{t} =
		\begin{aligned}[t]
			& \hmat{\Lambda}_{\theta} + (\mat{\Sigma}_{xu, t} \otimes \mat{V}^{\theta}_{t+1}) \\
			& + \frac{2}{\beta} (\vec{s}_{xu, t}^{\top} \otimes \mat{I}_{x})^{\top} \mat{V}^{\theta}_{t+1} (\vec{s}_{xu, t}^{\top} \otimes \mat{I}_{x}),
		\end{aligned}
	\end{gathered}
\end{equation*}
where $\hvec{\mu}_{\theta}$ and $\hmat{\Lambda}_{\theta}$ are the mean and precision of the nominal distribution $\hat{p}(\vec{\theta})$, $\mat{V}^{\theta}$ and $v^{\theta}$ are the quadratic and linear terms of the adversarial state-value function $V^{\theta}$ and $\vec{s}_{xu}$ is the state-action distribution mean. Considering that $\beta \leq 0$ and $V^{\theta} \geq 0$, depending on $\hmat{\Lambda}_{\theta}$, there exists a value of $\beta$, for which $\mat{\Omega}$ becomes a negative-definite matrix and the distribution $p_{t}^{*}(\vec{\theta})$ does not exist anymore in a Gaussian form.
To overcome such issues, we propose a variant of our algorithm that mimics the trust-region sequential quadratic programming method \cite[Chapter 4, 18]{nocedal2006numerical}. Instead of the $p$-update in \eqref{eq:worst_param}, we iteratively update the worst-case distribution over smaller trust regions, i.e., we perform
\begin{align*}
	p^{k+1} & =\max_{p\in\mathcal{B}_{\delta}(\hat{p})\cap \mathcal{B}_{\delta_{k}}({p^{k}})}J(\pi^{k},p),
\end{align*}
where $\mathcal{B}_{\delta_{k}}$ denotes the KL-divergence trust region $\mathcal{B}_{\delta_{k}}(p^k) = \{p \ \vert \ \kl (p \| p^{k}) \le \delta_{k}\}$.
In practice, this iterative update is performed until the constraint in \eqref{eq:worst_param} becomes active.


\begin{figure*}[!h]
	\begin{minipage}[t]{0.45\textwidth}
		\centering
		\input{figures/linear_kl_over_time.tex}
	\end{minipage}\hspace{1cm}
	\begin{minipage}[t]{0.45\textwidth}
		\centering
		\input{figures/linear_cost_over_distance.tex}
	\end{minipage}
	\vspace{-0.5cm}
	\caption{Uncertain linear system experiment. Right, the worst-case KL budget allocation over the whole trajectory. Notice that most of the deviation happens in the first part of the trajectory. Left, the expected cost of the uncertainty-aware (blue) and robust (red) controllers evaluated on a range of distributions inter- and extrapolated between and beyond the nominal and worst-case distribution. The robust controller shows lower sensitivity to changes in the disturbance. Note the double logarithmic scale.}
	\label{fig:linear_kl}
	\vspace{-0.25cm}
\end{figure*}
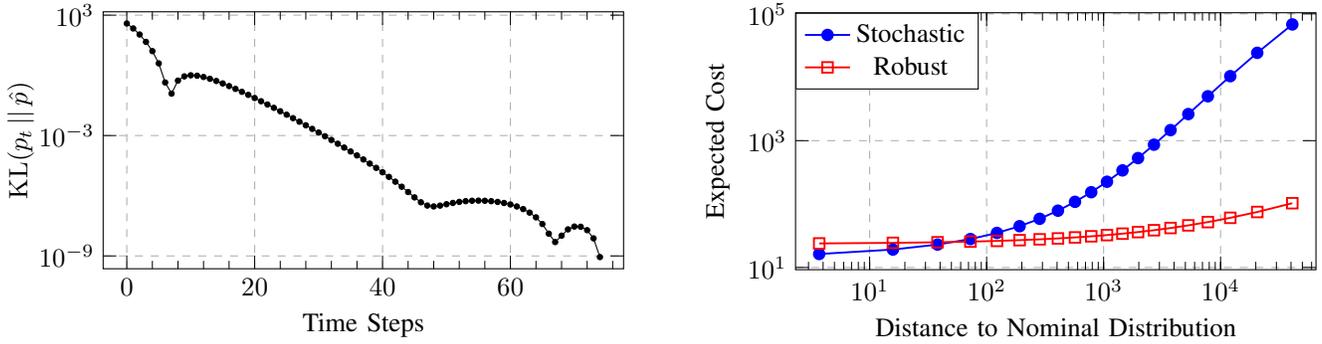

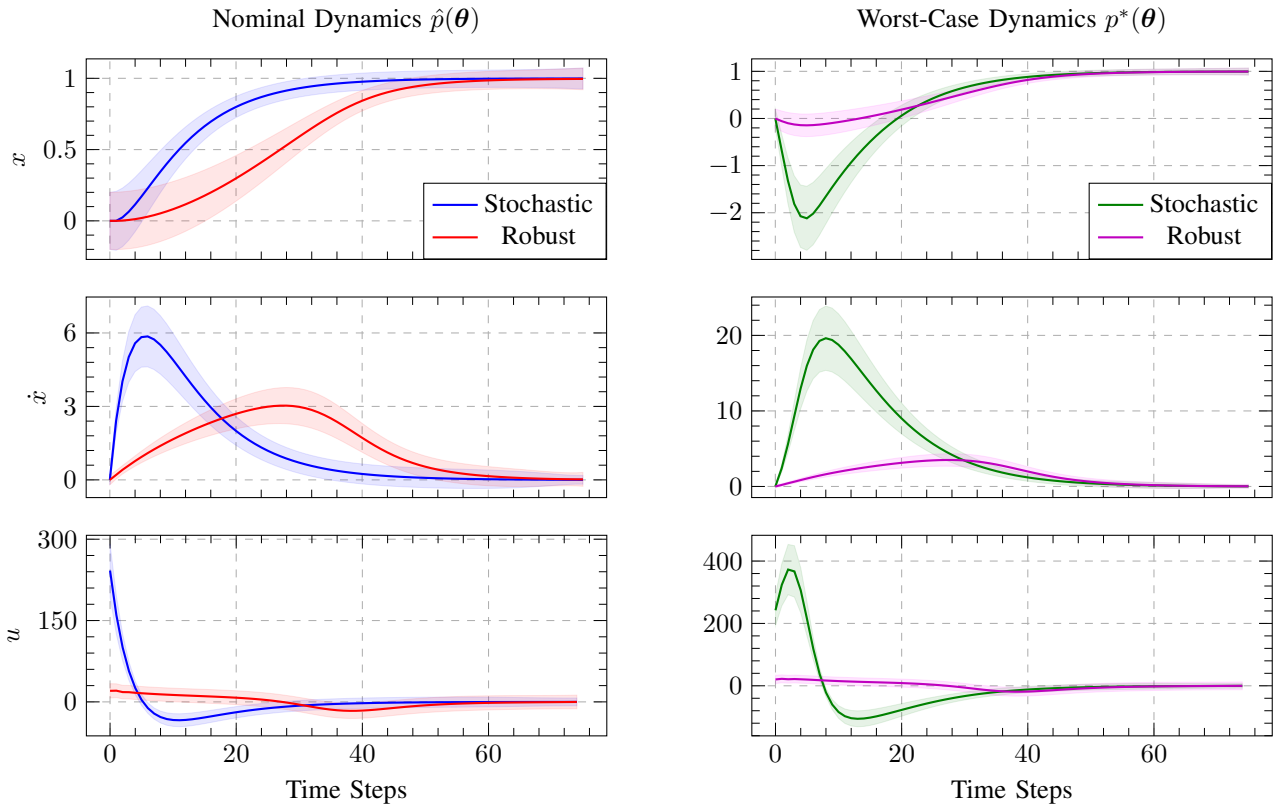
\begin{figure*}
	\begin{minipage}{0.42\textwidth}
		\centering
		\input{figures/linear_trajectories_nominal.tex}
	\end{minipage}\hspace{1.5cm}
	\begin{minipage}{0.42\textwidth}
		\centering
		\input{figures/linear_trajectories_adversarial.tex}
	\end{minipage}
	\vspace{-0.5cm}
	\caption{Uncertain linear system experiment. Comparison between the uncertainty-aware and distributionally robust controllers. Left, the trajectory distributions induced by uncertainty-aware (blue) and robust (red) controllers evaluated under the nominal dynamics distribution. The uncertainty-aware controller is aggressive and reaches the target faster. Right, the trajectory distributions induced by uncertainty-aware (green) and robust (magenta) controllers evaluated under the worst-case disturbance. The uncertainty-aware controller overshoots dramatically beyond the target, while the robust controller shows is barely affected.}
	\label{fig:linear_trajs}
	\vspace{-0.25cm}	
\end{figure*}

\section{Empirical Evaluation}\label{sec:eval}
We empirically evaluate the proposed distributionally robust control on a set of linear and nonlinear dynamical systems with uncertain dynamics. Without loss of generality, we limit the scope and assume the existence of a probabilistic dynamics model that has been won from data at an earlier stage. We linearize this model along a trajectory to deliver the probabilistic time-variant dynamics, i.e., the nominal distribution. Moreover, we limit the evaluation to a classic finite-horizon trajectory optimization scenario and do not consider a receding horizon control scheme.

The evaluation highlights the performance of the distributionally robust controller, iteratively optimized under its worst-case distribution, compared to an uncertainty-aware optimal controller, optimized under the nominal distributional dynamics using only the policy optimization stage of our approach. We perform this comparison by using the worst-case parameter optimization to compute an optimal disturbance on the uncertainty-aware controller and subsequently evaluate the performance of both controllers under this disturbance. This comparison allows the assessment of both controllers under previously unseen distributional disturbances since the worst-case attack on the uncertainty-aware controller may vary from the worst-case attack on the iteratively optimized robust controller. As evaluation criteria, we consider (1) the overall expected cost on a set of intermediate distributions between the nominal and worst-case, which we find using barycentric interpolation, (2) the induced trajectory distributions, and (3) the allocation strategy of the disturbance budget over the complete trajectory distribution. The source code of an efficient implementation can be found under \url{https://github.com/hanyas/trajopt}.

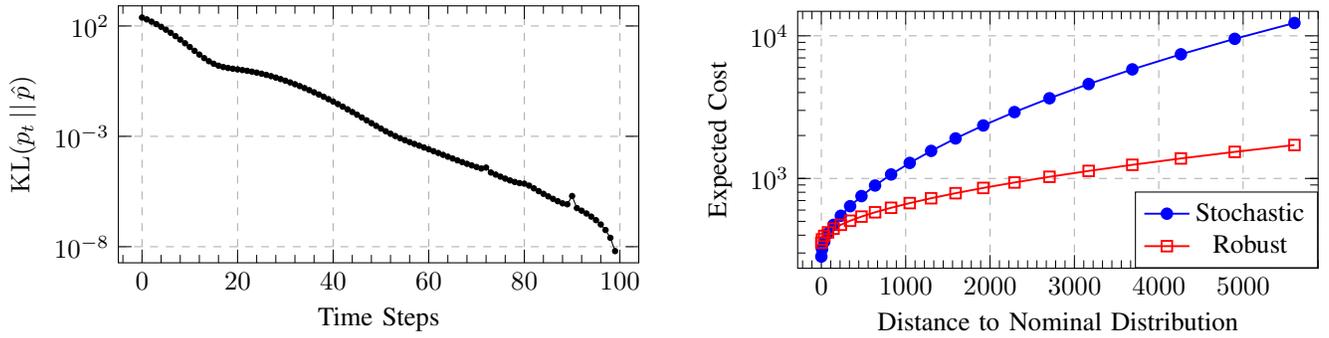
\begin{figure*}[t]
	\begin{minipage}[t]{0.45\textwidth}
		\centering
		\input{figures/robot_kl_over_time.tex}
	\end{minipage}\hspace{1cm}
	\begin{minipage}[t]{0.45\textwidth}
		\centering
		\input{figures/robot_cost_over_distance.tex}
	\end{minipage}
	\vspace{-0.5cm}
	\caption{Uncertain nonlinear robot experiment. Right, allocation of the worst-case KL budget over time steps. Most of the deviation is concentrated towards the early phase of the trajectory. Left, the expected cost of the uncertainty-aware (blue) and robust (red) controllers evaluated on a range of distributions inter- and extrapolated from the nominal and worst-case distribution: The robust controller shows much lower sensitivity to changes in the disturbance.}
	\label{fig:robot_kl}
	\vspace{-0.25cm}
\end{figure*}

\subsection{Uncertain Linear Dynamical System.}
We consider a simple actuated mass-spring-damper linear system with a mass $m=1 \, \si{\kilo\gram}$, a spring constant $k = 0.01 \, \si[per-mode=repeated-symbol]{\newton\per\meter}$, and a damping factor $d = 0.1 \, \si[per-mode=repeated-symbol]{\newton\second\per\meter}$. The linear differential equation has the form
\begin{equation*}
	\begin{bmatrix} \dot{x}_{t} \\ \ddot{x}_{t} \end{bmatrix} = \begin{bmatrix} 0 & 1 \\ -0.01 & -0.1 \end{bmatrix} \begin{bmatrix} x_{t} \\ \dot{x}_{t} \end{bmatrix} + \begin{bmatrix} 0 \\ 1 \end{bmatrix} u_{t},
\end{equation*}
which is integrated in time for a horizon $T = 75$ with a step size $\Delta t = 0.01 \, \si{\second}$. Moreover, we assume an initial distribution $\mu_{1}(\vec{x})$ centered at $\vec{x}_0 = \vec{0}$ with a diagonal standard deviation of $\sigma_{x_0} = \num{1e-1}$ and a discrete-time zero-mean process noise with a diagonal standard deviation $\sigma_{x} = \num{1e-2}$. The aim is to drive the system towards a goal state $\vec{x}_{g} = [1, \, 0]^{\top}$, under a quadratic state-action cost with the matrices $\mat{C}_{x} = \text{diag}([100, \, 0])$ and $\mat{C}_{u} = \text{diag}([0.001])$. As stated previously, we assume the existence of a nominal distribution $\hat{p}(\vec{\theta})$, which in this case is centered at the true linear dynamics with a diagonal standard deviation of $\sigma_{\theta} = \num{1e-4}$ to represent uncertainty over the parameters. We initialize a zero-mean controller with a diagonal standard deviation $\sigma_{\pi} = \num{10}$ and set the trust-region sizes $\varepsilon=0.25$ and $\delta=750$.

The comparison between the controllers on the linear system is depicted in Figure~\ref{fig:linear_trajs}. The plots on the left show the trajectory distribution induced by the uncertainty-aware (blue) and robust (red) policies under the nominal parameter distribution $\hat{p}(\vec{\theta})$. The figures show an aggressive uncertainty-aware controller that takes full advantage of the nominal dynamics to reach the goal as fast as possible, while the robust controller shows sub-optimal behavior. However, when evaluated on the worst-case dynamics, on the right, the uncertainty-aware controller (green) overshoots beyond the target incurring a massive cost, while the robust policy (magenta) maintains a  consistent behavior associated with much lower overall cost.

Furthermore, the right plot in Figure~\ref{fig:linear_kl} illustrates the worst-case KL allocation over the trajectory. A large portion of the overall deviation takes place in the first 20 time steps, leading to the sub-optimal performance of the uncertainty-aware controller in the same time window. Finally, the left plot highlights the superior performance of the robust policy (red) on a continuum of distributions interpolated between the nominal and worst-case distributions and beyond using the method described in Appendix~\ref{sec:barycentric}. The uncertainty-aware controller (blue) delivers better performance in a small region around the nominal distribution but very quickly worsens as the distance to that distribution increases.

\begin{figure*}[b]
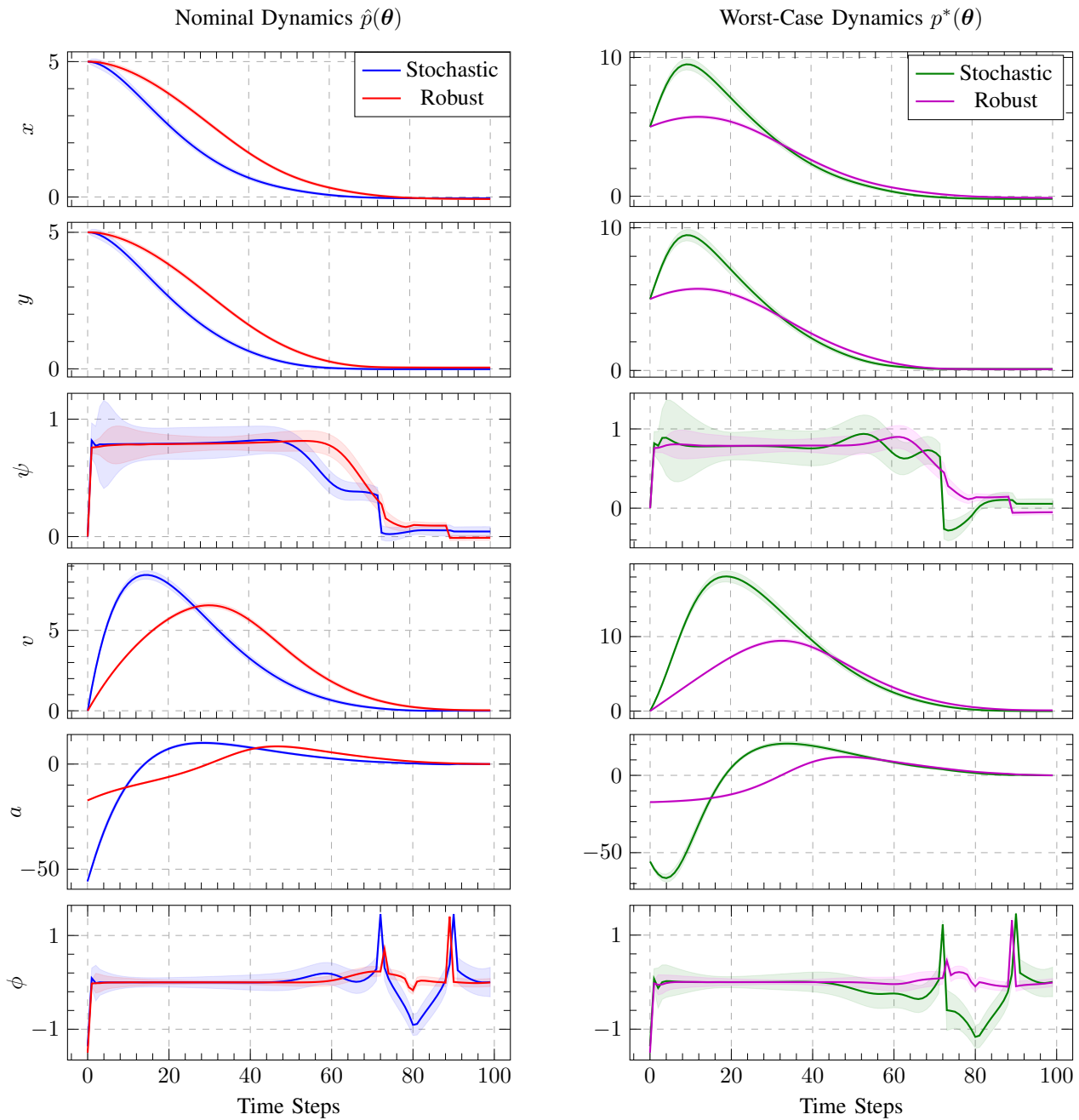

	\begin{minipage}[t]{0.42\textwidth}
		\centering
		\input{figures/robot_trajectories_nominal.tex}
	\end{minipage}\hspace{1.25cm}
	\begin{minipage}[t]{0.42\textwidth}
		\centering
		\input{figures/robot_trajectories_adversarial.tex}
	\end{minipage}
	\vspace{-0.5cm}
	\caption{Uncertain nonlinear robot experiment. Comparison of uncertainty-aware and distributionally robust controllers. Left, the trajectory induced by the uncertainty-aware (blue) and robust (red) controllers evaluated under the nominal dynamics distribution. The uncertainty-aware controller aggressively applies large controls to reach the target faster. Right, the trajectory distributions induced by uncertainty-aware (green) and robust (magenta) controllers evaluated under the worst-case disturbance. The uncertainty-aware controller shows clear sub-optimal behavior, while the robust controller is barely affected.}
	\label{fig:robot_trajs}
\end{figure*}

\subsection{Uncertain Nonlinear Robot Car}
This experiment validates our approach for general nonlinear dynamical systems via iterative linearization of the dynamics around a trust-region. We consider a nonholonomic robot moving in 2D-space. The state vector consists of the $x, y$-coordinates of the position, the speed $v$ and the orientation $\psi$, while the acceleration $a$ and the steering angle $\phi$ are used for actuation. The global dynamics is nonlinear in state and action and given by
\begin{equation*}
	\begin{bmatrix} \dot{x}_{t} \\ \dot{y}_{t} \\ \dot{\psi}_{t} \\ \dot{v}_{t} \end{bmatrix} = \begin{bmatrix} v_{t} \sin{\psi_{t}} \\ v_{t} \cos{\psi_{t}} \\ v_{t} \tan{(\phi_{t})} / d \\ a_{t} \end{bmatrix},
\end{equation*}
where the constant $d = 0.1 \, \si{\meter}$ is the car length. This ODE is integrated for a horizon $T = 100$ with a step size $\Delta t = 0.025 \, \si{\second}$. The initial state distribution is centered at $[5, \, 5, \, 0, \, 0]^{\top}$ with a diagonal standard deviation $\sigma_{x_0} = \num{1e-2}$ and the discrete-time process noise is zero-mean with a diagonal standard deviation $\sigma_{x} = \num{1e-4}$. The goal state is $g = [0, \, 0, \, 0, \, 0]^{\top}$ and the quadratic cost matrices are $\mat{C}_{x} = \text{diag}([10, \, 10, \, 1, \, 1])$ and $\mat{C}_{u} = \text{diag}([0.1, \, 0.1])$. Analogous to the previous experiment, we assume the nominal parameter distribution to be centered at the linearized dynamics with a diagonal standard deviation $\sigma_{\theta} =\num{1e-3}$. We initialize a zero-mean controller with a diagonal standard deviation $\sigma_{\pi} = \sqrt{0.1}$ and set the trust regions to $\varepsilon=0.25$ and $\delta=500$.

Figure~\ref{fig:robot_trajs} depicts the results in a similar fashion to what we presented in the last experiment. Here again, the uncertainty-aware controller (blue) acts aggressively under the nominal dynamics, while the robust controller (red) is slower and applies smaller controls. When evaluating the controllers under the uncertainty-aware controller's optimal adversary, the uncertainty-aware controller (green) overshoots and shows sub-optimal behavior while the trajectory distribution induced by the robust controller (magenta) is hardly affected. Lastly, the comparison of both controllers on a set of distributions interpolated between the nominal and the adversary highlights the overwhelming advantage of the robust controller, Figure~\ref{fig:robot_kl}.

\section{Discussion}
We have presented a technique to robustify data-driven stochastic optimal control approaches that rely on probabilistic models of the dynamics. Our approach consists of an iterative two-stage relative entropy trust-region optimization. The first stage optimizes the maximum entropy worst-case Gaussian distributional dynamics in a KL-ball around a nominal distribution, while the second stage optimizes the policy w.r.t. the worst-case dynamics.
We show that both stages admit closed-form backward value recursions and approximate cubature forward passes for probabilistic time-variant dynamics models. Furthermore, empirical results on linear and nonlinear dynamical systems validate the benefits of robustifying stochastic control against worst-case statistical model disturbances.

Despite the encouraging initial results, our approach still has multiple limitations. The assumption of Gaussian densities for the nominal and worst-case distributions is rather limiting. Similarly, although reasonable, the restriction of the adversary to a time-variant form does not reflect the statistical errors that arise while approximating stationary representations of dynamics. In addition, long-horizon trajectory optimization is often prone to get stuck in local minima. An investigation of a nonlinear model predictive control formulation can prove very beneficial, despite the additional computational load it may require. Finally, the KL divergence is not a proper distance metric in the space of distributions. Analyzing the drawbacks of this design choice can inspire better alternatives, e.g., using kernel methods and optimal transport.

\section{Appendix}

\subsection{Barycentric Distribution Interpolation}\label{sec:barycentric}
Interpolating between two reference distributions $p(x), q(x)$ in order to find an intermediate $h(x)$ is done by minimizing the weighted Kullback-Leibler divergence objective
\begin{align*}
	& \argmin_{h} & \, & \lambda \kl \left( h(x) \,||\, p(x) \right) + (1 - \lambda) \kl \left( h(x) \,||\, q(x) \right), \\
	& \st         & \, & \int h(x) \dif x = 1,
\end{align*}
which results in an optimal distribution $h^{*}(x)$
\begin{equation*}
	h^{*}(x) \propto q(x) \exp \left[ \lambda \left( \log p(x) - \log q(x) \right) \right],
\end{equation*}
which for Gaussians yields $h^{*}(x) = \N(x | \mu^{*}, \Sigma^{*})$
\begin{equation*}
	\begin{aligned}
		\Sigma^{*} & = \left(\lambda \Sigma^{-1}_{p} + (1 - \lambda) \Sigma^{-1}_{q} \right)^{-1}, \\
		\mu^{*}    & = \Sigma^{*} \left(\lambda \Sigma^{-1}_{p} \mu_{p} + (1 - \lambda) \Sigma^{-1}_{q} \mu_{q} \right).
	\end{aligned}
\end{equation*}

\subsection{Parameter Optimization}
The Lagrangian function of the worst-case parameter optimization problem $H(p, \mu, V^{\theta}, \beta, \gamma)$
\begin{equation*}\label{param_lagrangian}
	H =
	\begin{aligned}[t]
		& \sum_{t=1}^{T-1} \iint c_{t}(\vec{x},\vec{u}) \mu_{t}(\vec{x}) \pi^{k}_{t}(\vec{u} | \vec{x}) \dif \vec{u} \dif \vec{x} \\
		& + \int c_{T}(\vec{x}) \mu_{T}(\vec{x}) \dif \vec{x} + \int V_{1}^{\theta}(\vec{x}) \hat{\mu}_{1}(\vec{x}) \dif \vec{x} \\
		& + \sum_{t=1}^{T-1} \int V_{t+1}^{\theta}(\vec{x}\p)
		\begin{aligned}[t]
			\iiint & \mu_{t}(\vec{x}) \pi^{k}_{t}(\vec{u} | \vec{x}) f(\vec{x}\p  | \vec{x}, \vec{u}, \vec\theta) \\
			& \times p^{k+1}_{t}(\vec{\theta}) \dif \vec{u} \dif \vec{x} \dif \vec{\theta}\dif \vec{x}\p
		\end{aligned} \\
		& - \sum_{t=1}^{T-1} \int V_{t}^{\theta}(\vec{x}\p) \mu_{t}(\vec{x}\p) \dif \vec{x}\p + \int V_{T}^{\theta}(\vec{x}\p) \mu_{T}(\vec{x}\p) \dif \vec{x}\p \\
		& + \beta \sum_{t=1}^{T-1} \int p^{k+1}_{t}(\vec{\theta}) \log \frac{p^{k+1}_{t}(\vec{\theta})}{\hat{p}(\vec{\theta})} \dif \vec{\theta} \\
		& - \beta \delta + \sum_{t=1}^{T-1} \gamma_{t} \left( \int_\theta p^{k+1}_{t}(\vec{\theta}) \dif \vec{\theta} - 1 \right).
	\end{aligned}
\end{equation*}
Solving for optimal distribution $p_{t}^{k+1}(\vec{\theta})$, normalization variable $\gamma_t$ leads to a simplified dual formulation
\begin{equation*}\label{eq:worst_dual}
	F = \int V_{1}^{\theta}(\vec{x}) \hat{\mu}_{1}(\vec{x}) \dif \vec{x} + \beta \sum_{t=1}^{T-1} \kl \left( p^{k+1}_{t}(\vec{\theta}) \,||\, \hat{p}(\vec{\theta}) \right) -  \beta \delta.
\end{equation*}

\subsection{Policy Optimization}
The Lagrangian function of the robust policy optimization problem $L(\pi, \mu, V^{\pi}, \alpha, \lambda)$
\begin{equation*}\label{policy_lagrangian}
	L =
	\begin{aligned}[t]
		& \sum_{t=1}^{T-1} \iint c_{t}(\vec{x}, \vec{u}) \mu_{t}(\vec{x}) \pi_{t}^{k+1}(\vec{u} | \vec{x}) \dif \vec{u} \dif \vec{x} \\
		& + \int c_{T}(\vec{x}) \mu_{T}(\vec{x}) \dif \vec{x} + \int V^{\pi}_{1}(\vec{x}) \hat{\mu}_{1}(\vec{x}) \dif \vec{x} \\
		& + \sum_{t=1}^{T-1} \int V^{\pi}_{t+1}(\vec{x}\p)
		\begin{aligned}[t]
			\iiint & \mu_{t}(\vec{x}) \pi_{t}^{k+1}(\vec{u} | \vec{x}) f(\vec{x}\p  | \vec{x}, \vec{u}, \vec\theta) \\
			& \times p^{k+1}_{t}(\vec{\theta}) \dif \vec{u} \dif \vec{x} \dif \vec{\theta}\dif \vec{x}\p
		\end{aligned} \\
		& - \sum_{t=1}^{T-1} \int V^{\pi}_{t}(\vec{x}\p) \mu_{t}(\vec{x}\p) \dif \vec{x}\p - \int V^{\pi}_{T}(\vec{x}\p) \mu_{T}(\vec{x}\p) \dif \vec{x}\p \\
		& + \alpha \sum_{t=1}^{T-1} \iint \mu_{t}(\vec{x}) \pi_{t}^{k+1}(\vec{u} | \vec{x}) \log \frac{\pi_{t}^{k+1}(\vec{u} | \vec{x})}{\pi_{t}^{k}(\vec{u} | \vec{x})} \dif \vec{u} \dif \vec{x} \\
		& - \alpha \varepsilon + \sum_{t=1}^{T-1} \int \lambda_{t}(\vec{x}) \left( \int \pi_{t}^{k+1}(\vec{u} | \vec{x}) \dif \vec{u} - 1 \right) \dif \vec{x}.
	\end{aligned}
\end{equation*}
Solving for optimal distribution $\pi_{t}^{k+1}(\vec{u} | \vec{x})$ and normalization variable $\lambda_t$ leads to the simplified dual formulation
\begin{equation*}
	G = \int V^{\pi}_{1}(\vec{x}) \hat{\mu}_{1}(\vec{x}) \dif \vec{x} - \alpha \varepsilon.
\end{equation*}

\bibliographystyle{IEEEtran}
\bibliography{references.bib}

\end{document}

%% file: packages.tex
\usepackage{generic}

\usepackage{cite}

\usepackage{dsfont}
\usepackage{amsfonts}
\usepackage{textcomp}

\usepackage{amsmath}
\usepackage{amssymb}
\usepackage{commath}
\usepackage{derivative}

\usepackage{graphicx}
\usepackage{adjustbox}
\usepackage{caption}
\usepackage{svg}

\usepackage{pgfplots}
\usepackage{tikz}

\pgfplotsset{compat=newest}
\usetikzlibrary{decorations.pathmorphing}
\usetikzlibrary{fit}
\usetikzlibrary{backgrounds}
\usetikzlibrary{pgfplots.groupplots}
\usetikzlibrary{external}

\usepackage{url}
\usepackage{booktabs}
\usepackage{siunitx}
\usepackage{xcolor}

\usepackage[acronym, shortcuts]{glossaries}
\makeglossaries

\usepackage{hyperref}
\hypersetup{
	colorlinks=true,
	linkcolor=blue,
	filecolor=magenta,      
	urlcolor=blue,
}

\usepackage[ruled, vlined]{algorithm2e}
\makeatletter
\renewcommand{\algocf@caption@boxruled}{%
	\hrule
	\hbox to \hsize{%
		\vrule\hskip1pt
		\vbox{   
			\vskip\interspacetitleboxruled%
			\vspace{0.1em}
			\unhbox\algocf@capbox\hfill
			\vspace{0.1em}
			\vskip\interspacetitleboxruled
		}%
		\hskip-2pt\vrule%
	}\nointerlineskip%
}%

%% file: macros.tex
\usepackage{bm}

\renewcommand{\vec}[1]{\bm{\mathbf{#1}}}
\newcommand{\mat}[1]{\bm{\mathbf{#1}}}

\newcommand{\hvec}[1]{\hat{\bm{\mathbf{#1}}}}
\newcommand{\hmat}[1]{\hat{\bm{\mathbf{#1}}}}

\DeclareMathOperator*{\argmax}{arg\,max}
\DeclareMathOperator*{\argmin}{arg\,min}

\DeclareMathOperator*{\minimize}{minimize}
\DeclareMathOperator*{\maximize}{maximize}
\DeclareMathOperator*{\st}{subject\,\,to}
\DeclareMathOperator*{\vecop}{vec}

\DeclareMathOperator{\kl}{KL}

\newcommand*\p{\textnormal{\textquotesingle}}

\DeclareOldFontCommand{\bf}{\normalfont\bfseries}{\mathbf}

\newcommand{\op}[1]{\operatorname{#1}}

\newcommand{\N}{\op{N}}

%% file: acronyms.tex
\newacronym{EM}{EM}{expectation-maximization}
\newacronym{RL}{RL}{reinforcement learning}
\newacronym{Hb-REPS}{Hb-REPS}{hybrid relative entropy policy search}
\newacronym{SDS}{SDS}{switching dynamical systems}
\newacronym{SSM}{SSM}{switching state-space models}
\newacronym{PWL}{PWL}{piecewise linear}
\newacronym{ReLU}{ReLU}{rectified linear units}
\newacronym{PWARX}{PWARX}{piecewise autoregressive exogenous systems}
\newacronym{MIQP}{MIQP}{mixed-integer quadratic program}
\newacronym{PGM}{PGM}{probabilistic graphical models}
\newacronym{HDBN}{HDBN}{hybrid dynamic Bayesian networks}
\newacronym{BNP}{BNP}{Bayesian nonparametrics}
\newacronym{MCMC}{MCMC}{Markov chain Monte Carlo}
\newacronym{SVI}{SVI}{stochastic variational inference}
\newacronym{HRL}{HRL}{hierarchical reinforcement learning}
\newacronym{SMDP}{SMDP}{semi-Markov decision processes}
\newacronym{MDP}{MDP}{Markov decision processes}
\newacronym{rARHMM}{rARHMM}{recurrent autoregressive hidden Markov model}
\newacronym{ARHMM}{ARHMM}{autoregressive hidden Markov model}
\newacronym{HMM}{HMM}{hidden Markov model}
\newacronym{HSMM}{HSMM}{semi-hidden Markov model}
\newacronym{SLDS}{SLDS}{switching linear dynamical systems}
\newacronym{rSLDS}{rSLDS}{recurrent switching linear dynamical systems}
\newacronym{FHMM}{FHMM}{factorial hidden Markov model}
\newacronym{REPS}{REPS}{relative entropy policy search}
\newacronym{KL}{KL}{Kullback-Leibler divergence}
\newacronym{MAP}{MAP}{maximum a posteriori}
\newacronym{NW}{NW}{normal-Wishart}
\newacronym{MN}{MN}{matrix-normal}
\newacronym{MNW}{MNW}{matrix-normal-Wishart}
\newacronym{Dir}{Dir}{Dirichlet}
\newacronym{Cat}{Cat}{categorical}
\newacronym{E-step}{E-step}{expectation step}
\newacronym{M-step}{M-step}{maximization step}
\newacronym{EB-step}{EB-step}{empirical Bayes step}
\newacronym{FNN}{FNN}{feed-forward neural net}
\newacronym{GP}{GP}{Gaussian process}
\newacronym{LSTM}{LSTM}{long-short-term memory network}
\newacronym{RNN}{RNN}{recurrent neural network}
\newacronym{NMSE}{NMSE}{normalized mean squared error}
\newacronym{SAC}{SAC}{soft actor-critic}
\newacronym{RFF}{RFF}{random Fourier feature}
\newacronym{GPR}{GPR}{Gaussian process regression}
\newacronym{SGPR}{SGPR}{sparse Gaussian process regression}
\newacronym{ANN}{ANN}{artificial neural network}
\newacronym{LR}{LR}{local regression}
\newacronym{ILR}{ILR}{infinite local regression}
\newacronym{HILR}{HILR}{hierarchical infinite local regression}
\newacronym{CMB}{CMB}{cosmic microwave background}
\newacronym{BNN}{BNN}{Bayesian neural network}
\newacronym{RFWR}{RFWR}{receptive field weighted regression}
\newacronym{LWPR}{LWPR}{locally weighted projection regression}
\newacronym{LGR}{LGR}{local Gaussian regression}
\newacronym{MoE}{MoE}{mixture of experts}
\newacronym{VBEM}{VBEM}{variational Bayes expectation-maximization}
\newacronym{VB}{VB}{variational Bayes}
\newacronym{GMM}{GMM}{Gaussian mixture model}
\newacronym{DP}{DP}{Dirichlet process}
\newacronym{VI}{VI}{variational inference}
\newacronym{ELBO}{ELBO}{evidence lower bound}
\newacronym{MSE}{MSE}{mean squared error}

%% file: figures/linear_kl_over_time.tex
\begin{tikzpicture}

	\begin{axis}[
			height=5cm,
			width=8.5cm,
			tick pos=both,
			tick align=inside,
			minor tick num=4,
			xmajorticks=true,
			ymajorticks=true,
			xmajorgrids,
			ymajorgrids,
			xtick style={color=black},
			ytick style={color=black},
			x grid style={white!69.0196078431373!black},
			y grid style={white!69.0196078431373!black},
			ymin=2.29617684805982e-10, ymax=1467.93076506236,
			ymode=log,
			xmin=-3.7, xmax=77.7,
			ylabel={$\kl(p_{t} \,||\, \hat{p})$},
			y label style={yshift=-.5em},
			xlabel={Time Steps},
			xlabel near ticks,
			every major tick/.append style={major tick length=5pt, black},
			every minor tick/.append style={minor tick length=3pt, black},
			major grid style={dashed, gray},
		]

		\addplot [black, mark=*, mark size=1, mark options={solid}]
		table {%
				0 384.262752211864
				1 217.304896686366
				2 108.579719825534
				3 46.4570291602369
				4 16.085759962017
				5 3.94777989368743
				6 0.441422146438234
				7 0.123239985109285
				8 0.545918476533665
				9 0.888489955878074
				10 1.00696548678363
				11 0.959894042185797
				12 0.82924166760059
				13 0.673193877792981
				14 0.523890038921211
				15 0.395469891757605
				16 0.291739903038791
				17 0.211363416622677
				18 0.150902421448849
				19 0.106428785778707
				20 0.0742865975888636
				21 0.0513875599153719
				22 0.0352679174794561
				23 0.0240358205776401
				24 0.0162781563659422
				25 0.0109614735270425
				26 0.00734259033839013
				27 0.00489439256691515
				28 0.00324731864928651
				29 0.00214481234869179
				30 0.00141027884208267
				31 0.000923062688345411
				32 0.000601264802917001
				33 0.0003896141475348
				34 0.00025100241119258
				35 0.00016063133472688
				36 0.000101999292841271
				37 6.41685596445285e-05
				38 3.99160556154854e-05
				39 2.44887220510748e-05
				40 1.47698418357933e-05
				41 8.72300069776344e-06
				42 5.02264587787238e-06
				43 2.8094781630017e-06
				44 1.52901677719086e-06
				45 8.25378805302535e-07
				46 4.71592386475095e-07
				47 3.24010386520968e-07
				48 2.92578626215345e-07
				49 3.21506539613381e-07
				50 3.76746709385145e-07
				51 4.37921421791998e-07
				52 4.93148813340838e-07
				53 5.35759031805583e-07
				54 5.6224434974439e-07
				55 5.71019780792881e-07
				56 5.61722178993307e-07
				57 5.34872137691877e-07
				58 4.91780907552197e-07
				59 4.34612966948578e-07
				60 3.66520027661466e-07
				61 2.91749741698766e-07
				62 2.15611175136132e-07
				63 1.44167159987774e-07
				64 8.35470403970362e-08
				65 3.88638943249475e-08
				66 1.28941168853203e-08
				67 4.9076325225883e-09
				68 1.0227744873248e-08
				69 2.10950803491983e-08
				70 2.9062012352199e-08
				71 2.84183903076496e-08
				72 1.92009688149142e-08
				73 7.50853690334452e-09
				74 8.77167671831103e-10
			};
	\end{axis}

\end{tikzpicture}

%% file: figures/linear_cost_over_distance.tex
\begin{tikzpicture}

	\begin{axis}[
			height=5cm,
			width=8.5cm,
			tick pos=both,
			tick align=inside,
			minor tick num=8,
			xmajorticks=true,
			ymajorticks=true,
			xmajorgrids,
			ymajorgrids,
			xtick style={color=black},
			ytick style={color=black},
			x grid style={white!69.0196078431373!black},
			y grid style={white!69.0196078431373!black},
			ymin=9.22094882142207, ymax=104061.786617609,
			ymode=log,
			xmin=2.32531780839473, xmax=65388.1712997196,
			xmode=log,
			ylabel={Expected Cost},
			xlabel={Distance to Nominal Distribution},
			every major tick/.append style={major tick length=5pt, black},
			every minor tick/.append style={minor tick length=3pt, black},
			major grid style={dashed, gray},
			legend style={at={(0.,1.)},anchor=north west}
		]

		\addplot [blue, mark=*, mark size=2, mark options={solid}, line width=0.25mm]
		table {%
				0 14.0922672003203
				3.7043358730626 16.2989987572772
				15.8038904292809 19.1700230301312
				38.0581100235599 22.9640089666162
				72.6947272321226 28.0615651808084
				122.568775332883 35.032568327578
				191.389228265894 44.7463152161485
				284.048653067454 58.5556496873425
				407.113737373346 78.6120834645259
				569.574189495034 108.419353285229
				784.019952265338 153.834567614436
				1068.55457014986 224.939373908368
				1450.02752242257 339.671079550305
				1969.74689064706 531.182402052657
				2694.13004507868 863.542319208715
				3735.88019043409 1467.3534472974
				5299.55296204228 2626.88157562339
				7789.8776771282 5014.54047634275
				12105.0877227029 10406.3268126887
				20587.2676884847 24280.2100509575
				41046.0294076671 68090.4211527371
			};
		\addlegendentry{Stochastic}
			
		\addplot [red, mark=square, mark size=2, mark options={solid}, line width=0.25mm]
		table {%
				0 23.4897217743292
				3.7043358730626 23.9409671690821
				15.8038904292809 24.4332391104915
				38.0581100235599 24.9726312000013
				72.6947272321226 25.5665484732757
				122.568775332883 26.2240867678052
				191.389228265894 26.9565533405327
				284.048653067454 27.7781945706415
				407.113737373346 28.7072342889551
				569.574189495034 29.7673901923273
				784.019952265338 30.9901478686556
				1068.55457014986 32.4182761361761
				1450.02752242257 34.1114561455744
				1969.74689064706 36.155675637167
				2694.13004507868 38.6796971147013
				3735.88019043409 41.8856971182635
				5299.55296204228 46.1106229633717
				7789.8776771282 51.9610950458161
				12105.0877227029 60.6488317355576
				20587.2676884847 74.9796628492991
				41046.0294076671 103.119740647937
			};
		\addlegendentry{Robust}
			
	\end{axis}

\end{tikzpicture}

%% file: figures/linear_trajectories_nominal.tex
\begin{tikzpicture}

	\begin{groupplot}[group style={group size=1 by 3, vertical sep=0.5cm}]
		\nextgroupplot[
			height=4.25cm,
			width=8.5cm,
			tick pos=both,
			tick align=inside,
			minor tick num=4,
			xmajorticks=false,
			ymajorticks=true,
			xmajorgrids,
			ymajorgrids,
			xtick style={color=black},
			ytick style={color=black},
			x grid style={white!69.0196078431373!black},
			y grid style={white!69.0196078431373!black},
			xmin=-3.75, xmax=78.75,
			ymin=-0.270678566130648, ymax=1.13646799539441,
			every major tick/.append style={major tick length=5pt, black},
			every minor tick/.append style={minor tick length=3pt, black},
			major grid style={dashed, gray},
			ylabel=$x$,
			legend style={at={(1.,0.)},anchor=south east},
			title={Nominal Dynamics $\hat{p}(\vec{\theta})$},
		]

		\path [draw=blue, fill=blue, opacity=0.1]
		(axis cs:0,0.2)
		--(axis cs:0,-0.2)
		--(axis cs:1,-0.206758599609918)
		--(axis cs:2,-0.181382962615137)
		--(axis cs:3,-0.135267687120998)
		--(axis cs:4,-0.0763874581268242)
		--(axis cs:5,-0.0103836666970031)
		--(axis cs:6,0.0587651941037564)
		--(axis cs:7,0.128287159204468)
		--(axis cs:8,0.19629121390359)
		--(axis cs:9,0.261529623183358)
		--(axis cs:10,0.323222492082264)
		--(axis cs:11,0.380927649038012)
		--(axis cs:12,0.43444405071962)
		--(axis cs:13,0.483740177572215)
		--(axis cs:14,0.528901123331241)
		--(axis cs:15,0.570089674404466)
		--(axis cs:16,0.607517843512728)
		--(axis cs:17,0.641426196136745)
		--(axis cs:18,0.672068970397436)
		--(axis cs:19,0.699703495188032)
		--(axis cs:20,0.724582794593966)
		--(axis cs:21,0.74695055526659)
		--(axis cs:22,0.767037847683474)
		--(axis cs:23,0.785061148774462)
		--(axis cs:24,0.801221326430061)
		--(axis cs:25,0.815703327921404)
		--(axis cs:26,0.828676373834621)
		--(axis cs:27,0.840294503858914)
		--(axis cs:28,0.850697355489421)
		--(axis cs:29,0.860011084419768)
		--(axis cs:30,0.868349357858677)
		--(axis cs:31,0.875814370222104)
		--(axis cs:32,0.882497845269289)
		--(axis cs:33,0.888482000261713)
		--(axis cs:34,0.893840456577207)
		--(axis cs:35,0.898639087840615)
		--(axis cs:36,0.902936801450332)
		--(axis cs:37,0.906786252764053)
		--(axis cs:38,0.910234493486295)
		--(axis cs:39,0.913323557250341)
		--(axis cs:40,0.916090986230092)
		--(axis cs:41,0.918570303026324)
		--(axis cs:42,0.920791432178727)
		--(axis cs:43,0.922781075557251)
		--(axis cs:44,0.924563045654369)
		--(axis cs:45,0.926158560483714)
		--(axis cs:46,0.927586503424721)
		--(axis cs:47,0.928863650960801)
		--(axis cs:48,0.930004870855992)
		--(axis cs:49,0.931023292912853)
		--(axis cs:50,0.93193045406117)
		--(axis cs:51,0.932736419151298)
		--(axis cs:52,0.933449878478333)
		--(axis cs:53,0.934078222759559)
		--(axis cs:54,0.934627596051186)
		--(axis cs:55,0.935102926956807)
		--(axis cs:56,0.935507938500055)
		--(axis cs:57,0.93584513727888)
		--(axis cs:58,0.936115783083056)
		--(axis cs:59,0.936319841159198)
		--(axis cs:60,0.936455920888783)
		--(axis cs:61,0.936521206953955)
		--(axis cs:62,0.93651139223962)
		--(axis cs:63,0.936420625847364)
		--(axis cs:64,0.936241494678246)
		--(axis cs:65,0.935965062961618)
		--(axis cs:66,0.935581000649684)
		--(axis cs:67,0.935077838559779)
		--(axis cs:68,0.934443395612244)
		--(axis cs:69,0.933665432333542)
		--(axis cs:70,0.932732597306124)
		--(axis cs:71,0.931635754165134)
		--(axis cs:72,0.930369813967756)
		--(axis cs:73,0.928936261624526)
		--(axis cs:74,0.927346660628961)
		--(axis cs:75,0.92562751952472)
		--(axis cs:75,1.07241070576538)
		--(axis cs:75,1.07241070576538)
		--(axis cs:74,1.07062061008825)
		--(axis cs:73,1.06894806242695)
		--(axis cs:72,1.0674158608409)
		--(axis cs:71,1.06603233569129)
		--(axis cs:70,1.0647960207923)
		--(axis cs:69,1.06369900402288)
		--(axis cs:68,1.06272932992322)
		--(axis cs:67,1.06187273015175)
		--(axis cs:66,1.0611138640551)
		--(axis cs:65,1.06043718918598)
		--(axis cs:64,1.0598275454631)
		--(axis cs:63,1.05927051655618)
		--(axis cs:62,1.05875262016223)
		--(axis cs:61,1.05826137044967)
		--(axis cs:60,1.05778524878722)
		--(axis cs:59,1.05731361212171)
		--(axis cs:58,1.05683656196)
		--(axis cs:57,1.05634479106407)
		--(axis cs:56,1.05582941991272)
		--(axis cs:55,1.05528183084313)
		--(axis cs:54,1.05469350456487)
		--(axis cs:53,1.05405586135446)
		--(axis cs:52,1.0533601075564)
		--(axis cs:51,1.05259708688898)
		--(axis cs:50,1.05175713534281)
		--(axis cs:49,1.05082993804737)
		--(axis cs:48,1.0498043862728)
		--(axis cs:47,1.04866843266151)
		--(axis cs:46,1.04740894279462)
		--(axis cs:45,1.04601154126075)
		--(axis cs:44,1.0444604504856)
		--(axis cs:43,1.04273832069013)
		--(axis cs:42,1.04082604946547)
		--(axis cs:41,1.0387025895852)
		--(axis cs:40,1.03634474382175)
		--(axis cs:39,1.03372694569793)
		--(axis cs:38,1.03082102529414)
		--(axis cs:37,1.02759595945126)
		--(axis cs:36,1.02401760596625)
		--(axis cs:35,1.02004842167617)
		--(axis cs:34,1.01564716467035)
		--(axis cs:33,1.01076858126223)
		--(axis cs:32,1.00536307879142)
		--(axis cs:31,0.999376385818239)
		--(axis cs:30,0.992749201827553)
		--(axis cs:29,0.985416839214)
		--(axis cs:28,0.977308861157885)
		--(axis cs:27,0.96834872018765)
		--(axis cs:26,0.958453404066756)
		--(axis cs:25,0.947533098654117)
		--(axis cs:24,0.935490882368052)
		--(axis cs:23,0.922222474984266)
		--(axis cs:22,0.907616076240521)
		--(axis cs:21,0.891552348938692)
		--(axis cs:20,0.8739046289172)
		--(axis cs:19,0.854539482325209)
		--(axis cs:18,0.83331778070382)
		--(axis cs:17,0.81009652785552)
		--(axis cs:16,0.784731750922888)
		--(axis cs:15,0.757082864154033)
		--(axis cs:14,0.727019032509831)
		--(axis cs:13,0.69442821214757)
		--(axis cs:12,0.659229738778961)
		--(axis cs:11,0.621391590010457)
		--(axis cs:10,0.58095378412986)
		--(axis cs:9,0.538059816033979)
		--(axis cs:8,0.492998587003524)
		--(axis cs:7,0.446259957808997)
		--(axis cs:6,0.39860779591466)
		--(axis cs:5,0.3511750237519)
		--(axis cs:4,0.305585197771564)
		--(axis cs:3,0.264103113520999)
		--(axis cs:2,0.229808560274373)
		--(axis cs:1,0.206758599609918)
		--(axis cs:0,0.2)
		--cycle;

		\path [draw=red, fill=red, opacity=0.1]
		(axis cs:0,0.2)
		--(axis cs:0,-0.2)
		--(axis cs:1,-0.201048854591652)
		--(axis cs:2,-0.199668289112924)
		--(axis cs:3,-0.19580713756702)
		--(axis cs:4,-0.189784157515455)
		--(axis cs:5,-0.181617943539743)
		--(axis cs:6,-0.171445417708754)
		--(axis cs:7,-0.15936481912919)
		--(axis cs:8,-0.145466881772375)
		--(axis cs:9,-0.129834138565607)
		--(axis cs:10,-0.112542026113767)
		--(axis cs:11,-0.0936600741143612)
		--(axis cs:12,-0.0732530587960176)
		--(axis cs:13,-0.0513821021038156)
		--(axis cs:14,-0.0281057242414121)
		--(axis cs:15,-0.00348086812055665)
		--(axis cs:16,0.0224360804860598)
		--(axis cs:17,0.0495882505266914)
		--(axis cs:18,0.0779171928568784)
		--(axis cs:19,0.107361707994629)
		--(axis cs:20,0.137856565165752)
		--(axis cs:21,0.169331084251274)
		--(axis cs:22,0.201707555221271)
		--(axis cs:23,0.234899476214091)
		--(axis cs:24,0.268809603825064)
		--(axis cs:25,0.303327830528753)
		--(axis cs:26,0.33832893914033)
		--(axis cs:27,0.373670340771224)
		--(axis cs:28,0.409189993081083)
		--(axis cs:29,0.444704829739935)
		--(axis cs:30,0.480010200553038)
		--(axis cs:31,0.514880980880497)
		--(axis cs:32,0.54907507887969)
		--(axis cs:33,0.582339943098409)
		--(axis cs:34,0.614422254557706)
		--(axis cs:35,0.645080276285799)
		--(axis cs:36,0.674097504491022)
		--(axis cs:37,0.701295636491363)
		--(axis cs:38,0.726544740591841)
		--(axis cs:39,0.74976898976844)
		--(axis cs:40,0.770947253437407)
		--(axis cs:41,0.790108898007454)
		--(axis cs:42,0.807325980562824)
		--(axis cs:43,0.822703422609447)
		--(axis cs:44,0.83636871113971)
		--(axis cs:45,0.848462329924926)
		--(axis cs:46,0.859129660893386)
		--(axis cs:47,0.868514663777789)
		--(axis cs:48,0.876755321415277)
		--(axis cs:49,0.883980644741474)
		--(axis cs:50,0.890308945046587)
		--(axis cs:51,0.895847067951704)
		--(axis cs:52,0.900690312332799)
		--(axis cs:53,0.904922804662179)
		--(axis cs:54,0.908618150192329)
		--(axis cs:55,0.911840229171764)
		--(axis cs:56,0.914644045551667)
		--(axis cs:57,0.917076566812861)
		--(axis cs:58,0.919177517386802)
		--(axis cs:59,0.920980105982533)
		--(axis cs:60,0.92251168036918)
		--(axis cs:61,0.923794313053783)
		--(axis cs:62,0.924845328832458)
		--(axis cs:63,0.925677791084726)
		--(axis cs:64,0.926300968353309)
		--(axis cs:65,0.926720806405835)
		--(axis cs:66,0.926940433687891)
		--(axis cs:67,0.926960729964041)
		--(axis cs:68,0.926780989376507)
		--(axis cs:69,0.926399710986332)
		--(axis cs:70,0.925815553565212)
		--(axis cs:71,0.925028498820469)
		--(axis cs:72,0.924041279403407)
		--(axis cs:73,0.922861141836698)
		--(axis cs:74,0.921502014972648)
		--(axis cs:75,0.919987102500623)
		--(axis cs:75,1.07168185784922)
		--(axis cs:75,1.07168185784922)
		--(axis cs:74,1.06970938366955)
		--(axis cs:73,1.06784320476997)
		--(axis cs:72,1.06608691053017)
		--(axis cs:71,1.06443623316776)
		--(axis cs:70,1.06288090819057)
		--(axis cs:69,1.06140620314521)
		--(axis cs:68,1.05999410640377)
		--(axis cs:67,1.05862422371377)
		--(axis cs:66,1.05727443271704)
		--(axis cs:65,1.05592133405723)
		--(axis cs:64,1.05454052712734)
		--(axis cs:63,1.05310673223691)
		--(axis cs:62,1.05159377810525)
		--(axis cs:61,1.04997447242688)
		--(axis cs:60,1.04822037253063)
		--(axis cs:59,1.04630147225118)
		--(axis cs:58,1.04418582000143)
		--(axis cs:57,1.04183908199919)
		--(axis cs:56,1.03922406417475)
		--(axis cs:55,1.03630020705627)
		--(axis cs:54,1.03302307049407)
		--(axis cs:53,1.02934383002707)
		--(axis cs:52,1.02520881455336)
		--(axis cs:51,1.02055912616774)
		--(axis cs:50,1.01533039773974)
		--(axis cs:49,1.00945276165393)
		--(axis cs:48,1.00285112278594)
		--(axis cs:47,0.995445847359873)
		--(axis cs:46,0.987153991754694)
		--(axis cs:45,0.977891193914238)
		--(axis cs:44,0.96757432475958)
		--(axis cs:43,0.956124937352496)
		--(axis cs:42,0.943473450380004)
		--(axis cs:41,0.929563862193756)
		--(axis cs:40,0.914358630338582)
		--(axis cs:39,0.897843205558208)
		--(axis cs:38,0.880029626408901)
		--(axis cs:37,0.860958603123326)
		--(axis cs:36,0.840699664106558)
		--(axis cs:35,0.819349186753952)
		--(axis cs:34,0.797026438996998)
		--(axis cs:33,0.773868059718599)
		--(axis cs:32,0.750021644301905)
		--(axis cs:31,0.725639219067644)
		--(axis cs:30,0.70087134570492)
		--(axis cs:29,0.675862399361678)
		--(axis cs:28,0.650747276781391)
		--(axis cs:27,0.625649511273313)
		--(axis cs:26,0.60068057555825)
		--(axis cs:25,0.575940065190497)
		--(axis cs:24,0.551516456048525)
		--(axis cs:23,0.527488184016929)
		--(axis cs:22,0.50392486741521)
		--(axis cs:21,0.480888557296238)
		--(axis cs:20,0.458434947889376)
		--(axis cs:19,0.436614511133118)
		--(axis cs:18,0.415473540203717)
		--(axis cs:17,0.395055100655786)
		--(axis cs:16,0.375399896402211)
		--(axis cs:15,0.356547062834272)
		--(axis cs:14,0.338534902123618)
		--(axis cs:13,0.321401577029994)
		--(axis cs:12,0.305185779930591)
		--(axis cs:11,0.289927393576561)
		--(axis cs:10,0.275668159267279)
		--(axis cs:9,0.262452366392969)
		--(axis cs:8,0.25032757391725)
		--(axis cs:7,0.239345367681753)
		--(axis cs:6,0.229562142686308)
		--(axis cs:5,0.221039841477116)
		--(axis cs:4,0.213845611782728)
		--(axis cs:3,0.208071122484496)
		--(axis cs:2,0.203737924061551)
		--(axis cs:1,0.201048854591652)
		--(axis cs:0,0.2)
		--cycle;

		\addplot [blue, thick]
		table {%
				0 0
				1 -1.15648231731787e-18
				2 0.024212798829618
				3 0.0644177132000001
				4 0.11459886982237
				5 0.170395678527449
				6 0.228686495009208
				7 0.287273558506732
				8 0.344644900453557
				9 0.399794719608668
				10 0.452088138106062
				11 0.501159619524234
				12 0.546836894749291
				13 0.589084194859892
				14 0.627960077920536
				15 0.663586269279249
				16 0.696124797217808
				17 0.725761361996133
				18 0.752693375550628
				19 0.77712148875662
				20 0.799243711755583
				21 0.819251452102641
				22 0.837326961961998
				23 0.853641811879364
				24 0.868356104399057
				25 0.88161821328776
				26 0.893564888950689
				27 0.904321612023282
				28 0.914003108323653
				29 0.922713961816884
				30 0.930549279843115
				31 0.937595378020172
				32 0.943930462030353
				33 0.94962529076197
				34 0.954743810623781
				35 0.959343754758391
				36 0.963477203708293
				37 0.967191106107656
				38 0.970527759390218
				39 0.973525251474136
				40 0.976217865025921
				41 0.978636446305763
				42 0.980808740822097
				43 0.982759698123691
				44 0.984511748069984
				45 0.986085050872231
				46 0.987497723109673
				47 0.988766041811153
				48 0.989904628564394
				49 0.99092661548011
				50 0.991843794701992
				51 0.992666753020137
				52 0.993404993017368
				53 0.994067042057011
				54 0.994660550308026
				55 0.995192378899968
				56 0.995668679206386
				57 0.996094964171474
				58 0.99647617252153
				59 0.996816726640453
				60 0.997120584838001
				61 0.997391288701813
				62 0.997632006200925
				63 0.997845571201771
				64 0.998034520070674
				65 0.998201126073801
				66 0.998347432352394
				67 0.998475284355764
				68 0.99858636276773
				69 0.998682218178213
				70 0.99876430904921
				71 0.998834044928213
				72 0.998892837404327
				73 0.998942162025739
				74 0.998983635358604
				75 0.999019112645051
			};
		\addlegendentry{Stochastic}

		\addplot [red, thick]
		table {%
				0 0
				1 8.67361737988404e-19
				2 0.00203481747431322
				3 0.0061319924587382
				4 0.0120307271336367
				5 0.0197109489686861
				6 0.029058362488777
				7 0.0399902742762815
				8 0.0524303460724376
				9 0.0663091139136812
				10 0.0815630665767557
				11 0.0981336597311001
				12 0.115966360567287
				13 0.135009737463089
				14 0.155214588941103
				15 0.176533097356858
				16 0.198917988444135
				17 0.222321675591239
				18 0.246695366530298
				19 0.271988109563873
				20 0.298145756527564
				21 0.325109820773756
				22 0.35281621131824
				23 0.38119383011551
				24 0.410163029936795
				25 0.439633947859625
				26 0.46950475734929
				27 0.499659926022269
				28 0.529968634931237
				29 0.560283614550807
				30 0.590440773128979
				31 0.62026009997407
				32 0.649548361590798
				33 0.678104001408504
				34 0.705724346777352
				35 0.732214731519875
				36 0.75739858429879
				37 0.781127119807344
				38 0.803287183500371
				39 0.823806097663324
				40 0.842652941887995
				41 0.859836380100605
				42 0.875399715471414
				43 0.889414179980971
				44 0.901971517949645
				45 0.913176761919582
				46 0.92314182632404
				47 0.931980255568831
				48 0.939803222100609
				49 0.946716703197702
				50 0.952819671393165
				51 0.958203097059721
				52 0.962949563443081
				53 0.967133317344626
				54 0.970820610343198
				55 0.974070218114019
				56 0.976934054863207
				57 0.979457824406025
				58 0.981681668694118
				59 0.983640789116855
				60 0.985366026449906
				61 0.986884392740331
				62 0.988219553468855
				63 0.989392261660819
				64 0.990420747740326
				65 0.991321070231532
				66 0.992107433202465
				67 0.992792476838908
				68 0.993387547890139
				69 0.993902957065771
				70 0.99434823087789
				71 0.994732365994116
				72 0.995064094966789
				73 0.995352173303332
				74 0.995605699321099
				75 0.99583448017492
			};
		\addlegendentry{Robust}

		\nextgroupplot[
			height=4.25cm,
			width=8.5cm,
			tick pos=both,
			tick align=inside,
			minor tick num=4,
			xmajorticks=false,
			ymajorticks=true,
			xmajorgrids,
			ymajorgrids,
			xtick style={color=black},
			ytick style={color=black},
			x grid style={white!69.0196078431373!black},
			y grid style={white!69.0196078431373!black},
			xmin=-3.75, xmax=78.75,
			ymin=-0.723153681572885, ymax=7.46756729604455,
			ytick={0, 3, 6},
			every major tick/.append style={major tick length=5pt, black},
			every minor tick/.append style={minor tick length=3pt, black},
			major grid style={dashed, gray},
			ylabel=$\dot{x}$
		]

		\path [draw=blue, fill=blue, opacity=0.1]
		(axis cs:0,0.2)
		--(axis cs:0,-0.2)
		--(axis cs:1,1.91301501791112)
		--(axis cs:2,3.1975075488068)
		--(axis cs:3,3.98376832096461)
		--(axis cs:4,4.41781796484913)
		--(axis cs:5,4.60288555839533)
		--(axis cs:6,4.61418604797827)
		--(axis cs:7,4.50670173312461)
		--(axis cs:8,4.32054231022759)
		--(axis cs:9,4.08483194376805)
		--(axis cs:10,3.82056985057904)
		--(axis cs:11,3.54274766265421)
		--(axis cs:12,3.26192117926274)
		--(axis cs:13,2.9853790298044)
		--(axis cs:14,2.71801246161949)
		--(axis cs:15,2.46296296373724)
		--(axis cs:16,2.22210439703886)
		--(axis cs:17,1.99640158920617)
		--(axis cs:18,1.7861765142626)
		--(axis cs:19,1.59130518558607)
		--(axis cs:20,1.41136250769018)
		--(axis cs:21,1.24572800982969)
		--(axis cs:22,1.09366221196025)
		--(axis cs:23,0.954361035757008)
		--(axis cs:24,0.826993930085421)
		--(axis cs:25,0.710730054050802)
		--(axis cs:26,0.604755828087772)
		--(axis cs:27,0.508286345261104)
		--(axis cs:28,0.420572483837408)
		--(axis cs:29,0.340905050226156)
		--(axis cs:30,0.268616889031184)
		--(axis cs:31,0.203083606514669)
		--(axis cs:32,0.143723346965115)
		--(axis cs:33,0.0899959197500443)
		--(axis cs:34,0.041401480948651)
		--(axis cs:35,-0.00252108725811595)
		--(axis cs:36,-0.042195995189574)
		--(axis cs:37,-0.0780124832829379)
		--(axis cs:38,-0.110326816771285)
		--(axis cs:39,-0.139464287571205)
		--(axis cs:40,-0.165721182640655)
		--(axis cs:41,-0.189366685219158)
		--(axis cs:42,-0.210644681075351)
		--(axis cs:43,-0.229775446413408)
		--(axis cs:44,-0.24695719756239)
		--(axis cs:45,-0.262367485086641)
		--(axis cs:46,-0.276164416616952)
		--(axis cs:47,-0.288487693658151)
		--(axis cs:48,-0.299459448092166)
		--(axis cs:49,-0.309184864378978)
		--(axis cs:50,-0.317752574018183)
		--(axis cs:51,-0.325234810335088)
		--(axis cs:52,-0.33168731505414)
		--(axis cs:53,-0.337148994775884)
		--(axis cs:54,-0.34164133726439)
		--(axis cs:55,-0.345167616918824)
		--(axis cs:56,-0.347711949238505)
		--(axis cs:57,-0.349238299554964)
		--(axis cs:58,-0.349689616453683)
		--(axis cs:59,-0.348987349931315)
		--(axis cs:60,-0.34703173248231)
		--(axis cs:61,-0.343703350026446)
		--(axis cs:62,-0.338866707548586)
		--(axis cs:63,-0.332376696307654)
		--(axis cs:64,-0.324089088678958)
		--(axis cs:65,-0.313876423217591)
		--(axis cs:66,-0.301650922682021)
		--(axis cs:67,-0.287396498662483)
		--(axis cs:68,-0.271212634356389)
		--(axis cs:69,-0.253374338636019)
		--(axis cs:70,-0.23441477114287)
		--(axis cs:71,-0.215240003077753)
		--(axis cs:72,-0.197283791429053)
		--(axis cs:73,-0.18268600964746)
		--(axis cs:74,-0.174397223593909)
		--(axis cs:75,-0.176013940302675)
		--(axis cs:75,0.182354054620736)
		--(axis cs:75,0.182354054620736)
		--(axis cs:74,0.181492680883349)
		--(axis cs:73,0.190980676220599)
		--(axis cs:72,0.207148715711397)
		--(axis cs:71,0.226998498300596)
		--(axis cs:70,0.248361946943429)
		--(axis cs:69,0.26979251283551)
		--(axis cs:68,0.29038371645283)
		--(axis cs:67,0.309612181055844)
		--(axis cs:66,0.327221323355968)
		--(axis cs:65,0.343137678936277)
		--(axis cs:64,0.357410289304334)
		--(axis cs:63,0.370166470088271)
		--(axis cs:62,0.381579707717802)
		--(axis cs:61,0.391846849848821)
		--(axis cs:60,0.401172505244636)
		--(axis cs:59,0.409758989440812)
		--(axis cs:58,0.417800440238399)
		--(axis cs:57,0.425479969566123)
		--(axis cs:56,0.432968942256061)
		--(axis cs:55,0.44042767820239)
		--(axis cs:54,0.44800705565293)
		--(axis cs:53,0.455850644978775)
		--(axis cs:52,0.464097122982736)
		--(axis cs:51,0.47288280978141)
		--(axis cs:50,0.482344237647021)
		--(axis cs:49,0.492620708755512)
		--(axis cs:48,0.503856831235295)
		--(axis cs:47,0.516205044306352)
		--(axis cs:46,0.529828156913009)
		--(axis cs:45,0.544901932574968)
		--(axis cs:44,0.561617758011925)
		--(axis cs:43,0.580185435671905)
		--(axis cs:42,0.600836141394221)
		--(axis cs:41,0.623825588485895)
		--(axis cs:40,0.649437438609092)
		--(axis cs:39,0.677986997928143)
		--(axis cs:38,0.709825233554914)
		--(axis cs:37,0.745343139795337)
		--(axis cs:36,0.784976475062262)
		--(axis cs:35,0.829210877238373)
		--(axis cs:34,0.878587345973428)
		--(axis cs:33,0.933708052612096)
		--(axis cs:32,0.995242399358299)
		--(axis cs:31,1.06393319552166)
		--(axis cs:30,1.14060274638018)
		--(axis cs:29,1.22615855502001)
		--(axis cs:28,1.32159821480884)
		--(axis cs:27,1.42801291481302)
		--(axis cs:26,1.54658878643094)
		--(axis cs:25,1.67860507853483)
		--(axis cs:24,1.82542784765533)
		--(axis cs:23,1.98849746818157)
		--(axis cs:22,2.169307771513)
		--(axis cs:21,2.36937396204168)
		--(axis cs:20,2.59018556172136)
		--(axis cs:19,2.8331394142065)
		--(axis cs:18,3.09944612693588)
		--(axis cs:17,3.39000112169293)
		--(axis cs:16,3.70520855862605)
		--(axis cs:15,4.04474262397455)
		--(axis cs:14,4.40722581012311)
		--(axis cs:13,4.78979758232433)
		--(axis cs:12,5.18753884285764)
		--(axis cs:11,5.59270738235702)
		--(axis cs:10,5.99372643305541)
		--(axis cs:9,6.37385175571065)
		--(axis cs:8,6.70942152079464)
		--(axis cs:7,6.96756665624037)
		--(axis cs:6,7.10322665152662)
		--(axis cs:5,7.05527773795655)
		--(axis cs:4,6.74154377616664)
		--(axis cs:3,6.05246300350931)
		--(axis cs:2,4.84347532526961)
		--(axis cs:1,2.92954474801249)
		--(axis cs:0,0.2)
		--cycle;

		\path [draw=red, fill=red, opacity=0.1]
		(axis cs:0,0.2)
		--(axis cs:0,-0.2)
		--(axis cs:1,0.0309248197048322)
		--(axis cs:2,0.237173789009302)
		--(axis cs:3,0.402519248141242)
		--(axis cs:4,0.55761548587013)
		--(axis cs:5,0.698361301531399)
		--(axis cs:6,0.829595801918528)
		--(axis cs:7,0.952865741272116)
		--(axis cs:8,1.06927786482417)
		--(axis cs:9,1.1796279300285)
		--(axis cs:10,1.28450909359507)
		--(axis cs:11,1.38436829273202)
		--(axis cs:12,1.47953725469584)
		--(axis cs:13,1.57025057454558)
		--(axis cs:14,1.65665637716722)
		--(axis cs:15,1.73882213571003)
		--(axis cs:16,1.8167369268954)
		--(axis cs:17,1.89031081225574)
		--(axis cs:18,1.95937175461623)
		--(axis cs:19,2.0236603613259)
		--(axis cs:20,2.08282273861872)
		--(axis cs:21,2.13640183674576)
		--(axis cs:22,2.18382787850684)
		--(axis cs:23,2.22440882433072)
		--(axis cs:24,2.25732237637974)
		--(axis cs:25,2.28161183794845)
		--(axis cs:26,2.29618938861226)
		--(axis cs:27,2.29985220183969)
		--(axis cs:28,2.2913190781816)
		--(axis cs:29,2.26929665763519)
		--(axis cs:30,2.23258285064279)
		--(axis cs:31,2.18020916499363)
		--(axis cs:32,2.11161228232886)
		--(axis cs:33,2.02680988535559)
		--(axis cs:34,1.92654192452985)
		--(axis cs:35,1.81233497574698)
		--(axis cs:36,1.68645988009676)
		--(axis cs:37,1.55177885212494)
		--(axis cs:38,1.41150757247109)
		--(axis cs:39,1.26893882846942)
		--(axis cs:40,1.12717987271504)
		--(axis cs:41,0.988945752900786)
		--(axis cs:42,0.856431303808142)
		--(axis cs:43,0.73126352507177)
		--(axis cs:44,0.61452037244448)
		--(axis cs:45,0.5067944351491)
		--(axis cs:46,0.408279753875042)
		--(axis cs:47,0.318864351974496)
		--(axis cs:48,0.238216937482833)
		--(axis cs:49,0.16586163115178)
		--(axis cs:50,0.101238561717312)
		--(axis cs:51,0.0437506484815021)
		--(axis cs:52,-0.00720182772340933)
		--(axis cs:53,-0.0521967821706993)
		--(axis cs:54,-0.0917739140456885)
		--(axis cs:55,-0.126424450733781)
		--(axis cs:56,-0.156585504177431)
		--(axis cs:57,-0.1826378549062)
		--(axis cs:58,-0.204906333598742)
		--(axis cs:59,-0.223662302946097)
		--(axis cs:60,-0.239128049567558)
		--(axis cs:61,-0.251483177444817)
		--(axis cs:62,-0.260873353868193)
		--(axis cs:63,-0.26742200165132)
		--(axis cs:64,-0.27124576815573)
		--(axis cs:65,-0.272474854551261)
		--(axis cs:66,-0.271279596596155)
		--(axis cs:67,-0.267905099473578)
		--(axis cs:68,-0.262716246764658)
		--(axis cs:69,-0.256255797531031)
		--(axis cs:70,-0.249317614320523)
		--(axis cs:71,-0.243032792196837)
		--(axis cs:72,-0.238953958464972)
		--(axis cs:73,-0.239099803010749)
		--(axis cs:74,-0.245905326583784)
		--(axis cs:75,-0.262070718204966)
		--(axis cs:75,0.30506667397311)
		--(axis cs:75,0.30506667397311)
		--(axis cs:74,0.291661497347824)
		--(axis cs:73,0.289805006564181)
		--(axis cs:72,0.296569625773642)
		--(axis cs:71,0.309378586731464)
		--(axis cs:70,0.326144637565714)
		--(axis cs:69,0.345310559954711)
		--(axis cs:68,0.365798081891021)
		--(axis cs:67,0.386919309719939)
		--(axis cs:66,0.408288323884631)
		--(axis cs:65,0.429747448737922)
		--(axis cs:64,0.451310266397023)
		--(axis cs:63,0.473119217552803)
		--(axis cs:62,0.495414992260954)
		--(axis cs:61,0.518515323149581)
		--(axis cs:60,0.542801307652546)
		--(axis cs:59,0.568709769556343)
		--(axis cs:58,0.596730418146117)
		--(axis cs:57,0.627406712524874)
		--(axis cs:56,0.661339412740947)
		--(axis cs:55,0.699191800571368)
		--(axis cs:54,0.741695468210082)
		--(axis cs:53,0.789655381885078)
		--(axis cs:52,0.843952608032393)
		--(axis cs:51,0.905542628190373)
		--(axis cs:50,0.975446571593857)
		--(axis cs:49,1.05473200794093)
		--(axis cs:48,1.14447928193569)
		--(axis cs:47,1.24572895438113)
		--(axis cs:46,1.35940609508316)
		--(axis cs:45,1.48621844574261)
		--(axis cs:44,1.62652842154283)
		--(axis cs:43,1.78020406866287)
		--(axis cs:42,1.94646159810336)
		--(axis cs:41,2.12372132126108)
		--(axis cs:40,2.30950776980694)
		--(axis cs:39,2.50043001646473)
		--(axis cs:38,2.69227526011947)
		--(axis cs:37,2.88023388648041)
		--(axis cs:36,3.05924722161412)
		--(axis cs:35,3.22443558003597)
		--(axis cs:34,3.37153502397487)
		--(axis cs:33,3.497259188414)
		--(axis cs:32,3.5995156812124)
		--(axis cs:31,3.67744315835183)
		--(axis cs:30,3.7312825183755)
		--(axis cs:29,3.76213505799921)
		--(axis cs:28,3.77167684573235)
		--(axis cs:27,3.76188957995396)
		--(axis cs:26,3.73484434598347)
		--(axis cs:25,3.69255005998452)
		--(axis cs:24,3.63686120818639)
		--(axis cs:23,3.56943113992625)
		--(axis cs:22,3.49169588094708)
		--(axis cs:21,3.40487627215106)
		--(axis cs:20,3.30999011061969)
		--(axis cs:19,3.20786903141219)
		--(axis cs:18,3.09917685209894)
		--(axis cs:17,2.98442737555607)
		--(axis cs:16,2.86400050252524)
		--(axis cs:15,2.7381560817455)
		--(axis cs:14,2.6070453059837)
		--(axis cs:13,2.47071972105725)
		--(axis cs:12,2.32913812446463)
		--(axis cs:11,2.1821718745053)
		--(axis cs:10,2.0296095372738)
		--(axis cs:9,1.87116260258641)
		--(axis cs:8,1.70647570342455)
		--(axis cs:7,1.5351486179591)
		--(axis cs:6,1.35678655558236)
		--(axis cs:5,1.17112140248679)
		--(axis cs:4,0.978428881139749)
		--(axis cs:3,0.777227686838458)
		--(axis cs:2,0.582261207875695)
		--(axis cs:1,0.376038675157811)
		--(axis cs:0,0.2)
		--cycle;

		\addplot [blue, thick]
		table {%
				0 0
				1 2.4212798829618
				2 4.0204914370382
				3 5.01811566223696
				4 5.57968087050788
				5 5.82908164817594
				6 5.85870634975245
				7 5.73713419468249
				8 5.51498191551112
				9 5.22934184973935
				10 4.90714814181723
				11 4.56772752250561
				12 4.22473001106019
				13 3.88758830606437
				14 3.5626191358713
				15 3.25385279385589
				16 2.96365647783246
				17 2.69320135544955
				18 2.44281132059924
				19 2.21222229989628
				20 2.00077403470577
				21 1.80755098593569
				22 1.63148499173663
				23 1.47142925196929
				24 1.32621088887038
				25 1.19466756629282
				26 1.07567230725936
				27 0.96814963003706
				28 0.871085349323126
				29 0.783531802623083
				30 0.704609817705681
				31 0.633508401018164
				32 0.569482873161707
				33 0.51185198618107
				34 0.459994413461039
				35 0.413344894990129
				36 0.371390239936344
				37 0.333665328256199
				38 0.299749208391814
				39 0.269261355178469
				40 0.241858127984219
				41 0.217229451633369
				42 0.195095730159435
				43 0.175204994629249
				44 0.157330280224768
				45 0.141267223744163
				46 0.126831870148028
				47 0.113858675324101
				48 0.102198691571564
				49 0.0917179221882669
				50 0.0822958318144188
				51 0.0738239997231608
				52 0.066204903964298
				53 0.0593508251014455
				54 0.0531828591942702
				55 0.047630030641783
				56 0.0426284965087777
				57 0.0381208350055797
				58 0.0340554118923576
				59 0.0303858197547481
				60 0.0270703863811632
				61 0.0240717499111874
				62 0.0213565000846082
				63 0.0188948868903081
				64 0.0166606003126882
				65 0.0146306278593429
				66 0.0127852003369733
				67 0.0111078411966804
				68 0.00958554104822073
				69 0.00820908709974525
				70 0.00697358790027969
				71 0.0058792476114213
				72 0.00493246214117204
				73 0.0041473332865695
				74 0.00354772864471996
				75 0.00317005715903056
			};
		\addplot [red, thick]
		table {%
				0 0
				1 0.203481747431322
				2 0.409717498442498
				3 0.58987346748985
				4 0.768022183504939
				5 0.934741352009096
				6 1.09319117875045
				7 1.24400717961561
				8 1.38787678412436
				9 1.52539526630745
				10 1.65705931543443
				11 1.78327008361866
				12 1.90433768958023
				13 2.02048514780141
				14 2.13185084157546
				15 2.23848910872776
				16 2.34036871471032
				17 2.4373690939059
				18 2.52927430335758
				19 2.61576469636905
				20 2.69640642461921
				21 2.77063905444841
				22 2.83776187972696
				23 2.89691998212849
				24 2.94709179228306
				25 2.98708094896649
				26 3.01551686729787
				27 3.03087089089683
				28 3.03149796195697
				29 3.0157158578172
				30 2.98193268450914
				31 2.92882616167273
				32 2.85556398177063
				33 2.76203453688479
				34 2.64903847425236
				35 2.51838527789148
				36 2.37285355085544
				37 2.21600636930267
				38 2.05189141629528
				39 1.88468442246708
				40 1.71834382126099
				41 1.55633353708093
				42 1.40144645095575
				43 1.25573379686732
				44 1.12052439699365
				45 0.996506440445855
				46 0.883842924479102
				47 0.782296653177812
				48 0.691348109709264
				49 0.610296819546357
				50 0.538342566655584
				51 0.474646638335937
				52 0.418375390154492
				53 0.36872929985719
				54 0.324960777082197
				55 0.286383674918794
				56 0.252376954281758
				57 0.222384428809337
				58 0.195912042273688
				59 0.172523733305123
				60 0.151836629042494
				61 0.133516072852382
				62 0.11727081919638
				63 0.102848607950742
				64 0.0900322491206463
				65 0.0786362970933306
				66 0.0685043636442383
				67 0.0595071051231808
				68 0.0515409175631816
				69 0.0445273812118397
				70 0.0384135116225956
				71 0.0331728972673137
				72 0.0288078336543349
				73 0.0253526017767159
				74 0.0228780853820199
				75 0.0214979778840721
			};

		\nextgroupplot[
			height=4.25cm,
			width=8.5cm,
			tick pos=both,
			tick align=inside,
			minor tick num=4,
			xmajorticks=true,
			ymajorticks=true,
			xmajorgrids,
			ymajorgrids,
			xtick style={color=black},
			ytick style={color=black},
			x grid style={white!69.0196078431373!black},
			y grid style={white!69.0196078431373!black},
			xmin=-3.75, xmax=78.75,
			ymin=-62.6877902547486, ymax=307.202666517506,
			ytick={0, 150, 300},
			every major tick/.append style={major tick length=5pt, black},
			every minor tick/.append style={minor tick length=3pt, black},
			major grid style={dashed, gray},
			ylabel=$u$,
			xlabel={Time Steps},
		]

		\path [draw=blue, fill=blue, opacity=0.1]
		(axis cs:0,291.39288465128)
		--(axis cs:0,192.863091941081)
		--(axis cs:1,124.735834030069)
		--(axis cs:2,75.3807679110053)
		--(axis cs:3,39.5497322779348)
		--(axis cs:4,13.229781412291)
		--(axis cs:5,-6.52521411095339)
		--(axis cs:6,-21.3483587957925)
		--(axis cs:7,-31.9441379729253)
		--(axis cs:8,-38.9585253238326)
		--(axis cs:9,-43.1398109587364)
		--(axis cs:10,-45.192434530222)
		--(axis cs:11,-45.7036303670234)
		--(axis cs:12,-45.135619218848)
		--(axis cs:13,-43.8411323519077)
		--(axis cs:14,-42.0837794808728)
		--(axis cs:15,-40.0570847768155)
		--(axis cs:16,-37.9005415090687)
		--(axis cs:17,-35.7125726320155)
		--(axis cs:18,-33.5607588168336)
		--(axis cs:19,-31.4897959871762)
		--(axis cs:20,-29.5276190392436)
		--(axis cs:21,-27.6900674992785)
		--(axis cs:22,-25.9844045222066)
		--(axis cs:23,-24.4119428957827)
		--(axis cs:24,-22.969982524193)
		--(axis cs:25,-21.6532226299049)
		--(axis cs:26,-20.4547775750775)
		--(axis cs:27,-19.3668968381569)
		--(axis cs:28,-18.3814665375418)
		--(axis cs:29,-17.4903513105089)
		--(axis cs:30,-16.6856206934999)
		--(axis cs:31,-15.9596927809697)
		--(axis cs:32,-15.3054192637493)
		--(axis cs:33,-14.7161294173309)
		--(axis cs:34,-14.1856457509987)
		--(axis cs:35,-13.708280445618)
		--(axis cs:36,-13.2788190843133)
		--(axis cs:37,-12.8924962690895)
		--(axis cs:38,-12.5449663290239)
		--(axis cs:39,-12.2322713206227)
		--(axis cs:40,-11.950807793724)
		--(axis cs:41,-11.697293270546)
		--(axis cs:42,-11.4687330059684)
		--(axis cs:43,-11.2623873248311)
		--(axis cs:44,-11.075739640138)
		--(axis cs:45,-10.9064651275453)
		--(axis cs:46,-10.7523999577688)
		--(axis cs:47,-10.6115109685902)
		--(axis cs:48,-10.4818656990407)
		--(axis cs:49,-10.3616028259523)
		--(axis cs:50,-10.2489032632041)
		--(axis cs:51,-10.1419625436355)
		--(axis cs:52,-10.0389656506688)
		--(axis cs:53,-9.93806625688288)
		--(axis cs:54,-9.83737341379685)
		--(axis cs:55,-9.73495015060685)
		--(axis cs:56,-9.62883014285232)
		--(axis cs:57,-9.51706042816121)
		--(axis cs:58,-9.39777964354292)
		--(axis cs:59,-9.26934159179812)
		--(axis cs:60,-9.13049167785529)
		--(axis cs:61,-8.98059675373552)
		--(axis cs:62,-8.81991449806282)
		--(axis cs:63,-8.64986422554485)
		--(axis cs:64,-8.47322685827324)
		--(axis cs:65,-8.2941636411142)
		--(axis cs:66,-8.11791826872562)
		--(axis cs:67,-7.95008771491323)
		--(axis cs:68,-7.79545603777281)
		--(axis cs:69,-7.6566154575304)
		--(axis cs:70,-7.5329438950696)
		--(axis cs:71,-7.42093731290393)
		--(axis cs:72,-7.31751118425744)
		--(axis cs:73,-7.22929020385989)
		--(axis cs:74,-7.19431374032538)
		--(axis cs:74,7.13946866162362)
		--(axis cs:74,7.13946866162362)
		--(axis cs:73,7.13017758538782)
		--(axis cs:72,7.18144976251326)
		--(axis cs:71,7.25273274927492)
		--(axis cs:70,7.33544584105897)
		--(axis cs:69,7.43113107942081)
		--(axis cs:68,7.54205408354271)
		--(axis cs:67,7.66781875914774)
		--(axis cs:66,7.80497042938149)
		--(axis cs:65,7.94796828473363)
		--(axis cs:64,8.09052517806813)
		--(axis cs:63,8.22674279882298)
		--(axis cs:62,8.35181579934375)
		--(axis cs:61,8.46230896417595)
		--(axis cs:60,8.55612087283312)
		--(axis cs:59,8.63226841556489)
		--(axis cs:58,8.69060182184992)
		--(axis cs:57,8.73152187180133)
		--(axis cs:56,8.75573691509861)
		--(axis cs:55,8.76407317771215)
		--(axis cs:54,8.75733748614442)
		--(axis cs:53,8.73622458130924)
		--(axis cs:52,8.70125895875151)
		--(axis cs:51,8.65276152686798)
		--(axis cs:50,8.59083288720948)
		--(axis cs:49,8.5153468679299)
		--(axis cs:48,8.4259496532668)
		--(axis cs:47,8.32206127398396)
		--(axis cs:46,8.20287732147503)
		--(axis cs:45,8.06736955408462)
		--(axis cs:44,7.91428463502356)
		--(axis cs:43,7.74214063682322)
		--(axis cs:42,7.54922122077943)
		--(axis cs:41,7.33356759501211)
		--(axis cs:40,7.09296850645131)
		--(axis cs:39,6.82494865783781)
		--(axis cs:38,6.52675608322098)
		--(axis cs:37,6.19534918398593)
		--(axis cs:36,5.82738434034583)
		--(axis cs:35,5.41920528895427)
		--(axis cs:34,4.96683581572122)
		--(axis cs:33,4.46597777637617)
		--(axis cs:32,3.91201705149481)
		--(axis cs:31,3.30004079744235)
		--(axis cs:30,2.62487030513456)
		--(axis cs:29,1.88111496678936)
		--(axis cs:28,1.06325432956435)
		--(axis cs:27,0.165757053617808)
		--(axis cs:26,-0.816752110150878)
		--(axis cs:25,-1.88926329926262)
		--(axis cs:24,-3.05607269145689)
		--(axis cs:23,-4.32037103736854)
		--(axis cs:22,-5.68369989367459)
		--(axis cs:21,-7.14523611430441)
		--(axis cs:20,-8.70085103359642)
		--(axis cs:19,-10.3418701611724)
		--(axis cs:18,-12.0534291921266)
		--(axis cs:17,-13.812278839717)
		--(axis cs:16,-15.5838291760013)
		--(axis cs:15,-17.3181361437152)
		--(axis cs:14,-18.9444058934756)
		--(axis cs:13,-20.3634023415954)
		--(axis cs:12,-21.4368390402101)
		--(axis cs:11,-21.9723032251694)
		--(axis cs:10,-21.7012179409753)
		--(axis cs:9,-20.245066361349)
		--(axis cs:8,-17.0595985494091)
		--(axis cs:7,-11.333145551242)
		--(axis cs:6,-1.78975721834859)
		--(axis cs:5,13.6193786694597)
		--(axis cs:4,37.7686022728192)
		--(axis cs:3,73.7682208629607)
		--(axis cs:2,124.948659672131)
		--(axis cs:1,195.590732761803)
		--(axis cs:0,291.39288465128)
		--cycle;

		\path [draw=red, fill=red, opacity=0.1]
		(axis cs:0,33.5617124795809)
		--(axis cs:0,7.13463700668349)
		--(axis cs:1,7.93815019988172)
		--(axis cs:2,5.61717643338084)
		--(axis cs:3,5.43613610604209)
		--(axis cs:4,4.35332079108895)
		--(axis cs:5,3.54369476031713)
		--(axis cs:6,2.78401668041039)
		--(axis cs:7,2.08516930088672)
		--(axis cs:8,1.44110468765549)
		--(axis cs:9,0.843620458428601)
		--(axis cs:10,0.284096742517116)
		--(axis cs:11,-0.245999334839128)
		--(axis cs:12,-0.755008214155836)
		--(axis cs:13,-1.25114604186252)
		--(axis cs:14,-1.74265387416846)
		--(axis cs:15,-2.23798722161679)
		--(axis cs:16,-2.74602690492862)
		--(axis cs:17,-3.27630151339415)
		--(axis cs:18,-3.83921409762382)
		--(axis cs:19,-4.44626320712197)
		--(axis cs:20,-5.11024172629184)
		--(axis cs:21,-5.84538583285204)
		--(axis cs:22,-6.66742997098684)
		--(axis cs:23,-7.59350057932637)
		--(axis cs:24,-8.64174640423651)
		--(axis cs:25,-9.83054357023701)
		--(axis cs:26,-11.1770223599815)
		--(axis cs:27,-12.6945886933859)
		--(axis cs:28,-14.3891794077217)
		--(axis cs:29,-16.2543001721203)
		--(axis cs:30,-18.2654531529395)
		--(axis cs:31,-20.3753083165459)
		--(axis cs:32,-22.5116830580067)
		--(axis cs:33,-24.5805047665635)
		--(axis cs:34,-26.4748626921963)
		--(axis cs:35,-28.0891329466949)
		--(axis cs:36,-29.3349716182826)
		--(axis cs:37,-30.154905517271)
		--(axis cs:38,-30.5298975214061)
		--(axis cs:39,-30.4792967718862)
		--(axis cs:40,-30.0540180359234)
		--(axis cs:41,-29.325575813173)
		--(axis cs:42,-28.3741525927131)
		--(axis cs:43,-27.2783006767252)
		--(axis cs:44,-26.1077060704169)
		--(axis cs:45,-24.9192783637572)
		--(axis cs:46,-23.7560392748502)
		--(axis cs:47,-22.6479436988459)
		--(axis cs:48,-21.6137725357766)
		--(axis cs:49,-20.66342274957)
		--(axis cs:50,-19.8001534455355)
		--(axis cs:51,-19.0225506855006)
		--(axis cs:52,-18.3261202758549)
		--(axis cs:53,-17.7045078051167)
		--(axis cs:54,-17.1503917641523)
		--(axis cs:55,-16.6561134627879)
		--(axis cs:56,-16.2141087443926)
		--(axis cs:57,-15.8171994289756)
		--(axis cs:58,-15.4587918101345)
		--(axis cs:59,-15.1330177284821)
		--(axis cs:60,-14.8348415250426)
		--(axis cs:61,-14.5601436053664)
		--(axis cs:62,-14.3057785518023)
		--(axis cs:63,-14.0695936992489)
		--(axis cs:64,-13.8503854018687)
		--(axis cs:65,-13.6477693365924)
		--(axis cs:66,-13.461953956064)
		--(axis cs:67,-13.2934379966728)
		--(axis cs:68,-13.1427058976783)
		--(axis cs:69,-13.0100663237759)
		--(axis cs:70,-12.8958650232638)
		--(axis cs:71,-12.8014154892449)
		--(axis cs:72,-12.7311868771008)
		--(axis cs:73,-12.6971885893856)
		--(axis cs:74,-12.7271403066794)
		--(axis cs:74,12.4756065381527)
		--(axis cs:74,12.4756065381527)
		--(axis cs:73,12.2272628742678)
		--(axis cs:72,12.0658033502072)
		--(axis cs:71,11.9549319934225)
		--(axis cs:70,11.8753118191494)
		--(axis cs:69,11.8160759413107)
		--(axis cs:68,11.7701745618804)
		--(axis cs:67,11.7319577552344)
		--(axis cs:66,11.6960452732454)
		--(axis cs:65,11.6569363275973)
		--(axis cs:64,11.6090098611845)
		--(axis cs:63,11.5466795000531)
		--(axis cs:62,11.4645548575832)
		--(axis cs:61,11.3575337765914)
		--(axis cs:60,11.2208049333577)
		--(axis cs:59,11.0497744383995)
		--(axis cs:58,10.8399460582503)
		--(axis cs:57,10.5867881640957)
		--(axis cs:56,10.285617721862)
		--(axis cs:55,9.9315274747268)
		--(axis cs:54,9.51937989909496)
		--(axis cs:53,9.04389177643649)
		--(axis cs:52,8.49983628569421)
		--(axis cs:51,7.88239443881979)
		--(axis cs:50,7.18769268836513)
		--(axis cs:49,6.41356586938864)
		--(axis cs:48,5.56058018957923)
		--(axis cs:47,4.63333394088307)
		--(axis cs:46,3.64201643601466)
		--(axis cs:45,2.60413999373401)
		--(axis cs:44,1.54625907061495)
		--(axis cs:43,0.505355744965257)
		--(axis cs:42,-0.470580940472811)
		--(axis cs:41,-1.3233779768453)
		--(axis cs:40,-1.98751697699721)
		--(axis cs:39,-2.39541046288513)
		--(axis cs:38,-2.48505721730593)
		--(axis cs:37,-2.2092612679508)
		--(axis cs:36,-1.54474601041368)
		--(axis cs:35,-0.498891110304109)
		--(axis cs:34,0.888145601806015)
		--(axis cs:33,2.54726122748169)
		--(axis cs:32,4.38989784442545)
		--(axis cs:31,6.3210427704594)
		--(axis cs:30,8.25234393802196)
		--(axis cs:29,10.1120143543632)
		--(axis cs:28,11.8496575448572)
		--(axis cs:27,13.4361702821151)
		--(axis cs:26,14.8603205483795)
		--(axis cs:25,16.123936105264)
		--(axis cs:24,17.2371993599763)
		--(axis cs:23,18.2148704832697)
		--(axis cs:22,19.0736591514635)
		--(axis cs:21,19.8305808958672)
		--(axis cs:20,20.5020118921878)
		--(axis cs:19,21.1032015586186)
		--(axis cs:18,21.6480814679188)
		--(axis cs:17,22.1492636560234)
		--(axis cs:16,22.6181548467557)
		--(axis cs:15,23.0651369018208)
		--(axis cs:14,23.4997817647228)
		--(axis cs:13,23.931082020982)
		--(axis cs:12,24.3676867235189)
		--(axis cs:11,24.8181372170719)
		--(axis cs:10,25.2911000187469)
		--(axis cs:9,25.7955946025076)
		--(axis cs:8,26.3412157127083)
		--(axis cs:7,26.9383528422728)
		--(axis cs:6,27.5984028956223)
		--(axis cs:5,28.333613077334)
		--(axis cs:4,29.144357960986)
		--(axis cs:3,30.3117044303229)
		--(axis cs:2,30.4960015721275)
		--(axis cs:1,33.3496963518398)
		--(axis cs:0,33.5617124795809)
		--cycle;

		\addplot [blue, thick]
		table {%
				0 242.12798829618
				1 160.163283395936
				2 100.164713791568
				3 56.6589765704477
				4 25.4991918425551
				5 3.54708227925317
				6 -11.5690580070706
				7 -21.6386417620836
				8 -28.0090619366208
				9 -31.6924386600427
				10 -33.4468262355986
				11 -33.8379667960964
				12 -33.2862291295291
				13 -32.1022673467515
				14 -30.5140926871742
				15 -28.6876104602653
				16 -26.742185342535
				17 -24.7624257358663
				18 -22.8070940044801
				19 -20.9158330741743
				20 -19.11423503642
				21 -17.4176518067914
				22 -15.8340522079406
				23 -14.3661569665756
				24 -13.013027607825
				25 -11.7712429645838
				26 -10.6357648426142
				27 -9.60056989226953
				28 -8.65910610398871
				29 -7.80461817185977
				30 -7.03037519418265
				31 -6.32982599176367
				32 -5.69670110612722
				33 -5.12507582047735
				34 -4.60940496763874
				35 -4.14453757833186
				36 -3.72571737198373
				37 -3.3485735425518
				38 -3.00910512290147
				39 -2.70366133139242
				40 -2.42891964363633
				41 -2.18186283776697
				42 -1.95975589259447
				43 -1.76012334400394
				44 -1.58072750255724
				45 -1.41954778673036
				46 -1.27476131814689
				47 -1.14472484730311
				48 -1.02795802288693
				49 -0.923127979011184
				50 -0.829035187997334
				51 -0.744600508383769
				52 -0.668853345958638
				53 -0.600920837786816
				54 -0.540017963826216
				55 -0.48543848644735
				56 -0.436546613876857
				57 -0.392769278179941
				58 -0.353588910846497
				59 -0.318536588116615
				60 -0.287185402511085
				61 -0.259143894779783
				62 -0.234049349359537
				63 -0.211560713360939
				64 -0.191350840102558
				65 -0.173097678190288
				66 -0.156473919672067
				67 -0.141134477882744
				68 -0.12670097711505
				69 -0.112742189054799
				70 -0.0987490270053186
				71 -0.0841022818145021
				72 -0.0680307108720939
				73 -0.0495563092360385
				74 -0.0274225393508828
			};
		\addplot [red, thick]
		table {%
				0 20.3481747431322
				1 20.6439232758608
				2 18.0565890027542
				3 17.8739202681825
				4 16.7488393760375
				5 15.9386539188255
				6 15.1912097880164
				7 14.5117610715797
				8 13.8911602001819
				9 13.3196075304681
				10 12.787598380632
				11 12.2860689411164
				12 11.8063392546815
				13 11.3399679895597
				14 10.8785639452772
				15 10.413574840102
				16 9.93606397091354
				17 9.43648107131462
				18 8.9044336851475
				19 8.3284691757483
				20 7.69588508294798
				21 6.99259753150758
				22 6.20311459023833
				23 5.31068495197167
				24 4.29772647786987
				25 3.14669626751351
				26 1.841649094199
				27 0.370790794364638
				28 -1.26976093143224
				29 -3.07114290887856
				30 -5.00655460745878
				31 -7.02713277304323
				32 -9.06089260679065
				33 -11.0166217695409
				34 -12.7933585451952
				35 -14.2940120284995
				36 -15.4398588143482
				37 -16.1820833926109
				38 -16.507477369356
				39 -16.4373536173856
				40 -16.0207675064603
				41 -15.3244768950092
				42 -14.422366766593
				43 -13.38647246588
				44 -12.280723499901
				45 -11.1575691850116
				46 -10.0570114194178
				47 -9.00730487898142
				48 -8.02659617309871
				49 -7.12492844009068
				50 -6.30623037858518
				51 -5.57007812334038
				52 -4.91314199508034
				53 -4.33030801434012
				54 -3.81550593252865
				55 -3.36229299403055
				56 -2.96424551126529
				57 -2.61520563243994
				58 -2.30942287594212
				59 -2.04162164504129
				60 -1.80701829584243
				61 -1.6013049143875
				62 -1.42061184710954
				63 -1.26145709959786
				64 -1.1206877703421
				65 -0.995416504497584
				66 -0.882954341409302
				67 -0.780740120719216
				68 -0.686265667898965
				69 -0.596995191232573
				70 -0.510276602057154
				71 -0.423241747911206
				72 -0.332691763446798
				73 -0.23496285755889
				74 -0.125766884263368
			};
	\end{groupplot}

\end{tikzpicture}

%% file: figures/linear_trajectories_adversarial.tex
\begin{tikzpicture}

	\definecolor{color0}{rgb}{1,0,1}

	\begin{groupplot}[group style={group size=1 by 3, vertical sep=0.5cm}]
		\nextgroupplot[
			height=4.25cm,
			width=8.5cm,
			tick pos=both,
			tick align=inside,
			minor tick num=4,
			xmajorticks=false,
			ymajorticks=true,
			xmajorgrids,
			ymajorgrids,
			xtick style={color=black},
			ytick style={color=black},
			x grid style={white!69.0196078431373!black},
			y grid style={white!69.0196078431373!black},
			xmin=-3.75, xmax=78.75,
			ymin=-2.99087359662115, ymax=1.26445431315246,
			every major tick/.append style={major tick length=5pt, black},
			every minor tick/.append style={minor tick length=3pt, black},
			major grid style={dashed, gray},
			legend style={at={(1.,0.)},anchor=south east},
			title={Worst-Case Dynamics $p^{*}(\vec{\theta})$},
		]

		\path [draw=green!50.1960784313725!black, fill=green!50.1960784313725!black, opacity=0.1]
		(axis cs:0,0.2)
		--(axis cs:0,-0.2)
		--(axis cs:1,-1.01177155188598)
		--(axis cs:2,-1.80466390897936)
		--(axis cs:3,-2.410957108609)
		--(axis cs:4,-2.73631013134086)
		--(axis cs:5,-2.79744960072235)
		--(axis cs:6,-2.68056205632227)
		--(axis cs:7,-2.47546130177028)
		--(axis cs:8,-2.24112980450837)
		--(axis cs:9,-2.0059841093695)
		--(axis cs:10,-1.78045783410021)
		--(axis cs:11,-1.56722816338655)
		--(axis cs:12,-1.36641948630212)
		--(axis cs:13,-1.17759625508949)
		--(axis cs:14,-1.00033581770154)
		--(axis cs:15,-0.834310673706299)
		--(axis cs:16,-0.679246036489061)
		--(axis cs:17,-0.534868987799659)
		--(axis cs:18,-0.400875789584254)
		--(axis cs:19,-0.276917281074546)
		--(axis cs:20,-0.162596792200274)
		--(axis cs:21,-0.0574754148643164)
		--(axis cs:22,0.0389190334329373)
		--(axis cs:23,0.12708171120523)
		--(axis cs:24,0.207520549081062)
		--(axis cs:25,0.280746953262309)
		--(axis cs:26,0.347267828178028)
		--(axis cs:27,0.407579070846443)
		--(axis cs:28,0.462160523446933)
		--(axis cs:29,0.511472267483633)
		--(axis cs:30,0.55595208367056)
		--(axis cs:31,0.596013874277352)
		--(axis cs:32,0.632046841329507)
		--(axis cs:33,0.664415229130102)
		--(axis cs:34,0.693458468084917)
		--(axis cs:35,0.719491593429357)
		--(axis cs:36,0.742805851222959)
		--(axis cs:37,0.763669438957031)
		--(axis cs:38,0.782328354591221)
		--(axis cs:39,0.799007343437725)
		--(axis cs:40,0.813910937496612)
		--(axis cs:41,0.827224579346127)
		--(axis cs:42,0.839115816313803)
		--(axis cs:43,0.849735543959874)
		--(axis cs:44,0.859219273318936)
		--(axis cs:45,0.867688394891052)
		--(axis cs:46,0.875251413911488)
		--(axis cs:47,0.882005135142968)
		--(axis cs:48,0.888035780300059)
		--(axis cs:49,0.893420026320027)
		--(axis cs:50,0.898225957390754)
		--(axis cs:51,0.902513927573124)
		--(axis cs:52,0.906337333888406)
		--(axis cs:53,0.909743301920518)
		--(axis cs:54,0.912773287445253)
		--(axis cs:55,0.915463598528663)
		--(axis cs:56,0.917845843138036)
		--(axis cs:57,0.919947307789101)
		--(axis cs:58,0.921791273321318)
		--(axis cs:59,0.9233972747594)
		--(axis cs:60,0.924781313594686)
		--(axis cs:61,0.925956032909026)
		--(axis cs:62,0.926930868745614)
		--(axis cs:63,0.927712195129329)
		--(axis cs:64,0.928303485182248)
		--(axis cs:65,0.928705516774277)
		--(axis cs:66,0.928916657894042)
		--(axis cs:67,0.928933274231812)
		--(axis cs:68,0.928750309436552)
		--(axis cs:69,0.92836209803603)
		--(axis cs:70,0.927763484510806)
		--(axis cs:71,0.926951344355636)
		--(axis cs:72,0.925926642268714)
		--(axis cs:73,0.924697229526603)
		--(axis cs:74,0.923281682708861)
		--(axis cs:75,0.921714592416909)
		--(axis cs:75,1.06853573634091)
		--(axis cs:75,1.06853573634091)
		--(axis cs:74,1.06659735616224)
		--(axis cs:73,1.06475512764855)
		--(axis cs:72,1.0630239808637)
		--(axis cs:71,1.06140548893634)
		--(axis cs:70,1.05989211496721)
		--(axis cs:69,1.05847023126941)
		--(axis cs:68,1.05712226071352)
		--(axis cs:67,1.05582819755818)
		--(axis cs:66,1.0545666754801)
		--(axis cs:65,1.05331569139964)
		--(axis cs:64,1.05205306022787)
		--(axis cs:63,1.05075665767065)
		--(axis cs:62,1.04940449794531)
		--(axis cs:61,1.04797468609676)
		--(axis cs:60,1.04644527832774)
		--(axis cs:59,1.04479407764264)
		--(axis cs:58,1.04299838616366)
		--(axis cs:57,1.04103472998905)
		--(axis cs:56,1.03887856769543)
		--(axis cs:55,1.03650398968621)
		--(axis cs:54,1.03388341256907)
		--(axis cs:53,1.0309872705186)
		--(axis cs:52,1.02778370399703)
		--(axis cs:51,1.02423824509146)
		--(axis cs:50,1.02031349789959)
		--(axis cs:49,1.0159688116807)
		--(axis cs:48,1.01115994371549)
		--(axis cs:47,1.00583870782684)
		--(axis cs:46,0.999952603160348)
		--(axis cs:45,0.993444415996271)
		--(axis cs:44,0.986251785020475)
		--(axis cs:43,0.978306717694635)
		--(axis cs:42,0.969535042402972)
		--(axis cs:41,0.959855778447977)
		--(axis cs:40,0.949180404578932)
		--(axis cs:39,0.937412007725962)
		--(axis cs:38,0.924444298295308)
		--(axis cs:37,0.910160487905056)
		--(axis cs:36,0.894432040304784)
		--(axis cs:35,0.877117325784386)
		--(axis cs:34,0.858060231578164)
		--(axis cs:33,0.837088802357096)
		--(axis cs:32,0.814014002251839)
		--(axis cs:31,0.788628700185111)
		--(axis cs:30,0.760706982762772)
		--(axis cs:29,0.730003894993873)
		--(axis cs:28,0.696255701696531)
		--(axis cs:27,0.659180754787144)
		--(axis cs:26,0.618481045569204)
		--(axis cs:25,0.57384451606583)
		--(axis cs:24,0.524948196145394)
		--(axis cs:23,0.471462218052373)
		--(axis cs:22,0.413054729301003)
		--(axis cs:21,0.349397668837327)
		--(axis cs:20,0.280173276716519)
		--(axis cs:19,0.205081055190393)
		--(axis cs:18,0.123844659841342)
		--(axis cs:17,0.0362178287751314)
		--(axis cs:16,-0.0580121042293024)
		--(axis cs:15,-0.159025369841352)
		--(axis cs:14,-0.266978810279)
		--(axis cs:13,-0.382026460308829)
		--(axis cs:12,-0.504352540681719)
		--(axis cs:11,-0.634203102484746)
		--(axis cs:10,-0.771851513433484)
		--(axis cs:9,-0.917286462905333)
		--(axis cs:8,-1.06906575755466)
		--(axis cs:7,-1.22120340438317)
		--(axis cs:6,-1.35676982358706)
		--(axis cs:5,-1.43940763419945)
		--(axis cs:4,-1.41171563905152)
		--(axis cs:3,-1.21574773362604)
		--(axis cs:2,-0.835753308689071)
		--(axis cs:1,-0.328096819510274)
		--(axis cs:0,0.2)
		--cycle;

		\path [draw=color0, fill=color0, opacity=0.1]
		(axis cs:0,0.2)
		--(axis cs:0,-0.2)
		--(axis cs:1,-0.271536922925765)
		--(axis cs:2,-0.327535081118224)
		--(axis cs:3,-0.361302996676127)
		--(axis cs:4,-0.380031836885674)
		--(axis cs:5,-0.385061363628259)
		--(axis cs:6,-0.380014235085861)
		--(axis cs:7,-0.367796833842594)
		--(axis cs:8,-0.350626922216059)
		--(axis cs:9,-0.330009380268805)
		--(axis cs:10,-0.306885188892345)
		--(axis cs:11,-0.281812871001342)
		--(axis cs:12,-0.255115441908266)
		--(axis cs:13,-0.226979358491773)
		--(axis cs:14,-0.197515069152286)
		--(axis cs:15,-0.166792379266871)
		--(axis cs:16,-0.134860796878329)
		--(axis cs:17,-0.101761336583251)
		--(axis cs:18,-0.0675335806901673)
		--(axis cs:19,-0.0322201501148579)
		--(axis cs:20,0.0041302040051808)
		--(axis cs:21,0.0414601998502409)
		--(axis cs:22,0.0797019174096616)
		--(axis cs:23,0.118774751147411)
		--(axis cs:24,0.158583292293225)
		--(axis cs:25,0.19901508462847)
		--(axis cs:26,0.239938266833488)
		--(axis cs:27,0.281199196649873)
		--(axis cs:28,0.3226202654579)
		--(axis cs:29,0.363998269951796)
		--(axis cs:30,0.40510390291471)
		--(axis cs:31,0.445683111891108)
		--(axis cs:32,0.485461164387378)
		--(axis cs:33,0.524150129446955)
		--(axis cs:34,0.561460024400371)
		--(axis cs:35,0.597113073214278)
		--(axis cs:36,0.630859567724911)
		--(axis cs:37,0.662493071697539)
		--(axis cs:38,0.691862510224964)
		--(axis cs:39,0.71887918774905)
		--(axis cs:40,0.743517828496931)
		--(axis cs:41,0.765811965040772)
		--(axis cs:42,0.785845006928289)
		--(axis cs:43,0.803738837850474)
		--(axis cs:44,0.819641779477174)
		--(axis cs:45,0.833717371077875)
		--(axis cs:46,0.846134866100341)
		--(axis cs:47,0.857061826095192)
		--(axis cs:48,0.86665880185174)
		--(axis cs:49,0.875075855904871)
		--(axis cs:50,0.882450576015258)
		--(axis cs:51,0.888907214626083)
		--(axis cs:52,0.894556625504463)
		--(axis cs:53,0.899496726762804)
		--(axis cs:54,0.903813281153662)
		--(axis cs:55,0.90758084049255)
		--(axis cs:56,0.91086374752631)
		--(axis cs:57,0.913717125031566)
		--(axis cs:58,0.916187809496848)
		--(axis cs:59,0.918315207102667)
		--(axis cs:60,0.920132064601346)
		--(axis cs:61,0.921665158630949)
		--(axis cs:62,0.922935915170347)
		--(axis cs:63,0.923960977104985)
		--(axis cs:64,0.924752742750489)
		--(axis cs:65,0.925319901937972)
		--(axis cs:66,0.92566799901116)
		--(axis cs:67,0.925800053949658)
		--(axis cs:68,0.925717274188186)
		--(axis cs:69,0.925419891409596)
		--(axis cs:70,0.924908161139446)
		--(axis cs:71,0.924183570255513)
		--(axis cs:72,0.92325030967024)
		--(axis cs:73,0.922117083431609)
		--(axis cs:74,0.920799326628144)
		--(axis cs:75,0.919321854103423)
		--(axis cs:75,1.07103031725366)
		--(axis cs:75,1.07103031725366)
		--(axis cs:74,1.06902200280124)
		--(axis cs:73,1.0671162871142)
		--(axis cs:72,1.06531522937587)
		--(axis cs:71,1.06361315700162)
		--(axis cs:70,1.06199846492204)
		--(axis cs:69,1.06045509201448)
		--(axis cs:68,1.05896366509135)
		--(axis cs:67,1.05750235610998)
		--(axis cs:66,1.05604750069827)
		--(axis cs:65,1.05457401471029)
		--(axis cs:64,1.05305563526506)
		--(axis cs:63,1.05146500678132)
		--(axis cs:62,1.04977362995081)
		--(axis cs:61,1.04795169068487)
		--(axis cs:60,1.04596778555336)
		--(axis cs:59,1.04378855953279)
		--(axis cs:58,1.04137827099301)
		--(axis cs:57,1.03869829815954)
		--(axis cs:56,1.03570660136468)
		--(axis cs:55,1.03235715689462)
		--(axis cs:54,1.02859938181151)
		--(axis cs:53,1.02437757541554)
		--(axis cs:52,1.01963041258506)
		--(axis cs:51,1.01429053750145)
		--(axis cs:50,1.00828432328706)
		--(axis cs:49,1.00153188322323)
		--(axis cs:48,0.99394744065696)
		--(axis cs:47,0.985440183818734)
		--(axis cs:46,0.975915742512553)
		--(axis cs:45,0.96527841728562)
		--(axis cs:44,0.95343425756403)
		--(axis cs:43,0.940295013120657)
		--(axis cs:42,0.925782867897316)
		--(axis cs:41,0.909835712529946)
		--(axis cs:40,0.892412544061008)
		--(axis cs:39,0.873498436522792)
		--(axis cs:38,0.853108449841511)
		--(axis cs:37,0.831289873495501)
		--(axis cs:36,0.808122348357936)
		--(axis cs:35,0.783715659776185)
		--(axis cs:34,0.758205310442585)
		--(axis cs:33,0.731746312226979)
		--(axis cs:32,0.704505917272326)
		--(axis cs:31,0.676656165391398)
		--(axis cs:30,0.648367097289305)
		--(axis cs:29,0.61980126873432)
		--(axis cs:28,0.591109873823225)
		--(axis cs:27,0.562430461920169)
		--(axis cs:26,0.533886004032957)
		--(axis cs:25,0.505584959748141)
		--(axis cs:24,0.477621995137504)
		--(axis cs:23,0.450079064363356)
		--(axis cs:22,0.423026651773987)
		--(axis cs:21,0.396525048806347)
		--(axis cs:20,0.370625601188375)
		--(axis cs:19,0.345371909990252)
		--(axis cs:18,0.320801013022387)
		--(axis cs:17,0.296944620728913)
		--(axis cs:16,0.273830546848615)
		--(axis cs:15,0.251484581666085)
		--(axis cs:14,0.229933245725398)
		--(axis cs:13,0.209208208637085)
		--(axis cs:12,0.189353790451143)
		--(axis cs:11,0.170440084971843)
		--(axis cs:10,0.152586091400567)
		--(axis cs:9,0.135999798406191)
		--(axis cs:8,0.121044225233582)
		--(axis cs:7,0.108335494328941)
		--(axis cs:6,0.0988613903058718)
		--(axis cs:5,0.0940680757945961)
		--(axis cs:4,0.0958071412180051)
		--(axis cs:3,0.106274649614838)
		--(axis cs:2,0.125950381461196)
		--(axis cs:1,0.15891510905982)
		--(axis cs:0,0.2)
		--cycle;

		\addplot [green!50!black, thick]
		table {%
				0 0
				1 -0.669934185698127
				2 -1.32020860883421
				3 -1.81335242111752
				4 -2.07401288519619
				5 -2.1184286174609
				6 -2.01866593995467
				7 -1.84833235307673
				8 -1.65509778103151
				9 -1.46163528613742
				10 -1.27615467376685
				11 -1.10071563293565
				12 -0.93538601349192
				13 -0.779811357699158
				14 -0.633657313990268
				15 -0.496668021773825
				16 -0.368629070359182
				17 -0.249325579512264
				18 -0.138515564871456
				19 -0.0359181129420762
				20 0.0587882422581225
				21 0.145961126986505
				22 0.22598688136697
				23 0.299271964628802
				24 0.366234372613228
				25 0.427295734664069
				26 0.482874436873616
				27 0.533379912816794
				28 0.579208112571732
				29 0.620738081238753
				30 0.658329533216666
				31 0.692321287231232
				32 0.723030421790673
				33 0.750752015743599
				34 0.77575934983154
				35 0.798304459606872
				36 0.818618945763872
				37 0.836914963431044
				38 0.853386326443264
				39 0.868209675581844
				40 0.881545671037772
				41 0.893540178897052
				42 0.904325429358387
				43 0.914021130827254
				44 0.922735529169706
				45 0.930566405443662
				46 0.937602008535918
				47 0.943921921484906
				48 0.949597862007776
				49 0.954694419000365
				50 0.95926972764517
				51 0.963376086332292
				52 0.967060518942718
				53 0.970365286219558
				54 0.973328350007159
				55 0.975983794107437
				56 0.978362205416735
				57 0.980491018889076
				58 0.982394829742487
				59 0.984095676201019
				60 0.985613295961213
				61 0.986965359502893
				62 0.98816768334546
				63 0.989234426399991
				64 0.990178272705057
				65 0.991010604086958
				66 0.991741666687072
				67 0.992380735894997
				68 0.992936285075036
				69 0.99341616465272
				70 0.993827799739008
				71 0.994178416645987
				72 0.994475311566209
				73 0.994726178587578
				74 0.994939519435552
				75 0.995125164378909
			};
		\addlegendentry{Stochastic}

		\addplot [color0!75!black, thick]
		table {%
				0 0
				1 -0.0563109069329729
				2 -0.100792349828514
				3 -0.127514173530644
				4 -0.142112347833835
				5 -0.145496643916831
				6 -0.140576422389995
				7 -0.129730669756826
				8 -0.114791348491238
				9 -0.0970047909313067
				10 -0.0771495487458892
				11 -0.0556863930147496
				12 -0.0328808257285617
				13 -0.00888557492734385
				14 0.016209088286556
				15 0.0423461011996071
				16 0.0694848749851432
				17 0.0975916420728306
				18 0.12663371616611
				19 0.156575879937697
				20 0.187377902596778
				21 0.218992624328294
				22 0.251364284591824
				23 0.284426907755384
				24 0.318102643715365
				25 0.352300022188305
				26 0.386912135433223
				27 0.421814829285021
				28 0.456865069640563
				29 0.491899769343058
				30 0.526735500102007
				31 0.561169638641253
				32 0.594983540829852
				33 0.627948220836967
				34 0.659832667421478
				35 0.690414366495232
				36 0.719490958041424
				37 0.74689147259652
				38 0.772485480033238
				39 0.796188812135921
				40 0.817965186278969
				41 0.837823838785359
				42 0.855813937412802
				43 0.872016925485566
				44 0.886538018520602
				45 0.899497894181747
				46 0.911025304306447
				47 0.921251004956963
				48 0.93030312125435
				49 0.938303869564049
				50 0.945367449651158
				51 0.951598876063767
				52 0.957093519044761
				53 0.961937151089174
				54 0.966206331482584
				55 0.969968998693586
				56 0.973285174445497
				57 0.976207711595555
				58 0.978783040244929
				59 0.98105188331773
				60 0.983049925077354
				61 0.984808424657908
				62 0.986354772560579
				63 0.987712991943153
				64 0.988904189007774
				65 0.989946958324129
				66 0.990857749854717
				67 0.991651205029819
				68 0.992340469639768
				69 0.99293749171204
				70 0.993453313030742
				71 0.993898363628567
				72 0.994282769523058
				73 0.994616685272904
				74 0.994910664714691
				75 0.99517608567854
			};
        \addlegendentry{Robust}
        
		\nextgroupplot[
			height=4.25cm,
			width=8.5cm,
			tick pos=both,
			tick align=inside,
			minor tick num=4,
			xmajorticks=false,
			ymajorticks=true,
			xmajorgrids,
			ymajorgrids,
			xtick style={color=black},
			ytick style={color=black},
			x grid style={white!69.0196078431373!black},
			y grid style={white!69.0196078431373!black},
			xmin=-3.75, xmax=78.75,
			ymin=-1.46755034172417, ymax=25.0834249743893,
			every major tick/.append style={major tick length=5pt, black},
			every minor tick/.append style={minor tick length=3pt, black},
			major grid style={dashed, gray},
		]

		\path [draw=green!50.1960784313725!black, fill=green!50.1960784313725!black, opacity=0.1]
		(axis cs:0,0.2)
		--(axis cs:0,-0.2)
		--(axis cs:1,1.88015843229982)
		--(axis cs:2,4.44791369979024)
		--(axis cs:3,7.37782095366622)
		--(axis cs:4,10.2424205569958)
		--(axis cs:5,12.613707805124)
		--(axis cs:6,14.2469873598997)
		--(axis cs:7,15.1233133658759)
		--(axis cs:8,15.3712782593163)
		--(axis cs:9,15.1648987232962)
		--(axis cs:10,14.6602652820886)
		--(axis cs:11,13.9747139245239)
		--(axis cs:12,13.188921614916)
		--(axis cs:13,12.355972544642)
		--(axis cs:14,11.5101350655864)
		--(axis cs:15,10.6733849013395)
		--(axis cs:16,9.85976742066175)
		--(axis cs:17,9.07818245461564)
		--(axis cs:18,8.33413278200164)
		--(axis cs:19,7.63082373446827)
		--(axis cs:20,6.96986514665767)
		--(axis cs:21,6.35173121382859)
		--(axis cs:22,5.77607240221505)
		--(axis cs:23,5.24193572625727)
		--(axis cs:24,4.7479268871087)
		--(axis cs:25,4.2923341684324)
		--(axis cs:26,3.87322595669137)
		--(axis cs:27,3.48852905625264)
		--(axis cs:28,3.13609224884988)
		--(axis cs:29,2.81373798400314)
		--(axis cs:30,2.51930418962685)
		--(axis cs:31,2.25067766675819)
		--(axis cs:32,2.00582020811451)
		--(axis cs:33,1.78278836264247)
		--(axis cs:34,1.57974761116146)
		--(axis cs:35,1.39498160483404)
		--(axis cs:36,1.22689704708204)
		--(axis cs:37,1.0740247730137)
		--(axis cs:38,0.935017595268626)
		--(axis cs:39,0.808645527576097)
		--(axis cs:40,0.693789043376397)
		--(axis cs:41,0.589431049156792)
		--(axis cs:42,0.494648229332059)
		--(axis cs:43,0.408602344117952)
		--(axis cs:44,0.330531942644136)
		--(axis cs:45,0.259744811177071)
		--(axis cs:46,0.195611335169067)
		--(axis cs:47,0.137558834121564)
		--(axis cs:48,0.0850668413259262)
		--(axis cs:49,0.0376632484150227)
		--(axis cs:50,-0.0050787877289224)
		--(axis cs:51,-0.0435432795392639)
		--(axis cs:52,-0.0780732784966215)
		--(axis cs:53,-0.108972260600983)
		--(axis cs:54,-0.13650509898605)
		--(axis cs:55,-0.160898695371924)
		--(axis cs:56,-0.182342356449381)
		--(axis cs:57,-0.200988034085964)
		--(axis cs:58,-0.216950603691389)
		--(axis cs:59,-0.230308437249676)
		--(axis cs:60,-0.241104639646954)
		--(axis cs:61,-0.249349460216996)
		--(axis cs:62,-0.25502456377168)
		--(axis cs:63,-0.258090040794102)
		--(axis cs:64,-0.258495247364699)
		--(axis cs:65,-0.256194791026207)
		--(axis cs:66,-0.251171244533412)
		--(axis cs:67,-0.243466561253093)
		--(axis cs:68,-0.233224878634595)
		--(axis cs:69,-0.22075076441766)
		--(axis cs:70,-0.206589322890414)
		--(axis cs:71,-0.191637386373078)
		--(axis cs:72,-0.177293358792407)
		--(axis cs:73,-0.16562895833507)
		--(axis cs:74,-0.159484837150035)
		--(axis cs:75,-0.162293855349383)
		--(axis cs:75,0.196249650071779)
		--(axis cs:75,0.196249650071779)
		--(axis cs:74,0.196614090747262)
		--(axis cs:73,0.208298076434347)
		--(axis cs:72,0.227468656405202)
		--(axis cs:71,0.251019159402181)
		--(axis cs:70,0.276716075691979)
		--(axis cs:69,0.303081248920824)
		--(axis cs:68,0.329203780114707)
		--(axis cs:67,0.354578296525422)
		--(axis cs:66,0.378985307270786)
		--(axis cs:65,0.402405302431474)
		--(axis cs:64,0.424956787299188)
		--(axis cs:63,0.446851398945656)
		--(axis cs:62,0.468361727836746)
		--(axis cs:61,0.489798921867246)
		--(axis cs:60,0.511497929007706)
		--(axis cs:59,0.533808677313558)
		--(axis cs:58,0.557091795818166)
		--(axis cs:57,0.581717735624493)
		--(axis cs:56,0.608068384816261)
		--(axis cs:55,0.636540484458491)
		--(axis cs:54,0.667550339487208)
		--(axis cs:53,0.701539472480007)
		--(axis cs:52,0.738980988517214)
		--(axis cs:51,0.78038650705439)
		--(axis cs:50,0.826313575203851)
		--(axis cs:49,0.877373509992211)
		--(axis cs:48,0.934239628946691)
		--(axis cs:47,0.997655822550104)
		--(axis cs:46,1.06844540262532)
		--(axis cs:45,1.14752013217514)
		--(axis cs:44,1.23588931042502)
		--(axis cs:43,1.33466875897722)
		--(axis cs:42,1.44508953902703)
		--(axis cs:41,1.5685062324811)
		--(axis cs:40,1.70640464467409)
		--(axis cs:39,1.86040882918107)
		--(axis cs:38,2.03228738256618)
		--(axis cs:37,2.22395898671789)
		--(axis cs:36,2.43749716278985)
		--(axis cs:35,2.67513412213972)
		--(axis cs:34,2.93926344725954)
		--(axis cs:33,3.23244111803157)
		--(axis cs:32,3.55738413939233)
		--(axis cs:31,3.91696575638152)
		--(axis cs:30,4.31420598903222)
		--(axis cs:29,4.7522559984027)
		--(axis cs:28,5.23437460610831)
		--(axis cs:27,5.76389511557817)
		--(axis cs:26,6.3441803874039)
		--(axis cs:25,6.97856384123186)
		--(axis cs:24,7.67027358902049)
		--(axis cs:23,8.42233607550135)
		--(axis cs:22,9.23745411988608)
		--(axis cs:21,10.1178516316267)
		--(axis cs:20,11.0650727058243)
		--(axis cs:19,12.0797149620134)
		--(axis cs:18,13.1610636958911)
		--(axis cs:17,14.3065711245986)
		--(axis cs:16,15.5110880516015)
		--(axis cs:15,16.7656949050021)
		--(axis cs:14,18.0558828979819)
		--(axis cs:13,19.3586899847615)
		--(axis cs:12,20.6381966124277)
		--(axis cs:11,21.8385818438699)
		--(axis cs:10,22.8739541425501)
		--(axis cs:9,23.6150839675185)
		--(axis cs:8,23.8765624600206)
		--(axis cs:7,23.4161761771954)
		--(axis cs:6,21.9710742842172)
		--(axis cs:5,19.3588835393018)
		--(axis cs:4,15.633485923651)
		--(axis cs:3,11.1961399170675)
		--(axis cs:2,6.71982333892156)
		--(axis cs:1,2.88170068301646)
		--(axis cs:0,0.2)
		--cycle;

		\path [draw=color0, fill=color0, opacity=0.1]
		(axis cs:0,0.2)
		--(axis cs:0,-0.2)
		--(axis cs:1,0.0278523768880691)
		--(axis cs:2,0.246870281424628)
		--(axis cs:3,0.434681301966817)
		--(axis cs:4,0.618553952139343)
		--(axis cs:5,0.790748144388678)
		--(axis cs:6,0.953397988226922)
		--(axis cs:7,1.10607928633546)
		--(axis cs:8,1.24891170358504)
		--(axis cs:9,1.38246435144151)
		--(axis cs:10,1.50756177366256)
		--(axis cs:11,1.62508760785545)
		--(axis cs:12,1.73585270128538)
		--(axis cs:13,1.84052812789109)
		--(axis cs:14,1.9396218814318)
		--(axis cs:15,2.03347862190449)
		--(axis cs:16,2.12228854094152)
		--(axis cs:17,2.20609754342139)
		--(axis cs:18,2.28481500572519)
		--(axis cs:19,2.35821768612206)
		--(axis cs:20,2.42594958530627)
		--(axis cs:21,2.48751821554881)
		--(axis cs:22,2.5422881859484)
		--(axis cs:23,2.58947346808631)
		--(axis cs:24,2.62813031441946)
		--(axis cs:25,2.65715369574509)
		--(axis cs:26,2.67528151967995)
		--(axis cs:27,2.68111302657705)
		--(axis cs:28,2.67315036810548)
		--(axis cs:29,2.6498740398808)
		--(axis cs:30,2.60986129686953)
		--(axis cs:31,2.55194990965257)
		--(axis cs:32,2.47543660749484)
		--(axis cs:33,2.38028174877783)
		--(axis cs:34,2.26727553034771)
		--(axis cs:35,2.1381165314668)
		--(axis cs:36,1.99536745786546)
		--(axis cs:37,1.84228277730205)
		--(axis cs:38,1.68253692795966)
		--(axis cs:39,1.51990646569069)
		--(axis cs:40,1.35796645925444)
		--(axis cs:41,1.19985035751379)
		--(axis cs:42,1.04810010144507)
		--(axis cs:43,0.904608948414764)
		--(axis cs:44,0.77064112922623)
		--(axis cs:45,0.646903525501809)
		--(axis cs:46,0.533644205571116)
		--(axis cs:47,0.430757664527233)
		--(axis cs:48,0.337883486178201)
		--(axis cs:49,0.254491431684314)
		--(axis cs:50,0.179950555777952)
		--(axis cs:51,0.113582756702531)
		--(axis cs:52,0.0547025743648153)
		--(axis cs:53,0.00264555633812502)
		--(axis cs:54,-0.0432124887696492)
		--(axis cs:55,-0.0834432786544401)
		--(axis cs:56,-0.118559163725653)
		--(axis cs:57,-0.149010210394342)
		--(axis cs:58,-0.175184292965418)
		--(axis cs:59,-0.197409579288013)
		--(axis cs:60,-0.215959122998879)
		--(axis cs:61,-0.231057576574627)
		--(axis cs:62,-0.242890316465221)
		--(axis cs:63,-0.25161552904296)
		--(axis cs:64,-0.257380054520635)
		--(axis cs:65,-0.26034004757191)
		--(axis cs:66,-0.260687827355372)
		--(axis cs:67,-0.258686704878326)
		--(axis cs:68,-0.254716095478768)
		--(axis cs:69,-0.249329619050283)
		--(axis cs:70,-0.243328219204931)
		--(axis cs:71,-0.237846058035673)
		--(axis cs:72,-0.234434437967103)
		--(axis cs:73,-0.235105781813467)
		--(axis cs:74,-0.242283112196053)
		--(axis cs:75,-0.258647885684962)
		--(axis cs:75,0.308560313611268)
		--(axis cs:75,0.308560313611268)
		--(axis cs:74,0.295367570000817)
		--(axis cs:73,0.293902639538634)
		--(axis cs:72,0.301219574843838)
		--(axis cs:71,0.314730236647819)
		--(axis cs:70,0.332342038626047)
		--(axis cs:69,0.352497734468382)
		--(axis cs:68,0.374123809143423)
		--(axis cs:67,0.396541581686198)
		--(axis cs:66,0.419378645169608)
		--(axis cs:65,0.442495136789853)
		--(axis cs:64,0.465926895256149)
		--(axis cs:63,0.48984334084696)
		--(axis cs:62,0.514517278102784)
		--(axis cs:61,0.540304233762322)
		--(axis cs:60,0.567629458342296)
		--(axis cs:59,0.596981102963105)
		--(axis cs:58,0.628908327638967)
		--(axis cs:57,0.664023238936336)
		--(axis cs:56,0.703005615123968)
		--(axis cs:55,0.746609355186157)
		--(axis cs:54,0.795669467549551)
		--(axis cs:53,0.851108169655271)
		--(axis cs:52,0.913938274352235)
		--(axis cs:51,0.985261476471148)
		--(axis cs:50,1.06625843071602)
		--(axis cs:49,1.15816668743099)
		--(axis cs:48,1.26224176686294)
		--(axis cs:47,1.3796961698431)
		--(axis cs:46,1.51161136350805)
		--(axis cs:45,1.65881931945616)
		--(axis cs:44,1.82175367764769)
		--(axis cs:43,2.00027662430376)
		--(axis cs:42,2.19349624029995)
		--(axis cs:41,2.39959962104534)
		--(axis cs:40,2.61573726567457)
		--(axis cs:39,2.83800014550609)
		--(axis cs:38,3.06152741760988)
		--(axis cs:37,3.28076570027639)
		--(axis cs:36,3.48986985637364)
		--(axis cs:35,3.68319661662479)
		--(axis cs:34,3.85580877190112)
		--(axis cs:33,4.0038936591422)
		--(axis cs:32,4.12501461293153)
		--(axis cs:31,4.21815599021276)
		--(axis cs:30,4.28357679860782)
		--(axis cs:29,4.32253339869922)
		--(axis cs:28,4.33695086159759)
		--(axis cs:27,4.32911244357097)
		--(axis cs:26,4.30140931871676)
		--(axis cs:25,4.25616437407452)
		--(axis cs:24,4.19552402196034)
		--(axis cs:23,4.12140226306094)
		--(axis cs:22,4.03545963647339)
		--(axis cs:21,3.93910286831348)
		--(axis cs:20,3.83349541390596)
		--(axis cs:19,3.71957254744507)
		--(axis cs:18,3.59805686824055)
		--(axis cs:17,3.46947148979154)
		--(axis cs:16,3.33414911357939)
		--(axis cs:15,3.1922358750872)
		--(axis cs:14,3.04368949741823)
		--(axis cs:13,2.88827221802997)
		--(axis cs:12,2.72554069348284)
		--(axis cs:11,2.55483849484504)
		--(axis cs:10,2.37530314094308)
		--(axis cs:9,2.18591021167569)
		--(axis cs:8,1.98559192053287)
		--(axis cs:7,1.7734819420325)
		--(axis cs:6,1.54933966139607)
		--(axis cs:5,1.3141865149442)
		--(axis cs:4,1.07129795478488)
		--(axis cs:3,0.824003368896965)
		--(axis cs:2,0.594880658999614)
		--(axis cs:1,0.372327874331345)
		--(axis cs:0,0.2)
		--cycle;

		\addplot [green!50!black, thick]
		table {%
				0 0
				1 2.38092955765814
				2 5.5838685193559
				3 9.28698043536688
				4 12.9379532403234
				5 15.9862956722129
				6 18.1090308220585
				7 19.2697447715356
				8 19.6239203596684
				9 19.3899913454074
				10 18.7671097123193
				11 17.9066478841969
				12 16.9135591136718
				13 15.8573312647018
				14 14.7830089817842
				15 13.7195399031708
				16 12.6854277361316
				17 11.6923767896071
				18 10.7475982389464
				19 9.85526934824083
				20 9.01746892624097
				21 8.23479142272765
				22 7.50676326105057
				23 6.83213590087931
				24 6.2091002380646
				25 5.63544900483213
				26 5.10870317204763
				27 4.6262120859154
				28 4.1852334274791
				29 3.78299699120292
				30 3.41675508932954
				31 3.08382171156986
				32 2.78160217375342
				33 2.50761474033702
				34 2.2595055292105
				35 2.03505786348688
				36 1.83219710493594
				37 1.6489918798658
				38 1.48365248891741
				39 1.33452717837858
				40 1.20009684402525
				41 1.07896864081895
				42 0.969868884179543
				43 0.871635551547587
				44 0.78321062653458
				45 0.703632471676106
				46 0.632028368897196
				47 0.567607328335834
				48 0.509653235136308
				49 0.457518379203617
				50 0.410617393737464
				51 0.368421613757563
				52 0.330453855010296
				53 0.296283605939512
				54 0.265522620250579
				55 0.237820894543283
				56 0.21286301418344
				57 0.190364850769264
				58 0.170070596063389
				59 0.151750120031941
				60 0.135196644680376
				61 0.120224730825125
				62 0.106668582032533
				63 0.0943806790757769
				64 0.0832307699672444
				65 0.0731052557026337
				66 0.063907031368687
				67 0.0555558676361649
				68 0.047989450740056
				69 0.0411652422515823
				70 0.0350633764007829
				71 0.0296908865145517
				72 0.0250876488063974
				73 0.0213345590496383
				74 0.0185646267986134
				75 0.016977897361198
			};
		\addplot [color0!75!black, thick]
		table {%
				0 0
				1 0.200090125609707
				2 0.420875470212121
				3 0.629342335431891
				4 0.844925953462114
				5 1.05246732966644
				6 1.25136882481149
				7 1.43978061418398
				8 1.61725181205896
				9 1.7841872815586
				10 1.94143245730282
				11 2.08996305135025
				12 2.23069669738411
				13 2.36440017296053
				14 2.49165568942501
				15 2.61285724849585
				16 2.72821882726045
				17 2.83778451660647
				18 2.94143593698287
				19 3.03889511678357
				20 3.12972249960612
				21 3.21331054193114
				22 3.28887391121089
				23 3.35543786557362
				24 3.4118271681899
				25 3.4566590349098
				26 3.48834541919836
				27 3.50511273507401
				28 3.50505061485153
				29 3.48620371929001
				30 3.44671904773867
				31 3.38505294993267
				32 3.30022561021318
				33 3.19208770396001
				34 3.06154215112442
				35 2.91065657404579
				36 2.74261865711955
				37 2.56152423878922
				38 2.37203217278477
				39 2.17895330559839
				40 1.98685186246451
				41 1.79972498927957
				42 1.62079817087251
				43 1.45244278635926
				44 1.29619740343696
				45 1.15286142247898
				46 1.02262778453958
				47 0.905226917185166
				48 0.800062626520572
				49 0.706329059557649
				50 0.623104493246984
				51 0.549422116586839
				52 0.484320424358525
				53 0.426876862996698
				54 0.376228489389951
				55 0.331583038265859
				56 0.292223225699158
				57 0.257506514270997
				58 0.226862017336774
				59 0.199785761837546
				60 0.175835167671709
				61 0.154623328593848
				62 0.135813480818782
				63 0.119113905902
				64 0.104273420367757
				65 0.0910775446089716
				66 0.0793454089071181
				67 0.0689274384039362
				68 0.0597038568323276
				69 0.0515840577090496
				70 0.0445069097105581
				71 0.0384420893060731
				72 0.0333925684383679
				73 0.0293984288625838
				74 0.0265422289023823
				75 0.0249562139631529
			};

		\nextgroupplot[
			height=4.25cm,
			width=8.5cm,
			tick pos=both,
			tick align=inside,
			minor tick num=4,
			xmajorticks=true,
			ymajorticks=true,
			xmajorgrids,
			ymajorgrids,
			xtick style={color=black},
			ytick style={color=black},
			x grid style={white!69.0196078431373!black},
			y grid style={white!69.0196078431373!black},
			xmin=-3.75, xmax=78.75,
			ymin=-160.592046061477, ymax=482.611538031708,
			every major tick/.append style={major tick length=5pt, black},
			every minor tick/.append style={minor tick length=3pt, black},
			major grid style={dashed, gray},
			xlabel={Time Steps},
		]

		\path [draw=green!50.1960784313725!black, fill=green!50.1960784313725!black, opacity=0.1]
		(axis cs:0,291.39288465128)
		--(axis cs:0,192.863091941081)
		--(axis cs:1,255.925272088969)
		--(axis cs:2,292.12528582992)
		--(axis cs:3,284.385618460883)
		--(axis cs:4,233.993354971004)
		--(axis cs:5,159.391991696812)
		--(axis cs:6,82.7885326147092)
		--(axis cs:7,17.7482101912132)
		--(axis cs:8,-34.7247933652273)
		--(axis cs:9,-76.469159076271)
		--(axis cs:10,-104.471009622893)
		--(axis cs:11,-120.744879307326)
		--(axis cs:12,-128.784706299097)
		--(axis cs:13,-131.355519511787)
		--(axis cs:14,-130.364995293551)
		--(axis cs:15,-127.079694341273)
		--(axis cs:16,-122.335737239962)
		--(axis cs:17,-116.690657440926)
		--(axis cs:18,-110.523104676248)
		--(axis cs:19,-104.095858142631)
		--(axis cs:20,-97.5950700675007)
		--(axis cs:21,-91.154678391057)
		--(axis cs:22,-84.8717401896476)
		--(axis cs:23,-78.8162619563131)
		--(axis cs:24,-73.0377087837416)
		--(axis cs:25,-67.5695073126572)
		--(axis cs:26,-62.4323264715269)
		--(axis cs:27,-57.6365985768807)
		--(axis cs:28,-53.1845504916912)
		--(axis cs:29,-49.0718998543942)
		--(axis cs:30,-45.2893038472849)
		--(axis cs:31,-41.8236087337819)
		--(axis cs:32,-38.6589262605345)
		--(axis cs:33,-35.7775514319645)
		--(axis cs:34,-33.1607312046669)
		--(axis cs:35,-30.7892927797118)
		--(axis cs:36,-28.6441415482637)
		--(axis cs:37,-26.7066409523585)
		--(axis cs:38,-24.9588885313107)
		--(axis cs:39,-23.3839036218457)
		--(axis cs:40,-21.9657423438803)
		--(axis cs:41,-20.6895547111597)
		--(axis cs:42,-19.5415972071772)
		--(axis cs:43,-18.5092122671655)
		--(axis cs:44,-17.5807840833436)
		--(axis cs:45,-16.7456782111295)
		--(axis cs:46,-15.9941707308734)
		--(axis cs:47,-15.3173712816133)
		--(axis cs:48,-14.7071431561798)
		--(axis cs:49,-14.1560228336151)
		--(axis cs:50,-13.6571408219283)
		--(axis cs:51,-13.2041454934618)
		--(axis cs:52,-12.7911317296814)
		--(axis cs:53,-12.4125766763188)
		--(axis cs:54,-12.0632857703796)
		--(axis cs:55,-11.7383534451522)
		--(axis cs:56,-11.4331444931676)
		--(axis cs:57,-11.1433037783964)
		--(axis cs:58,-10.8648033935198)
		--(axis cs:59,-10.5940365993852)
		--(axis cs:60,-10.3279655157917)
		--(axis cs:61,-10.064322409807)
		--(axis cs:62,-9.80184986388711)
		--(axis cs:63,-9.54054068241598)
		--(axis cs:64,-9.2818039821172)
		--(axis cs:65,-9.02844543577224)
		--(axis cs:66,-8.78432428759671)
		--(axis cs:67,-8.55356978312655)
		--(axis cs:68,-8.33934780968312)
		--(axis cs:69,-8.14239745163939)
		--(axis cs:70,-7.95990066838703)
		--(axis cs:71,-7.78567590856964)
		--(axis cs:72,-7.61330089340386)
		--(axis cs:73,-7.44517305494958)
		--(axis cs:74,-7.31392250326329)
		--(axis cs:74,7.02018833065588)
		--(axis cs:74,7.02018833065588)
		--(axis cs:73,6.91534803708076)
		--(axis cs:72,6.88758996707331)
		--(axis cs:71,6.89085008902011)
		--(axis cs:70,6.9122918819705)
		--(axis cs:69,6.95012559062956)
		--(axis cs:68,7.00396263626845)
		--(axis cs:67,7.07124506320172)
		--(axis cs:66,7.14670760471008)
		--(axis cs:65,7.22324159863755)
		--(axis cs:64,7.29315054639528)
		--(axis cs:63,7.34921930261894)
		--(axis cs:62,7.3853658680711)
		--(axis cs:61,7.39687632550608)
		--(axis cs:60,7.3803336424794)
		--(axis cs:59,7.33337263680078)
		--(axis cs:58,7.25436922395195)
		--(axis cs:57,7.1421344801693)
		--(axis cs:56,6.99565031846273)
		--(axis cs:55,6.81385967052074)
		--(axis cs:54,6.5955099103026)
		--(axis cs:53,6.33904146185523)
		--(axis cs:52,6.04251144694917)
		--(axis cs:51,5.7035427254821)
		--(axis cs:50,5.31929033756619)
		--(axis cs:49,4.88641933328823)
		--(axis cs:48,4.40108984801716)
		--(axis cs:47,3.85894687465886)
		--(axis cs:46,3.25511346374826)
		--(axis cs:45,2.58418707548729)
		--(axis cs:44,1.84023956294117)
		--(axis cs:43,1.01682182883043)
		--(axis cs:42,0.106974604511199)
		--(axis cs:41,-0.896752929221938)
		--(axis cs:40,-2.00227481087462)
		--(axis cs:39,-3.21792834820318)
		--(axis cs:38,-4.55242068563573)
		--(axis cs:37,-6.01476010128929)
		--(axis cs:36,-7.61417025660118)
		--(axis cs:35,-9.35998541986838)
		--(axis cs:34,-11.2615240808686)
		--(axis cs:33,-13.3279371730804)
		--(axis cs:32,-15.5680251939949)
		--(axis cs:31,-17.9900158874994)
		--(axis cs:30,-20.6012910706916)
		--(axis cs:29,-23.4080481014123)
		--(axis cs:28,-26.4148788933682)
		--(axis cs:27,-29.6242475337439)
		--(axis cs:26,-33.0358460032203)
		--(axis cs:25,-36.6458046683063)
		--(axis cs:24,-40.4457269678283)
		--(axis cs:23,-44.4215008131919)
		--(axis cs:22,-48.5518044591842)
		--(axis cs:21,-52.8061591067036)
		--(axis cs:20,-57.1422635465807)
		--(axis cs:19,-61.5021434770052)
		--(axis cs:18,-65.8063021179188)
		--(axis cs:17,-69.9444721044972)
		--(axis cs:16,-73.7605783811032)
		--(axis cs:15,-77.0278618421281)
		--(axis cs:14,-79.4073717224561)
		--(axis cs:13,-80.3786290666942)
		--(axis cs:12,-79.1244326991873)
		--(axis cs:11,-74.3408643829644)
		--(axis cs:10,-63.9167286547897)
		--(axis cs:9,-44.2731008080027)
		--(axis cs:8,-8.1734254900192)
		--(axis cs:7,56.9040086536382)
		--(axis cs:6,152.930175223224)
		--(axis cs:5,268.35252718139)
		--(axis cs:4,378.627546659247)
		--(axis cs:3,449.198410560258)
		--(axis cs:2,453.375011482018)
		--(axis cs:1,391.539392709633)
		--(axis cs:0,291.39288465128)
		--cycle;

		\path [draw=color0, fill=color0, opacity=0.1]
		(axis cs:0,33.5617124795809)
		--(axis cs:0,7.13463700668349)
		--(axis cs:1,9.49782905397289)
		--(axis cs:2,8.4194677341221)
		--(axis cs:3,9.07780636060485)
		--(axis cs:4,8.30847797632792)
		--(axis cs:5,7.45334496503563)
		--(axis cs:6,6.40925909791726)
		--(axis cs:7,5.31616100117297)
		--(axis cs:8,4.26083392563207)
		--(axis cs:9,3.28842169822482)
		--(axis cs:10,2.41256881276016)
		--(axis cs:11,1.62767219438215)
		--(axis cs:12,0.918636965697328)
		--(axis cs:13,0.26693594806062)
		--(axis cs:14,-0.346310915243244)
		--(axis cs:15,-0.939135193556433)
		--(axis cs:16,-1.5285318514391)
		--(axis cs:17,-2.13074811655311)
		--(axis cs:18,-2.76175699913234)
		--(axis cs:19,-3.43776694858746)
		--(axis cs:20,-4.17569055400912)
		--(axis cs:21,-4.99352133005892)
		--(axis cs:22,-5.9105660854113)
		--(axis cs:23,-6.94746018676467)
		--(axis cs:24,-8.12585393090126)
		--(axis cs:25,-9.46758771155802)
		--(axis cs:26,-10.9930645432937)
		--(axis cs:27,-12.7184370940938)
		--(axis cs:28,-14.6512953293261)
		--(axis cs:29,-16.7848933313577)
		--(axis cs:30,-19.091602583246)
		--(axis cs:31,-21.51716293436)
		--(axis cs:32,-23.9781632393377)
		--(axis cs:33,-26.3653487227306)
		--(axis cs:34,-28.5541158764414)
		--(axis cs:35,-30.4210400577104)
		--(axis cs:36,-31.8626610948431)
		--(axis cs:37,-32.8114706483443)
		--(axis cs:38,-33.2448176721892)
		--(axis cs:39,-33.18486808556)
		--(axis cs:40,-32.6906294814725)
		--(axis cs:41,-31.8451349372245)
		--(axis cs:42,-30.7415127673573)
		--(axis cs:43,-29.4709778999536)
		--(axis cs:44,-28.114406760835)
		--(axis cs:45,-26.7377970940944)
		--(axis cs:46,-25.390991963145)
		--(axis cs:47,-24.1086536006442)
		--(axis cs:48,-22.9124809634694)
		--(axis cs:49,-21.8138837809181)
		--(axis cs:50,-20.8165994127606)
		--(axis cs:51,-19.9189774281733)
		--(axis cs:52,-19.1158277977834)
		--(axis cs:53,-18.3998332409389)
		--(axis cs:54,-17.762579985252)
		--(axis cs:55,-17.1952814518289)
		--(axis cs:56,-16.6892703988574)
		--(axis cs:57,-16.2363265115916)
		--(axis cs:58,-15.8288939905455)
		--(axis cs:59,-15.4602301337495)
		--(axis cs:60,-15.124512197082)
		--(axis cs:61,-14.8169160567463)
		--(axis cs:62,-14.5336665019165)
		--(axis cs:63,-14.2720462991901)
		--(axis cs:64,-14.0303420060773)
		--(axis cs:65,-13.8077032933366)
		--(axis cs:66,-13.6039050664029)
		--(axis cs:67,-13.4190332961387)
		--(axis cs:68,-13.2531682803677)
		--(axis cs:69,-13.1062102699976)
		--(axis cs:70,-12.9780783207283)
		--(axis cs:71,-12.8696253736724)
		--(axis cs:72,-12.7848084854276)
		--(axis cs:73,-12.7350530205041)
		--(axis cs:74,-12.7473988541899)
		--(axis cs:74,12.45540252448)
		--(axis cs:74,12.45540252448)
		--(axis cs:73,12.1895850450952)
		--(axis cs:72,12.0125447286466)
		--(axis cs:71,11.887287561311)
		--(axis cs:70,11.7938846453529)
		--(axis cs:69,11.7209561633048)
		--(axis cs:68,11.6609959331501)
		--(axis cs:67,11.6079353436103)
		--(axis cs:66,11.5559969919575)
		--(axis cs:65,11.4992903140787)
		--(axis cs:64,11.4317992414597)
		--(axis cs:63,11.3475257401083)
		--(axis cs:62,11.24064068402)
		--(axis cs:61,11.1055665546007)
		--(axis cs:60,10.9369714522063)
		--(axis cs:59,10.7296883211184)
		--(axis cs:58,10.4785895447026)
		--(axis cs:57,10.1784508923352)
		--(axis cs:56,9.82383644706725)
		--(axis cs:55,9.40903253013931)
		--(axis cs:54,8.92805673916921)
		--(axis cs:53,8.37476925464372)
		--(axis cs:52,7.7431174591524)
		--(axis cs:51,7.02755058742334)
		--(axis cs:50,6.22364658942393)
		--(axis cs:49,5.32899551240735)
		--(axis cs:48,4.34437783513948)
		--(axis cs:47,3.27525577690743)
		--(axis cs:46,2.13355225824268)
		--(axis cs:45,0.939616710967664)
		--(axis cs:44,-0.275837627901684)
		--(axis cs:43,-1.47019241938139)
		--(axis cs:42,-2.58831596684059)
		--(axis cs:41,-3.56356115521107)
		--(axis cs:40,-4.32105658512729)
		--(axis cs:39,-4.78375546611752)
		--(axis cs:38,-4.88115809425154)
		--(axis cs:37,-4.55976795320263)
		--(axis cs:36,-3.7933866308253)
		--(axis cs:35,-2.59069002688529)
		--(axis cs:34,-0.99758766885326)
		--(axis cs:33,0.907117364729594)
		--(axis cs:32,3.02242932287851)
		--(axis cs:31,5.23983635475826)
		--(axis cs:30,7.45817982476548)
		--(axis cs:29,9.59497407067257)
		--(axis cs:28,11.5920262381441)
		--(axis cs:27,13.4154703194952)
		--(axis cs:26,15.0519803582217)
		--(axis cs:25,16.5033474494017)
		--(axis cs:24,17.7811346403395)
		--(axis cs:23,18.9023691370799)
		--(axis cs:22,19.8865444504609)
		--(axis cs:21,20.7537598969362)
		--(axis cs:20,21.5236794511097)
		--(axis cs:19,22.2150409643571)
		--(axis cs:18,22.8455362232734)
		--(axis cs:17,23.4319439383213)
		--(axis cs:16,23.9904334395726)
		--(axis cs:15,24.5369767019657)
		--(axis cs:14,25.0878090714641)
		--(axis cs:13,25.6598553770171)
		--(axis cs:12,26.2709718726054)
		--(axis cs:11,26.9397223102978)
		--(axis cs:10,27.6841866155835)
		--(axis cs:9,28.5190134049024)
		--(axis cs:8,29.4497417223905)
		--(axis cs:7,30.4638297347112)
		--(axis cs:6,31.5197058002695)
		--(axis cs:5,32.5385125276959)
		--(axis cs:4,33.3936031158155)
		--(axis cs:3,34.2549094193073)
		--(axis cs:2,33.5694534575393)
		--(axis cs:1,35.1404180821136)
		--(axis cs:0,33.5617124795809)
		--cycle;

		\addplot [green!50!black, thick]
		table {%
				0 242.12798829618
				1 323.732332399301
				2 372.750148655969
				3 366.792014510571
				4 306.310450815126
				5 213.872259439101
				6 117.859353918967
				7 37.3261094224257
				8 -21.4491094276233
				9 -60.3711299421369
				10 -84.1938691388416
				11 -97.5428718451452
				12 -103.954569499142
				13 -105.867074289241
				14 -104.886183508004
				15 -102.053778091701
				16 -98.0481578105327
				17 -93.3175647727117
				18 -88.1647033970833
				19 -82.7990008098181
				20 -77.3686668070407
				21 -71.9804187488803
				22 -66.7117723244159
				23 -61.6188813847525
				24 -56.741717875785
				25 -52.1076559904817
				26 -47.7340862373736
				27 -43.6304230553123
				28 -39.7997146925297
				29 -36.2399739779032
				30 -32.9452974589883
				31 -29.9068123106406
				32 -27.1134757272647
				33 -24.5527443025225
				34 -22.2111276427678
				35 -20.0746390997901
				36 -18.1291559024324
				37 -16.3607005268239
				38 -14.7556546084732
				39 -13.3009159850245
				40 -11.9840085773774
				41 -10.7931538201908
				42 -9.71731130133298
				43 -8.74619521916753
				44 -7.8702722602012
				45 -7.08074556782108
				46 -6.36952863356257
				47 -5.72921220347723
				48 -5.15302665408132
				49 -4.63480175016343
				50 -4.16892524218107
				51 -3.75030138398984
				52 -3.37431014136612
				53 -3.03676760723178
				54 -2.73388793003849
				55 -2.46224688731573
				56 -2.21874708735245
				57 -2.00058464911356
				58 -1.80521708478392
				59 -1.63033198129219
				60 -1.47381593665617
				61 -1.33372304215047
				62 -1.208241997908
				63 -1.09566068989852
				64 -0.994326717860957
				65 -0.902601918567342
				66 -0.818808341443315
				67 -0.741162359962416
				68 -0.667692586707332
				69 -0.596135930504914
				70 -0.523804393208266
				71 -0.447412909774762
				72 -0.362855463165275
				73 -0.264912508934408
				74 -0.146867086303709
			};
		\addplot [color0!75!black, thick]
		table {%
				0 20.3481747431322
				1 22.3191235680433
				2 20.9944605958307
				3 21.6663578899561
				4 20.8510405460717
				5 19.9959287463658
				6 18.9644824490934
				7 17.8899953679421
				8 16.8552878240113
				9 15.9037175515636
				10 15.0483777141718
				11 14.28369725234
				12 13.5948044191514
				13 12.9633956625389
				14 12.3707490781104
				15 11.7989207542047
				16 11.2309507940667
				17 10.6505979108841
				18 10.0418896120705
				19 9.38863700788481
				20 8.6739944485503
				21 7.88011928343866
				22 6.98798918252481
				23 5.9774544751576
				24 4.82764035471914
				25 3.51787986892182
				26 2.02945790746401
				27 0.348516612700706
				28 -1.52963454559099
				29 -3.59495963034254
				30 -5.81671137924027
				31 -8.13866328980089
				32 -10.4778669582296
				33 -12.7291156790005
				34 -14.7758517726473
				35 -16.5058650422978
				36 -17.8280238628342
				37 -18.6856193007735
				38 -19.0629878832204
				39 -18.9843117758388
				40 -18.5058430332999
				41 -17.7043480462178
				42 -16.6649143670989
				43 -15.4705851596675
				44 -14.1951221943684
				45 -12.8990901915633
				46 -11.6287198524512
				47 -10.4166989118684
				48 -9.28405156416494
				49 -8.24244413425538
				50 -7.29647641166835
				51 -6.44571342037497
				52 -5.6863551693155
				53 -5.0125319931476
				54 -4.41726162304138
				55 -3.8931244608448
				56 -3.43271697589509
				57 -3.02893780962819
				58 -2.67515222292144
				59 -2.36527090631552
				60 -2.09377037243789
				61 -1.85567475107283
				62 -1.64651290894827
				63 -1.46226027954089
				64 -1.2992713823088
				65 -1.15420648962895
				66 -1.02395403722269
				67 -0.905548976264171
				68 -0.796086173608799
				69 -0.692627053346357
				70 -0.592096837687706
				71 -0.491168906180697
				72 -0.386131878390492
				73 -0.272733987704408
				74 -0.145998164854936
			};
	\end{groupplot}

\end{tikzpicture}

%% file: figures/robot_kl_over_time.tex
\begin{tikzpicture}

	\begin{axis}[
			height=5cm,
			width=8.5cm,
			tick pos=both,
			tick align=inside,
			minor tick num=4,
			xmajorticks=true,
			ymajorticks=true,
			xmajorgrids,
			ymajorgrids,
			xtick style={color=black},
			ytick style={color=black},
			x grid style={white!69.0196078431373!black},
			y grid style={white!69.0196078431373!black},
			ymin=1.86600252694796e-09, ymax=815.101186178018,
			ymode=log,
			xmin=-4.95, xmax=103.95,
			ylabel={$\kl(p_{t} \,||\, \hat{p})$},
			xlabel={Time Steps},
			every major tick/.append style={major tick length=5pt, black},
			every minor tick/.append style={minor tick length=3pt, black},
			major grid style={dashed, gray},
		]

		\addplot [black, mark=*, mark size=1, mark options={solid}]
		table {%
				0 241.049931063879
				1 198.0962372526
				2 156.925809163014
				3 121.404106975875
				4 91.7910079538807
				5 67.8558868803152
				6 49.0765392085339
				7 34.7607826961586
				8 24.1510749982688
				9 16.5036886069786
				10 11.1422932894169
				11 7.48733520721035
				12 5.06589808532749
				13 3.50796519281514
				14 2.53468791648975
				15 1.9431248781297
				16 1.59051735080899
				17 1.37990895853388
				18 1.24796179599096
				19 1.15518778406794
				20 1.07844827828149
				21 1.00540085941792
				22 0.930520170513709
				23 0.852336022879864
				24 0.771580174630852
				25 0.689991003351096
				26 0.609580994583496
				27 0.532220325649737
				28 0.459429268617733
				29 0.392302959080732
				30 0.331515399635789
				31 0.277366763037026
				32 0.229850459494186
				33 0.188725189064753
				34 0.153583252898391
				35 0.123910480976036
				36 0.0991358077668583
				37 0.0786702075171704
				38 0.0619356931597714
				39 0.0483856081189238
				40 0.0375176562182808
				41 0.0288811327426686
				42 0.0220797173352203
				43 0.0167710214334313
				44 0.0126638879488716
				45 0.00951424610581242
				46 0.00712014800316041
				47 0.00531646540022912
				48 0.00396960697613302
				49 0.00297252207592003
				50 0.00224017669586374
				51 0.00170561187195872
				52 0.00131662006387145
				53 0.00103300600263445
				54 0.000824345805520821
				55 0.000668133561395834
				56 0.00054821198025401
				57 0.000453413456385832
				58 0.000376369288936473
				59 0.000312461488904958
				60 0.000258903596055404
				61 0.0002139718947749
				62 0.000176448964914755
				63 0.000145319201598326
				64 0.000119669853349436
				65 9.86890926757411e-05
				66 8.16652652595451e-05
				67 6.79554750764311e-05
				68 5.69534033427743e-05
				69 4.8093015768913e-05
				70 4.08821377675395e-05
				71 3.49355164175336e-05
				72 3.83933081931076e-05
				73 2.29513816698557e-05
				74 1.85073754046527e-05
				75 1.49243097489205e-05
				76 1.21469792535578e-05
				77 1.00599494992082e-05
				78 8.56059621767713e-06
				79 7.6131261810275e-06
				80 7.1870104783045e-06
				81 5.86164608229467e-06
				82 4.38982774220165e-06
				83 3.29537802912228e-06
				84 2.49506027749646e-06
				85 1.90637305230723e-06
				86 1.47107868464502e-06
				87 1.14959245500756e-06
				88 9.20563067552393e-07
				89 8.35236676266504e-07
				90 1.98509625626286e-06
				91 5.61525981268574e-07
				92 4.28077129299709e-07
				93 3.20002055076429e-07
				94 2.31581466891839e-07
				95 1.59443439784468e-07
				96 1.01586669742915e-07
				97 5.69918938708724e-08
				98 2.52778473708304e-08
				99 6.30981666915886e-09
			};
	\end{axis}

\end{tikzpicture}

%% file: figures/robot_cost_over_distance.tex
\begin{tikzpicture}

	\begin{axis}[
			height=5cm,
			width=8.5cm,
			tick pos=both,
			tick align=inside,
			minor tick num=8,
			xmajorticks=true,
			ymajorticks=true,
			xmajorgrids,
			ymajorgrids,
			xtick style={color=black},
			ytick style={color=black},
			x grid style={white!69.0196078431373!black},
			y grid style={white!69.0196078431373!black},
			ymin=234.560883617653, ymax=14859.4058148725,
			ymode=log,
			xmin=-280.416624183068, xmax=5888.74910784443,
			ylabel={Expected Cost},
			xlabel={Distance to Nominal Distribution},
			every major tick/.append style={major tick length=5pt, black},
			every minor tick/.append style={minor tick length=3pt, black},
			major grid style={dashed, gray},
            /pgf/number format/.cd, use comma, 1000 sep={},
            legend style={at={(1.,0.)},anchor=south east}
        ]

		\addplot [blue, mark=*, mark size=2, mark options={solid}, line width=0.25mm]
		table {%
				0 283.239049202333
				8.44549677137342 318.329617894085
				34.5419581102756 360.379853961653
				79.4989257338104 411.051216946879
				144.626072872646 472.452965461084
				231.343299747019 547.274170476579
				341.192046137766 638.95757075451
				475.847993238198 751.929390591698
				637.135355107992 891.904284998463
				827.042993426617 1066.29152444415
				1047.74262896353 1284.73813576159
				1301.60947060527 1559.85804858692
				1591.2456395927 1908.21488292088
				1919.50683492556 2351.65203494597
				2289.53276831185 2919.10031155716
				2704.78199687625 3649.04505980177
				3169.07190325698 4592.90811332507
				3686.62472099235 5819.70453385485
				4262.12068495684 7422.48414827075
				4900.75961066858 9527.28405728039
				5608.33248366136 12305.6314720219
			};
		\addlegendentry{Stochastic}
		
		\addplot [red, mark=square, mark size=2, mark options={solid}, line width=0.25mm]
		table {%
				0 354.239874921799
				8.44549677137342 373.709632124062
				34.5419581102756 395.069701026232
				79.4989257338104 418.555869212145
				144.626072872646 444.439169778519
				231.343299747019 473.031976965846
				341.192046137766 504.695295477389
				475.847993238198 539.847504342232
				637.135355107992 578.974879136238
				827.042993426617 622.644295985512
				1047.74262896353 671.518621840762
				1301.60947060527 726.375424326689
				1591.2456395927 788.129799344778
				1919.50683492556 857.862326544134
				2289.53276831185 936.853436421683
				2704.78199687625 1026.6258277688
				3169.07190325698 1128.99703680532
				3686.62472099235 1246.14486527651
				4262.12068495684 1380.68917250484
				4900.75961066858 1535.79459217874
				5608.33248366136 1715.30013975374
			};
    \addlegendentry{Robust}
    
	\end{axis}

\end{tikzpicture}

%% file: main.bbl
\begin{thebibliography}{10}
\providecommand{\url}[1]{#1}
\csname url@samestyle\endcsname
\providecommand{\newblock}{\relax}
\providecommand{\bibinfo}[2]{#2}
\providecommand{\BIBentrySTDinterwordspacing}{\spaceskip=0pt\relax}
\providecommand{\BIBentryALTinterwordstretchfactor}{4}
\providecommand{\BIBentryALTinterwordspacing}{\spaceskip=\fontdimen2\font plus
\BIBentryALTinterwordstretchfactor\fontdimen3\font minus
  \fontdimen4\font\relax}
\providecommand{\BIBforeignlanguage}[2]{{%
\expandafter\ifx\csname l@#1\endcsname\relax
\typeout{** WARNING: IEEEtran.bst: No hyphenation pattern has been}%
\typeout{** loaded for the language `#1'. Using the pattern for}%
\typeout{** the default language instead.}%
\else
\language=\csname l@#1\endcsname
\fi
#2}}
\providecommand{\BIBdecl}{\relax}
\BIBdecl

\bibitem{mayne1966second}
D.~Mayne, ``A second-order gradient method for determining optimal trajectories
  of nonlinear discrete-time systems,'' \emph{International Journal of
  Control}, 1966.

\bibitem{kamthe2018data}
S.~Kamthe and M.~Deisenroth, ``Data-efficient reinforcement learning with
  probabilistic model predictive control,'' in \emph{International Conference
  on Artificial Intelligence and Statistics}, 2018.

\bibitem{hewing2018cautious}
L.~Hewing, A.~Liniger, and M.~N. Zeilinger, ``Cautious nmpc with {G}aussian
  process dynamics for autonomous miniature race cars,'' in \emph{IEEE European
  Control Conference}, 2018.

\bibitem{scarf1958min}
H.~Scarf, ``A min-max solution of an inventory problem,'' \emph{Studies in The
  Mathematical Theory of Inventory and Production}, 1958.

\bibitem{delage2010distributionally}
E.~Delage and Y.~Ye, ``Distributionally robust optimization under moment
  uncertainty with application to data-driven problems,'' \emph{Operations
  Research}, 2010.

\bibitem{van2015distributionally}
B.~P. Van~Parys, D.~Kuhn, P.~J. Goulart, and M.~Morari, ``Distributionally
  robust control of constrained stochastic systems,'' \emph{IEEE Transactions
  on Automatic Control}, 2015.

\bibitem{coulson2019regularized}
J.~Coulson, J.~Lygeros, and F.~D{\"o}rfler, ``Regularized and distributionally
  robust data-enabled predictive control,'' in \emph{IEEE Conference on
  Decision and Control}, 2019.

\bibitem{yang2020wasserstein}
I.~Yang, ``Wasserstein distributionally robust stochastic control: {A}
  data-driven approach,'' \emph{IEEE Transactions on Automatic Control}, 2020.

\bibitem{zhu20worst}
J.-J. Zhu, W.~Jitkrittum, M.~Diehl, and B.~Sch{\"o}lkopf, ``Worst-case risk
  quantification under distributional ambiguity using kernel mean embedding in
  moment problem,'' in \emph{IEEE Conference on Decision and Control}, 2020.

\bibitem{coppens2020data}
P.~Coppens, M.~Schuurmans, and P.~Patrinos, ``Data-driven distributionally
  robust lqr with multiplicative noise,'' in \emph{Learning for Dynamics and
  Control}, 2020.

\bibitem{rahimian2019distributionally}
H.~Rahimian and S.~Mehrotra, ``Distributionally robust optimization: {A}
  review,'' \emph{arXiv preprint arXiv:1908.05659}, 2019.

\bibitem{hu2013kullback}
Z.~Hu and L.~J. Hong, ``{K}ullback-{L}eibler divergence constrained
  distributionally robust optimization,'' \emph{Optimization Online}, 2013.

\bibitem{charalambous2007stochastic}
C.~D. Charalambous and F.~Rezaei, ``Stochastic uncertain systems subject to
  relative entropy constraints: Induced norms and monotonicity properties of
  minimax games,'' \emph{IEEE Transactions on Automatic Control}, 2007.

\bibitem{peters2010relative}
J.~Peters, K.~M{\"u}lling, and Y.~Altun, ``Relative entropy policy search,'' in
  \emph{AAAI Conference on Artificial Intelligence}, 2010.

\bibitem{kim2020minimax}
K.~Kim and I.~Yang, ``Minimax control of ambiguous linear stochastic systems
  using the {W}asserstein metric,'' in \emph{IEEE Conference on Decision and
  Control}, 2020.

\bibitem{petersen2000minimax}
I.~R. Petersen, M.~R. James, and P.~Dupuis, ``Minimax optimal control of
  stochastic uncertain systems with relative entropy constraints,'' \emph{IEEE
  Transactions on Automatic Control}, 2000.

\bibitem{nishimura2021rat}
H.~Nishimura, N.~Mehr, A.~Gaidon, and M.~Schwager, ``{RAT}-i{LQR}: {A} risk
  auto-tuning controller to optimally account for stochastic model mismatch,''
  \emph{IEEE Robotics and Automation Letters}, 2021.

\bibitem{farshidian2015risk}
F.~Farshidian and J.~Buchli, ``Risk sensitive, nonlinear optimal control:
  {I}terative linear exponential-quadratic optimal control with {G}aussian
  noise,'' \emph{arXiv preprint arXiv:1512.07173}, 2015.

\bibitem{nass2019entropic}
D.~Nass, B.~Belousov, and J.~Peters, ``Entropic risk measure in policy
  search,'' in \emph{IEEE/RSJ International Conference on Intelligent Robots
  and Systems}, 2019.

\bibitem{levine2013guided}
S.~Levine and V.~Koltun, ``Guided policy search,'' in \emph{International
  Conference on Machine Learning}, 2013.

\bibitem{todorov2005generalized}
E.~Todorov and W.~Li, ``A generalized iterative lqg method for locally-optimal
  feedback control of constrained nonlinear stochastic systems,'' in \emph{IEEE
  American Control Conference}, 2005.

\bibitem{arenz2016optimal}
O.~Arenz, H.~Abdulsamad, and G.~Neumann, ``Optimal control and inverse optimal
  control by distribution matching,'' in \emph{IEEE/RSJ International
  Conference on Intelligent Robots and Systems}, 2016.

\bibitem{beck2003mirror}
A.~Beck and M.~Teboulle, ``Mirror descent and nonlinear projected subgradient
  methods for convex optimization,'' \emph{Operations Research Letters},
  vol.~31, no.~3, pp. 167--175, 2003.

\bibitem{boyd2004convex}
S.~Boyd and L.~Vandenberghe, \emph{Convex Optimization}, 2004.

\bibitem{abdulsamad2017state}
H.~Abdulsamad, O.~Arenz, J.~Peters, and G.~Neumann, ``State-regularized policy
  search for linearized dynamical systems,'' in \emph{International Conference
  on Automated Planning and Scheduling}, 2017.

\bibitem{nocedal2006numerical}
J.~Nocedal and S.~Wright, \emph{Numerical Optimization}, 2006.

\bibitem{solin2010cubature}
A.~Solin, ``Cubature integration methods in nonlinear {K}alman filtering and
  smoothing,'' 2010.

\bibitem{wan2001unscented}
E.~A. Wan, R.~Van Der~Merwe, and S.~Haykin, ``The unscented {K}alman filter,''
  \emph{Kalman Filtering and Neural Networks}, 2001.

\end{thebibliography}
